\newif\ifpreambleRelatedVersionLinkShow
\newif\ifpreambleNoLineNumbers
\newif\ifpreambleLipicsHide
\newif\ifpreamblePdfoutputYes
    \newif\ifbodyRefLocalAppendix
    \newif\ifbodyCiteRelatedVersion
    \newif\ifappendix

\preambleRelatedVersionLinkShowfalse       
\preambleNoLineNumberstrue                 
\preambleLipicsHidetrue                    
\preamblePdfoutputYestrue                  
\bodyRefLocalAppendixtrue                  
\bodyCiteRelatedVersionfalse               
\appendixtrue

\documentclass[a4paper,UKenglish,cleveref, autoref, thm-restate]{lipics-v2021}

\pdfoutput=1 
\hideLIPIcs

\title{Relational Algebras for Subset Selection and Optimisation}

\author{David Robert Pratten}{University of Technology Sydney, Australia}{david.r.pratten@student.uts.edu.au}{https://orcid.org/0000-0001-9210-9529}{}

\author{Fahimeh Ramezani}{University of Technology Sydney, Australia}{Fahimeh.Ramezani@uts.edu.au}{https://orcid.org/0000-0002-0368-321X}{}
\author{Luke Mathieson}{University of Technology Sydney, Australia}{Luke.Mathieson@uts.edu.au}{https://orcid.org/0000-0001-6470-2296}{}
\authorrunning{Pratten D.R., Mathieson L., Ramezani, F.}

\Copyright{David Robert Pratten, Luke Mathieson, Fahimeh Ramezani}

\acknowledgements{David thanks David S. Warren, Kaustubh Beedkar, Peter J. Stuckey, and Philip Wadler  for their encouragement and idea-sharpening questions.}



\category{}

\nolinenumbers

\EventEditors{John Q. Open and Joan R. Access}
\EventNoEds{2}
\EventLongTitle{42nd Conference on Very Important Topics (CVIT 2016)}
\EventShortTitle{CVIT 2016}
\EventAcronym{CVIT}
\EventYear{2016}
\EventDate{December 24--27, 2016}
\EventLocation{Little Whinging, United Kingdom}
\EventLogo{}
\SeriesVolume{42}
\ArticleNo{0}

\usepackage{float}
\usepackage{tikz}
\usetikzlibrary{decorations.pathreplacing}
\usepackage{array}
\usetikzlibrary{positioning, shadows,calc,shapes.geometric}
\usepackage{tabularx}
\usepackage{comment}
\usepackage{amsfonts}
\usepackage{amsthm}
\usepackage[ruled,vlined]{algorithm2e}
\usepackage{xcolor}
\usepackage[disable]{todonotes}
\usepackage{dirtytalk}
\usepackage{algorithmic}
\usepackage{placeins}
\usepackage{multirow}

\usepackage{listings}
\lstdefinelanguage{RA}{
  basicstyle=\ttfamily\footnotesize,
  commentstyle=\color{gray}\itshape,
  mathescape=true,
  columns=fullflexible,
  keepspaces=true,
  breaklines=true,
  showstringspaces=false,
  comment=[l]{--}
}

\lstdefinelanguage{SQL}{
  basicstyle=\ttfamily\footnotesize,
  commentstyle=\color{gray}\itshape,
  mathescape=true,
  columns=fullflexible,
  keepspaces=true,
  breaklines=true,
  showstringspaces=false,
  comment=[l]{--}
}

\lstdefinelanguage{MiniZinc}{
  basicstyle=\ttfamily\footnotesize,
  commentstyle=\color{gray}\itshape,
  mathescape=true,
  columns=fullflexible,
  keepspaces=true,
  breaklines=true,
  showstringspaces=false,
  comment=[l]{\%}
}

\lstdefinelanguage{JSON}{
  basicstyle=\ttfamily\footnotesize,
  commentstyle=\color{gray}\itshape,
  mathescape=true,
  columns=fullflexible,
  keepspaces=true,
  breaklines=true,
  showstringspaces=false,
  string=[b]"
}

\usepackage{tikz}
\usetikzlibrary{decorations.pathreplacing}

\newcommand{\codett}[1]{\texttt{#1}}

\begin{document}

\maketitle

\begin{abstract}

The database community lacks a unified relational query language for subset selection and optimisation queries, limiting both user expression and query optimiser reasoning about such problems. Decades of research (latterly under the rubric of prescriptive analytics) have produced powerful evaluation algorithms with incompatible, ad-hoc SQL extensions that specify and filter through distinct mechanisms. We present the first unified algebraic foundation for these queries, introducing relational exponentiation to complete the fundamental algebraic operations alongside union (addition) and cross product (multiplication). First, we extend relational algebra to complete domain relations---relations defined by characteristic functions rather than explicit extensions---achieving the expressiveness of NP-complete/hard problems, while simultaneously providing query safety for finite inputs. Second, we introduce solution sets, a higher-order relational algebra over sets of relations that naturally expresses search spaces as functions $f: Base \rightarrow Decision$, yielding $|Decision|^{|Base|}$ candidate relations. Third, we provide structure-preserving translation semantics from solution sets to standard relational algebra, enabling mechanical translation to existing evaluation algorithms. This framework achieves the expressiveness of the most powerful prior approaches while providing the theoretical clarity and compositional properties absent in previous work. We demonstrate the capabilities these algebras open up through a polymorphic SQL where standard clauses seamlessly express data management, subset selection, and optimisation queries within a single paradigm.

\end{abstract}

\section{Introduction} \label{sec:introduction}
Enterprises regularly track \say{what is} by using data management systems grounded in relational algebra and SQL, managing global databases with millions to billions of tuples.  At the same time, these same enterprises choose between possible futures, considering what \say{might be} or \say{should be} using subset selection algorithms and constraint-solving (CP, LP, SMT, SAT) optimisation models with thousands to millions of variables.

These problems are common and diverse~\cite{moesmann_2024_data.driven.prescriptive.analytics.applications}. Analysts solve \textbf{subset selection} problems---such as Pareto-optimal portfolios or diverse recommendation sets---selecting what \say{should be} from exponentially many collections (packages) of \say{what is} in the database, where the subset as a whole must satisfy global properties like diversity or coverage. At scale, logistics planners solve \textbf{optimisation} problems choosing which packages to assign to which vehicles and routes to minimise total cost within available capacity. In each case, these decisions update the database to inform subsequent optimisation steps. Data management and combinatorial optimisation are inextricably linked, and research into this has been gathering momentum under the rubric of \say{Prescriptive Analytics}
(PSA)~\cite{meliou_2025_data.management.perspectives.prescriptive.analytics}.

This research thread in the database context has achieved considerable success in evaluation algorithms for subset selection and optimisation (e.g., search over infinite INT attributes~\cite{brucato_2016_scalable.package.queries.relational.database}, stochastic queries~\cite{brucato_2020_stochastic.packagequeries.in.probabilistic.databases}, dual simplex algorithm scaling to billions of tuples~\cite{mai_2023_scaling.package.queries.billion.tuples}, and integration with query optimisers and search over infinite FLOAT attributes~\cite{siksnys_2016_solvedb.optimization.problem.solvers.sql}). However, these powerful algorithms appear trapped in silos—the underlying semantic foundations remain ad hoc.

We believe it's time to address the fragmentation in subset selection and optimisation queries. Just as the recent formalisation of Graph Query Language (GQL) unified diverse graph query approaches~\cite{rogova_2023_a.researchers.digest.of.gql}, the database community can similarly make existing sophisticated evaluation algorithms more accessible to users. This paper provides a unified algebraic foundation for such a language.

Relational exponentiation is not just a metaphor; it is an algebraic operation that provides increased expressiveness and greater scope for the query optimiser. Consider that values and the union operator can specify sets of values. Values, union, and cross product give sets of tuples with a schema (relations), which are the basis for data queries. And those values, together with union, cross product, and exponentiation, build sets of relations with a shared schema (our term \say{solution sets}).  As we shall see, solution sets are a key to unlock the specification challenge of prescriptive analytics queries, naturally expressing search spaces as functions $f: Base \rightarrow Decision$. These relational algebraic foundations enable the query optimiser to systematically select evaluation strategies for subset selection and optimisation queries based on problem characteristics, leveraging set-based algorithms when data volume dominates or constraint-solving algorithms when combinatorial complexity dominates~\cite{meliou_2025_data.management.perspectives.prescriptive.analytics}.

Constructing solution sets via exponentiation requires two capabilities absent from traditional relational algebra: the ability to express unknowns in problem specifications (what constraint programming calls \say{variables}), and the ability to reason about potentially infinite domains when decision spaces are unbounded. This necessitates extending relational algebra. Our three interdependent contributions then are:
\begin{enumerate}
    \item \textbf{A principled generalisation of relational algebra} that highlights the ability of characteristic functions to express both the domain semantics of finite data relations (our term \say{active domain relations}) and potentially infinite constraint-solving relations (our term \say{complete domain relations}) within a single framework. We show that the complete-domain algebra achieves the expressiveness of NP-complete/hard problems and contains the active-domain algebra as a finite, domain‑independent fragment that preserves query safety.
    \item \textbf{A higher-order relational algebra over solution sets.} $RA_{sol}$ expresses and manipulates subset selection and optimisation search spaces created via relational exponentiation.
    \item \textbf{We fix the semantics of solution sets} through formal translation to relational algebra. This shows that evaluation of subset selection and optimisation queries does not require interpretive semantics beyond that available in relational algebra.
\end{enumerate}

Evaluating these algebras spans the competencies of multiple communities of practice—database systems, constraint programming, operations research, and satisfiability solving, each with a distinct vocabulary. We use \emph{query evaluation} to encompass all such computational approaches, including what various communities call query optimisation, solving, execution, or search.

\begin{figure}[!htbp]
    \centering
    \includegraphics[width=0.8\linewidth]{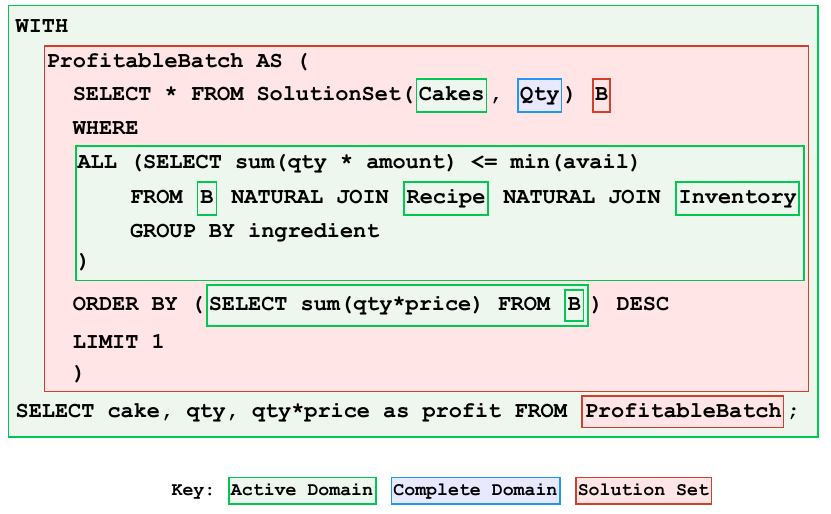}
    \caption{Profitable Cakes Batch demonstrated in polymorphic SQL}
    \label{fig:sql-cakes-intro}
\end{figure}

The algebras compose cleanly enough to demonstrate their integration through polymorphic SQL using only standard relational operators—no new keywords and no special clauses. To illustrate our destination, \autoref{fig:sql-cakes-intro} shows how active domain relations, complete domain relations and solution sets come together within a single polymorphic expression to solve a profitable cakes batch problem. For details, see \autoref{sec:validation-through-representative-problems}.

This paper formally and accessibly extends relational algebraic foundations to encompass subset selection and optimisation queries. Given this breadth, we prioritise precise formal definitions and clear connections showing how each component builds upon and relates to the others. 
Beyond proving query safety, we defer detailed proofs of properties and characterisation of computability, decidability, and termination—these can build on established theoretical results.

We structure the remainder of this paper as follows. We highlight the role of characteristic functions in traditional relational algebra and establish other foundational concepts for active domain relations in \autoref{sec:preliminaries-active-domain-relations}. We then extend this foundation in \autoref{sec:an-algebra-for-complete-domain-relations} by developing a principled algebra for complete domain relations that maintains query safety while achieving NP-complete/hard expressiveness. Building on these foundations, \autoref{sec:proposed-higher-order-algebra-for-solution-sets} introduces solution sets (sets of relations specified by exponentiation) and their higher-order relational algebra, with \autoref{sec:semantics-translating-solution-sets-to-relations} providing formal semantics through translation to relational algebra. \autoref{sec:validation-through-representative-problems} presents a worked example showing how a production planning problem may be parameterised by a database. \autoref{sec:related-work} reviews the foundational work from the last fifty years that our proposal builds on, and from that vantage point, we summarise the expressiveness of our unified framework in \autoref{sec:expressiveness-of-solution-sets}, demonstrating equivalence to the most expressive prior approaches while increasing theoretical clarity. We conclude in \autoref{sec:conclusion-and-future-work} with directions for future research and implementation.

\section{Preliminaries: Active Domain Relations (ADRs)} \label{sec:preliminaries-active-domain-relations}
For our purposes, a database $\mathfrak{D}$ is a set of relations. We use uppercase letters $R, S$ to denote database relations. A database relation $R \in \mathfrak{D}$ is a pair $\langle \alpha_R, \varepsilon_R\rangle$ where $\alpha_R$ is its attribute set with associated domain mappings and $\varepsilon_R$ its extension. We call elements $t \in \varepsilon_R$ tuples. $\text{dom}(a)$ is the domain of attribute $a$. Each tuple consists of $|\alpha_R|$ constants $c$, one for each attribute. The domain of $R$ equals $\text{dom}(R) = \prod_{a \in \alpha_R} \text{dom}(a)$. $\varepsilon_R$ is a finite subset of $\text{dom}(R)$ and this extension defines an active domain of values actually in the relation. In this paper, we refer to $R$ as an \textbf{active domain relation} due to its domain semantics~\cite{abiteboul_1994_foundations.of.databases}. $\mathfrak{A}$ is a set of aggregation functions. We write $agg \in \mathfrak{A}$ when the returned value type is not material, $boolAgg \in \mathfrak{A}$ for Boolean aggregation functions (\codett{BOOL\_OR}, \codett{BOOL\_AND}, \codett{AllDifferent}, \codett{hasSubset} etc.) 
 and $orderableAgg \in \mathfrak{A}$ for all aggregation operators that support the $<$ comparison operator.

We characterise combinatorial search spaces as the set of all functions ($f: Base \rightarrow Decision$), where the base set $Base$ represents the items that form the problem structure and the codomain $Decision$ represents possible choices for each item, which is a set of candidate solutions that is exponential in the cardinality of the base set---with cardinality $|Decision|^{|Base|}$~\cite{hrbacek_1978_introduction.to.set.theory}.

Through this lens, active domain relations (ADRs) exist within an exponentially large mathematical space. The size of $\text{dom}(R)$ is exponential in the size of the attribute set and potentially infinite if it includes domains such as INT and FLOAT. Within this space ADRs occupy the space consisting of all characteristic functions $f\colon \text{dom}(R) \to \{\text{True}, \text{False}\}$ where $|\{t \in \text{dom}(R) \colon f(t) = \text{True}\}| \leq k$ for some natural number $k$. We denote the characteristic function of relation $R$ as $\chi_R$, noting that $\chi_R(t) = \text{True}$ if and only if $t \in \varepsilon_R$. 

We assume the reader is familiar with relational algebra ($RA$), including natural join, cross product, intersection, difference, selection, projection, union, rename, aggregation, and limit denoted by $\bowtie, \times, \cap, -, \sigma, \pi, \cup, \rho, \gamma$ and $\lambda$ respectively. The ordering relational algebraic operator $\tau$ is an outer operator that returns a sequence rather than a set. We introduce the $\omega$ operator, a constructor patterned on SQL's \codett{CREATE TABLE} with immediate data insertion via \codett{VALUES}. Crucially, these operators maintain domain independence~\cite{abiteboul_1994_foundations.of.databases}—the result of each operation is not affected by extending the domains of the input relations beyond their active domains. This ensures the algebra can only express safe queries, returning finite results for finite input~\cite{gelder_1991_safety.and.translation.relational.calculus}. (We defer the notation for active domain $RA$ to Appendix \ref{app:additional-details-for-active-domain-relational-algebra}.)

We now turn our attention to our first challenge: increasing the expressivity of relational algebra without sacrificing query safety.

\section{Introducing an Algebra for Complete Domain Relations (CDRs)} \label{sec:an-algebra-for-complete-domain-relations}
This section extends relational algebra to complete domain relations,
achieving the expressiveness of NP-complete/hard problems while preserving
query safety for active domain operations.

\subparagraph*{Complete Domain Relations.}
Complete domain relations exist within the same exponentially large mathematical space as their active domain counterparts. This space consists of all characteristic functions $f\colon \text{dom}(R) \to \{\text{True}, \text{False}\}$ without a finiteness requirement. We term these \say{complete domain relations}  (CDRs) to emphasise the shift in domain semantics required to increase expressivity.

In parallel with our database $\mathfrak{D}$, we have $\mathfrak{C}$, a set of CDRs. We use uppercase letters $C, D$ to denote CDRs. A complete domain relation $C \in \mathfrak{C}$ is a pair $\langle \alpha_C, \chi_C \rangle$ where $\alpha_C$ is its attribute set and $\chi_C$ its characteristic function.

Notably, it is not the removal of the finiteness restriction that lifts the expressivity of CDRs to that of NP-Complete. The expressivity advantage arises because the relations start full (\say{everything is in unless excluded}). The full exponential (and possibly infinite) space is immediately accessible to be constrained in or out. Let's use 3-SAT to demonstrate NP-Completeness~\cite{garey_1979_computers.intractability.guide.np.completeness} using the $\omega$ constructor with a characteristic function instead of tuples.

Concerning the expressiveness of CDRs, consider the (NP-complete) 3-SAT formula $(x_1 \vee x_2 \vee x_3) \wedge (\neg x_1 \vee x_2 \vee \neg x_3) \wedge (x_1 \vee \neg x_2 \vee x_3)$. \autoref{fig:3-sat} shows how this formula can be transcribed as a complete domain relation with a characteristic function, in Conjunctive Normal Form (CNF)—the natural form for SAT problems, ready for an evaluation algorithm such as CDCL~\cite{marques.silva_2021_conflict.driven.clause.learning.sat}.

\begin{lstlisting}[language=RA, mathescape=true, caption={NP-Complete 3-SAT in Complete Domain $RA$},label=fig:3-sat, float=ht,
  abovecaptionskip=-\medskipamount]
ThreeSAT := $\omega$[x1:BOOL, x2:BOOL, x3:BOOL](
    (x1 OR x2 OR x3) AND (NOT x1 OR x2 OR NOT x3) AND (x1 OR NOT x2 OR x3)
)
\end{lstlisting}

Beyond the decision problems of SAT, this algebra can specify satisfaction and optimisation problems that are considered to be NP-Hard~\cite{garey_1979_computers.intractability.guide.np.completeness}. To give an intuition of this claim, consider a prototypical optimisation problem over domain $D$: $\text{min (or max)} f(x)$ subject to constraints $C(x)$ given $x \in D$. This translates naturally to $RA$ as: $\lambda[1](\tau[f(x)](\sigma[C(x)](D)))$ with $\tau$[f(x)] specifying the objective function. The limit $\lambda$ might be relaxed to specify a satisfaction problem. For more details, see Appendix \ref{subapp:constraint-solving-basis-in-relations}.

\subparagraph*{Relational Algebra for Complete Domain Relations.}
Here, we introduce a relational algebra over CDRs that inherits its properties directly from Boolean algebra.

Given CDRs $C$ and $D$, and recalling that we define them respectively as an attribute set and characteristic function pair $\langle \alpha_C, \chi_C \rangle$ and $\langle \alpha_D, \chi_D \rangle$, and given $\theta_C$ an arbitrary Boolean expression over $\alpha_C$, and $\alpha_\oplus$ a non-empty subset of attributes in $\alpha_C$ for operator $\oplus$: \autoref{tab:relational-operators-for-complete-relations} shows the algebra's operators along with their definitions. Projection is a special case and will be treated separately. Examples of the algebra in action are deferred to Appendix \ref{subapp:complete-domain-relational-algebra-example}.

The closure of the core operators over CDRs may be established by observing that each operator is defined to return a pair $\langle \alpha, \chi \rangle$ and that their inputs are closed over their types. A set operation on two attribute sets is an attribute set. A Boolean operation on two characteristic functions is a characteristic function. Thus, we may conclude that core operators are closed over CDRs.

The algebraic properties of these operators (Associativity, Distributivity, and Commutativity, ...) may be derived directly from their definitions and follow those of conjunction (natural join), disjunction (union) and conjunction with a negated conjunct(difference).

\begin{table}[!htbp]
    \centering
        \caption{Relational Operators for Complete Domain Relations}
    \begin{tabularx}{\textwidth}{|>{\raggedright\arraybackslash}X|>{\raggedright\arraybackslash}X|} \hline 
        \textbf{Operator $\oplus$ Notation}  & \textbf{Definition} \\ \hline
        Create Complete Domain Relation $\omega[\text{a}_1{:}\text{d}_1, $ $ \text{a}_2{:}\text{d}_2, ..., $ $ \text{a}_i{:}\text{d}_i](\chi=True) $ where $a_i$ are attribute identifiers and $d_i$ are domain specifications, and $\chi$ defaults to $True$. & $\langle A, \chi \rangle : A = \{\text{a}_1{:}\text{d}_1, $ $\text{a}_2{:}\text{d}_2, ... $ $\text{a}_i{:}\text{d}_i\}$\\ \hline
        Natural-Join $C \bowtie D$  & $\langle \alpha_C\cup \alpha_D, \chi_C \wedge \chi_D \rangle$    \\ \hline
        Cross-Product $C \times D : \alpha_C \cap \alpha_D = \varnothing  $ & $C \bowtie D$  \\ \hline
        Intersection $C \cap D : \alpha_C = \alpha_D $ &  $C \bowtie D$   \\ \hline
        Difference $C - D : \alpha_C = \alpha_D $ & $C \bowtie  \langle \alpha_C, \neg \chi_D \rangle$ which simplifies to $\langle \alpha_C, \chi_C \wedge \neg \chi_D \rangle$  \\ \hline
Selection $\sigma[\theta_C](C)$ &  $C \bowtie  \langle \alpha_C, \theta_C \rangle$ which simplifies to $\langle \alpha_C, \chi_C \wedge \, \theta_C \rangle$ \\[1pt] \hline
        Union $C \cup D : \alpha_C = \alpha_D $ & $\langle \alpha_C, \chi_C \vee \chi_D \rangle$  \\ \hline
        Rename $\rho[renamespec](C)$ &  $\langle \rho[renamespec](A_C),$ $ \rho[renamespec](\chi_C) \rangle$    \\ \hline
        (Outer) Order $\tau[\alpha_\tau] (C)$ & Yields a sequence (not a relation) that is ordered by the attributes in $\alpha_\tau \subseteq \alpha_C$ \\ \hline
        (Outer) Limit $\lambda[n](C)$ &  Restricts cardinality to at most $n$  \\ \hline
        \end{tabularx}

    \label{tab:relational-operators-for-complete-relations}
\end{table}

    \subparagraph*{Query Safety} Complete domain algebra is more expressive without compromising query safety properties: ADRs occupy a domain-independent region with characteristic functions in finite Disjunctive Normal Form (DNF) with every attribute in every tuple equated to a constant. And this region remains reachable in polynomial time under CDR operations (proof in Appendix~\ref{appsubsec:query-safety-in-the-complete-domain-algebra}). In this region, the domain is masked because every value is set by equality; changing an attribute $a$'s domain from \codett{1..10} to \codett{1..100} will not affect the value of the relation.

    \subparagraph*{Projection} plays a unique role as the transition operator from complete to active domains. While other operators preserve complete domain closure through Boolean operations on characteristic functions, projection requires evaluating the query to determine which values actually participate in tuples. We therefore define projection as the boundary where evaluation must
    occur, transitioning from complete domain to active domain semantics. Projection applied after the Limit $\lambda$ operator ensures finite results from potentially infinite domains and is closed over ADRs. (We defer detailed justification of projection's role to Appendix~\ref{appsubsec:projection-as-a-transition-operator}).

    While our upcoming second contribution does \textbf{not} depend on such joins, the interested reader may find a discussion of safely joining ADRs with CDRs and pointers to recent research in Appendix~\ref{appsubsec:safely-joining-active-domain-and-complete-domain-relations}.

With a full complement of relational algebraic operators that operate over CDRs, providing for safe queries if all the inputs are ADRs, and also expressing NP-complete/hard, satisfaction, and optimisation problems, we have now made our first contribution. From here, we proceed to building on both active and complete domain relations to express subset selection and optimisation problems with solution sets and to fix their semantics.

\section{Proposed Higher-Order Algebra for Solution Sets} \label{sec:proposed-higher-order-algebra-for-solution-sets}

This section introduces solution sets (sets of relations specified by exponentiation) and their higher-order relational algebra. Solution sets directly express the combinatorial search spaces fundamental to subset selection and optimisation: the set of all functions $f: Base \rightarrow Decision$, where Base represents problem structure and Decision represents choices, yielding $|Decision|^{|Base|}$ candidate solutions. 

We use a simple two-by-two Latin square puzzle as a pedagogical example throughout, containing essential elements while avoiding combinatorial complexity. The puzzle consists of just four cells arranged in a two-by-two pattern. Each cell may contain a value from the set $\{1,2\}$. All values in each row and each column must be different. We are required to find the solution with a $1$ in the top left corner. We model the Latin square as three relations in \autoref{fig:modelling-latin-square}: a decision (complete domain) relation \codett{Values} that models the choices for each cell, a base (active domain) relation \codett{Board} that models the board, and a required values active domain relation \codett{ReqValues} showing a $1$ in the top left $(1,1)$ cell. The decision relation \codett{Values} defines the domain {1,2} that constrains all other relations. In what follows \codett{IN }$\pi$\codett{[value](Values)} is analogous to SQL's \codett{REFERENCES Values(value)}.

\begin{lstlisting}[mathescape=true, language=RA, caption={Relational Model for Latin Square},label=fig:modelling-latin-square,float=ht,
  abovecaptionskip=-\medskipamount]
Values := $\omega$[value: 1..2](True)
-- For Latin squares, board dimensions equal |Values| by definition
Board := $\pi$[row,col]($\omega$[row:IN $\pi$[value](Values),col:IN $\pi$[value](Values)](True))
ReqValues := $\omega$[row: IN $\pi$[value](Values),col:IN $\pi$[value](Values),
    value:IN $\pi$[value](Values)]({$\langle$1,1,1$\rangle$})
\end{lstlisting}

Solution sets are introduced in three steps, preliminaries(\autoref{subsec:preliminaries-subset-selection-and-optimisation}), solution sets as a structure(\autoref{subsec:the-solution-set-abstraction}), and finally their algebra  $RA_{sol}$(\autoref{subsec:relational-algebra-over-solution-sets}).

\subsection{Preliminaries: Exponentiation, Aggregation and Global Constraints} \label{subsec:preliminaries-subset-selection-and-optimisation}

In the broader world of set theory, raising a set to the power of another is an ordinary operation, quoting Hrbacek \say{The \say{exponentiation} of sets is related to \say{multiplication} of sets in the same way as similar operations on numbers are related.}~\cite{hrbacek_1978_introduction.to.set.theory}. Let's apply the idea to our Latin square puzzle. The search space is all functions $f: Board \rightarrow Values$ , yielding $|Values|^{|Board|} = 2^4 = 16$ possible functions. This search space is shown in \autoref{fig:solution-set-for-latin-square} as a set of functions. Note, each function is represented by an active domain relation with attribute set $\alpha_{Board} \cup \alpha_{Values}$ in this case $\{row, col, value\}$, one relation for each possible functional extension of $Base$ with values from \codett{Values}.

\begin{figure}[!htbp]
\centering
\begin{tikzpicture}[
    table/.style={
        draw,
        thick,
        fill=white,
        drop shadow={shadow xshift=0.5ex, shadow yshift=-0.5ex, opacity=0.3},
        inner sep=0pt
    }
]

\newcommand{\tablecontentParam}[5]{
    \begin{tabular}{|c|c|c|}
        \hline
        \multicolumn{3}{|c|}{\textbf{Function #1}} \\
        \hline
        \makebox[0.3cm]{row} & \makebox[0.3cm]{col} & \makebox[0.3cm]{value} \\
        \hline
        1 & 1  & #2 \\
        1 & 2  & #3 \\
        2 & 1  & #4 \\
        2 & 2  & #5 \\

        \hline
    \end{tabular}
}

\node[table, scale=0.5, opacity=1] (table16) at (5.0,1.0) {
    \tablecontentParam{16}
            {2}
        {2}
        {2}
        {2}
};

\node[table, scale=0.52, opacity=1] (table15) at (4.84,0.97) {
    \tablecontentParam{15}
        {1}
        {1}
        {2}
        {1}

};

\node[table, scale=0.53, opacity=1] (table14) at (4.66,0.93) {
    \tablecontentParam{14}
        {1}
        {1}
        {2}
        {1}

};

\node[table, scale=0.55, opacity=1] (table13) at (4.46,0.89) {
    \tablecontentParam{13}
        {1}
        {1}
        {2}
        {1}

};

\node[table, scale=0.58, opacity=1] (table12) at (4.25,0.85) {
    \tablecontentParam{12}
        {1}
        {1}
        {2}
        {1}

};

\node[table, scale=0.6, opacity=1] (table11) at (4.01,0.8) {
    \tablecontentParam{11}
        {1}
        {1}
        {2}
        {1}

};

\node[table, scale=0.63, opacity=1] (table10) at (3.74,0.75) {
    \tablecontentParam{10}
        {1}
        {1}
        {2}
        {1}

};

\node[table, scale=0.65, opacity=1] (table9) at (3.45,0.69) {
    \tablecontentParam{9}
        {2}
        {1}
        {2}
        {1}

};

\node[table, scale=0.69, opacity=1] (table8) at (3.13,0.63) {
    \tablecontentParam{8}
        {1}
        {1}
        {2}
        {1}

};

\node[table, scale=0.72, opacity=1] (table7) at (2.78,0.56) {
    \tablecontentParam{7}
        {1}
        {1}
        {2}
        {1}

};
\node[table, scale=0.76, opacity=1] (table6) at (2.39,0.48) {
    \tablecontentParam{6}
        {1}
        {1}
        {2}
        {1}

};
\node[table, scale=0.8, opacity=1] (table5) at (1.97,0.39) {
    \tablecontentParam{5}
        {1}
        {2}
        {1}
        {1}

};

\node[table, scale=0.85, opacity=1] (table4) at (1.51,0.3) {
    \tablecontentParam{4}
        {1}
        {1}
        {2}
        {2}

};

\node[table, scale=0.9, opacity=1] (table3) at (1.03,0.21) {
    \tablecontentParam{3}
        {1}
        {1}
        {2}
        {1}

};

\node[table, scale=0.95, opacity=1] (table2) at (0.52,0.1) {
    \tablecontentParam{2}
        {1}
        {1}
        {1}
        {2}

};

\node[table, scale=1.0] (table1) at (0.0,0.0) {
    \tablecontentParam{1}
        {1}
        {1}
        {1}
        {1}

};

\end{tikzpicture}

\caption{Relational Exponentiation of \codett{Values} to the power \codett{Board}}
\label{fig:solution-set-for-latin-square}
\end{figure}

Note that filtering these sets of functions (relations) will require a different (but familiar) approach when compared to ordinary $RA$. When filtering tuples in a relation, we apply Boolean formulas tuple-by-tuple (e.g., $\sigma[\text{age} > 25](R)$). A set of relations like the Latin square example will need the Boolean formulas to be applied relation-by-relation. A key preliminary idea is to connect this use case with the ordinary $RA$ operator ($\gamma$), which does just that. In the Latin square example, we need constraints like \say{all values in each row are different} --- this can easily be expressed as a two-level aggregation. Given $I$ is one of the functions in the set for Latin square, $\gamma[][\text{Bool\_And}(res) \rightarrow res](\gamma[row][\text{AllDifferent}(Value)\rightarrow res](I))$ will return $True$ just when a function has all different values in its rows.

Finally, we observe that the constraint-solving community's global constraints~\cite{beldiceanu_2010_global.constraint.catalog} correspond exactly to the relational database community's aggregation and windowing functions~\cite{klug_1982_equivalence.relational.algebra.relational.calculus}. Global constraints are functions that take a possible solution to a (sub)problem as input and return a Boolean value. Typical examples are AllDifferent(), hasSubset(), and Max(). This parallels the database context where Boolean-valued aggregation functions take a relation (or partition of a relation) and return either True or False. We use global constraints frequently in aggregation contexts.

With these foundations in place, we can now formally define a solution set.

\subsection{Introducing Solution Sets} \label{subsec:the-solution-set-abstraction}
Solution sets operate within the exponentially large mathematical space $Decision^{Base}$, being all functions from an active domain $Base$ relation to a complete domain $Decision$ relation. A solution set is a set of ADRs, one for each candidate solution. \autoref{fig:solution-set-for-latin-square} above shows the solution set generated by raising the complete domain \codett{Values} relation to the power of the active domain Latin square \codett{Board} relation.

Following the pattern identified by Sakanashi and Sakai~\cite{sakanashi_2018_transformation.combinatorial.optimization.written.sql} the programming model is: specify candidate relations, restrict candidates, order, and limit. Note, however, this is a declarative specification model, not a query evaluation plan. Evaluation proceeds using the techniques from multiple disciplines that were called out in the introduction.

In parallel with our database $\mathfrak{D}$, and our set of CDRs $\mathfrak{C}$, we have $\mathfrak{U}$, a set of solution sets. We use uppercase letters $U, V$ to denote solution sets and individual candidate relations within them as $I$. A solution set $U \in \mathfrak{U}$ is constructed from an active domain relation $Base_U$ and a complete domain relation $Decision_U$ with disjoint attribute sets, giving $\text{dom}(U)$ with cardinality $|Decision_U|^{|Base_U|}$ as:
\begin{multline}
\text{dom}(U) = \left\{
  I_U \subseteq Base_U \times Decision_U
  \;\middle|\;
  \pi[\alpha_{Base_U}](I_U) = Base_U \right. \\
\left. \quad \ \wedge\
  \text{ the functional dependency }\alpha_{Base_U} \to \alpha_{Decision_U} \ \text{holds in } I_U
\right\}
\label{eq:solution-set-domain-1}
\end{multline}
where each $I_U$ is a solution candidate relation in $U$ with attribute set $\alpha_{I_U} =\alpha_{Base_U} \cup \alpha_{Decision_U}$ representing a complete assignment of decision values to base elements. The solution set inherits the characteristic function of the $Decision$ relation, which may restrict the allowable assignments. (We provide two equivalent definitions of solution set domains
useful for formalising translation in Appendix~\ref{subapp:solution-set-equivalent-definitions}.)

\begin{remark}
    We model solution sets as total functions by requiring $\pi[\alpha_{Base_U}](I_U) = Base_U$. This regularises the algebra without loss of generality-supporting subset and multiset queries through explicit cardinality attributes in $Decision_U$.
    \end{remark}

A solution set characteristic function $\chi_U$ is $f: \text{dom}(U) \rightarrow \{True, False\}$ and solution set $U$ may be denoted as the triple $\langle Base_U, Decision_U, \chi_U\rangle$. While structurally similar to active domain and complete domain relations, solution set characteristic functions differ fundamentally in their evaluation context: $\chi_U$ introduces the candidate relation $I_U$ into scope (not just a tuple). The characteristic function $\chi_U$ may also reference data by joining the candidate $I_U$ with relations from $\mathfrak{D}$, and to reference capabilities by joining it with CDRs in $\mathfrak{C}$.

Following the pattern with active domain and complete domain relations, we introduce a relational algebraic constructor for solution sets: $\omega_{sol}[Base, Decision](\chi)$. \autoref{fig:latinSquareSolutionraset} shows the solution for the Latin square, there we can see the solution set pattern: the constructor specifies
16 candidate relations, restrictions reduce these to valid Latin
squares, and projection triggers evaluation to return the single solution.

\begin{lstlisting}[mathescape=true, language=RA,caption={Latin square solution in $RA_{sol}$, demonstrating solution set
construction and filtering},label=fig:latinSquareSolutionraset,float=ht,
  abovecaptionskip=-\medskipamount]
-- Constructor
SearchSpace := $\omega_{sol}$(Board,Values)
-- Restrict all rows to have different values
UniqueValuesInRows := $\sigma_{sol}$[
    $\gamma$[$\varnothing$][Bool_And(ret) $\rightarrow$ ret](
        $\gamma$[row][AllDifferent(value) $\rightarrow$ ret](SearchSpace) -- candidate solution
    )](SearchSpace) -- solution set

-- Restrict all columns to have different values
EffectiveSearchSpace := $\sigma_{sol}$[
    $\gamma$[$\varnothing$][Bool_And(ret) $\rightarrow$ ret](
        $\gamma$[col][AllDifferent(value) $\rightarrow$ ret](UniqueValuesInRows)
    )](UniqueValuesInRows)

-- Restrict top-left corner to 1
LatinSquare := $\sigma_{sol}$[$\gamma$[$\varnothing$][hasSubset(ReqValues) $\rightarrow$ ret](EffectiveSearchSpace)](EffectiveSearchSpace)

-- Reduce back to a relation
solutionAsRelation := $\pi_{sol}$[$\varnothing$][row, col, value](LatinSquare)
\end{lstlisting}

The solution sets may be specified and manipulated by a higher-order algebra $RA_{sol}$.

\subsection{\texorpdfstring{$RA_{sol}$}{RA\_sol}: Relational Algebra over Solution Sets} \label{subsec:relational-algebra-over-solution-sets}
$RA_{sol}$ is a higher-order relational algebra that is closed over solution sets. Using these operators, solution sets may be filtered and composed. Relational algebraic operator $\oplus$ with a \say{sol} subscript $\oplus_{sol}$ denotes that it belongs to the solution set algebra.

\begin{figure}[!htbp]
    \centering
    \includegraphics[width=0.8\linewidth]{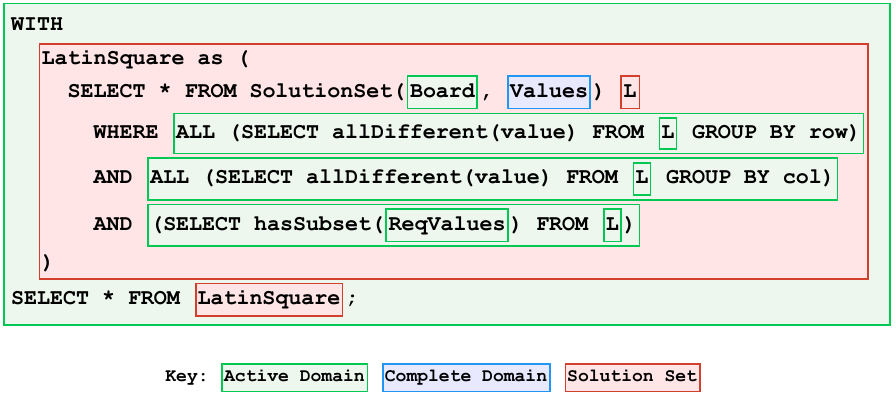}
    \caption{Latin Square demonstrated in polymorphic SQL}
    \label{fig:sql-latin-square}
\end{figure}

The Latin square solution (\autoref{fig:latinSquareSolutionraset}) provides an informal introduction to the restriction operator $\sigma_{sol}$ and projection $\pi_{sol}$. As expected, each $\sigma_{sol}$ in the figure has an ordinary Boolean-valued $\gamma$ operator as its restriction. In each $\sigma_{sol}$, the inner reference to the name of the solution set refers to the candidate solution, the outer reference to the solution set itself. This is analogous to SQL's \codett{SELECT * FROM Roles R WHERE R.A=2;} where \say{R} in \codett{R.A} is a reference to the tuple introduced into scope by the $\sigma$, and the \say{R} in \codett{Roles R} is a reference to the relation. The final restriction utilises a global constraint \text{hasSubset()} to verify that the required values are present in the candidate. Following our pragmatic approach, $\pi_{sol}$ is the operator that reduces a solution set back to being an active-domain relation. The three algebras compose naturally as illustrated by the solution in polymorphic SQL(\autoref{fig:sql-latin-square}).

Given solution sets $U$ and $V$, and recalling that they are defined respectively as triples $\langle Base_U, Decision_U, \chi_U\rangle$ and $\langle Base_V, Decision_V, \chi_V\rangle$, that candidate solutions $I_U\in U$ have attribute set $\alpha_{Base_U} \cup \alpha_{Decision_U}$, and given $\theta_U$ a Boolean expression of the form $\gamma[\varnothing][boolAgg() \rightarrow res](I\text{Expr}_U)$ where $I\text{Expr}_U$ may involve the candidate $I_U$ and joins with relations from $\mathfrak{D}$ and $\mathfrak{C}$, $\mu_U$ an orderable expression of the form $\gamma[\varnothing][orderableAgg() \rightarrow res](I\text{Expr}_U)$, and $\alpha_\oplus$ a non-empty subset of attributes in $\alpha_{I_U}$ for operator $\oplus$: \autoref{tab:relational-operators-for-solution-sets} shows the syntax and definitions of core and outer $RA_{sol}$ operators. Outer operators Order and Limit $RA_{sol}$ operators are closed over sequences of solution candidates, and $\pi_{sol}$, is closed over active-domain relations. (See Appendix~\ref{subapp:relational-operators-for-solution-sets} for further details on the operators.)

The closure of the core $RA_{sol}$ operators over solution sets may be established by observing that each operator is defined to return a triple $\langle Base_{result}, Decision_{result}, \chi_{result} \rangle$ where:

\begin{enumerate}
\item \textbf{Base and Decision components}: Constructed via natural join of active domain ($Base_U \bowtie Base_V$) and complete domain relations ($Decision_U \bowtie Decision_V$), which are closed over active and complete domains respectively.

\item \textbf{Characteristic function}: Constructed via Boolean combinations ($\wedge$, $\vee$, $\neg$) of existing characteristic functions, lifted as required to operate in a higher-dimensional solution set, preserving the $\gamma$-expression structure. (See Appendix~\ref{subapp:relational-operators-for-solution-sets} for details of the lifting function.) Since Boolean operations are closed over Boolean expressions, $\chi_{result}$ remains a valid characteristic function.
\end{enumerate}

As each component remains well-typed and the constructor $\omega_{sol}$ produces a solution set from these components, the core operators are closed over solution sets.

The algebraic properties of the core $RA_{sol}$ operators follow the same pattern as those for CDRs, inheriting from the Boolean algebra of their characteristic functions.

With the operators of $RA_{sol}$ defined, let's now consider how we may define a translation function $\Phi$ from the higher-order algebra to ordinary $RA$.

\begin{table}[!ht]
    \centering
        \caption{Relational Operators for Solution Sets}
    \begin{tabularx}{\textwidth}{|>{\raggedright\arraybackslash}p{5cm}|>{\raggedright\arraybackslash}X|} \hline 
        \textbf{Operator $\oplus_{sol}$}  & \textbf{Definition} \\ \hline
        Create Solution Set $\omega_{sol}(Base=\{\langle\rangle\}, Decision=\{\langle\rangle\},cf=True)$ & $\langle Base,Decision, cf \rangle$. The default values for $Base$ and $Decision$ are $\{\langle\rangle\}$ which is identity for natural join. \\ \hline 
        
        Natural-Join $U {\bowtie_{sol}} V$ & $\langle Base_U \bowtie Base_V,Decision_U \bowtie Decision_V,\chi_{U \bowtie_{sol} V}\rangle$ where $\chi_{U \bowtie_{sol} V}$ is $\chi_U$ and $\chi_V$ lifted to operate in a higher dimensional solution set. (See Appendix~\ref{subapp:relational-operators-for-solution-sets} for details.)  \\ \hline
        Cross-Product $U \times_{sol} V$ : $ \alpha_{I_U} \cap \alpha_{I_V} = \varnothing$ & $U {\bowtie_{sol}} V$   \\ \hline
         Intersection $U \cap_{sol} V : Base_U = Base_V \wedge \alpha_{Decision_U} = \alpha_{Decision_V} $ &  $U {\bowtie_{sol}} V$  which simplifies to $\langle Base_U,Decision_U \bowtie Decision_V,\chi_U \wedge \chi_V \rangle$   \\ \hline
        Difference $U -_{sol} V : Base_U = Base_V \wedge \alpha_{Decision_U} = \alpha_{Decision_V} $ & $U {\bowtie_{sol}} \langle Base_V, Decision_V,\neg \chi_V \rangle$ which simplifies to $\langle Base_U,Decision_U \bowtie Decision_V,\chi_U \wedge \neg \chi_V \rangle$  \\ \hline
Selection $\sigma_{sol}[\theta_U](U)$ &  $U {\bowtie_{sol}}  \langle Base_U, Decision_U,\theta_U\rangle$ which simplifies to $\langle Base_U,Decision_U ,\chi_U \wedge \theta \rangle$  \\[1pt] \hline

         Union $U \cup V : Base_U = Base_V,$ $ \alpha_{Decision_U} = \alpha_{Decision_V} $ & $\langle Base_U,Decision_U ,\chi_U \lor \chi_V \rangle$   \\ \hline

         Rename $\rho_{sol}[renamespec](U)$ &  $\langle  \rho[renamespec](Base_U), $ $ \rho[renamespec](Decision_U) ,$ $ \rho_{sol}[renamespec](\chi_U) \rangle$    \\ \hline

(Outer) Order $\tau_{sol}[\mu_U](U)$ & Yields a sequence (not a set) that is ordered by the expression $\mu_U$ over candidate expression $I\text{Expr}_U$  \\ \hline
  (Outer) Limit $\lambda_{sol}[n](U)$ &  Restricts cardinality to at most $n$   \\ \hline
  (Outer) Projection $\pi_{sol}[candRankAttr][\alpha_\pi](U)$ & Evaluates the solution set and returns the result as a single relation. The $candRankAttr$ generates a number starting at 1 and increasing monotonically for each candidate found. \\ \hline
  \end{tabularx}

    \label{tab:relational-operators-for-solution-sets}
\end{table}

\section{Semantics: Translating Solution Sets to Relations} \label{sec:semantics-translating-solution-sets-to-relations}

\begin{figure}[!htbp]
\centering
\begin{tikzpicture}[
    box/.style={rectangle, draw=black, fill=white, minimum width=1.5cm, minimum height=0.8cm},
    container/.style={rectangle, draw=black!50, fill=gray!10, inner sep=8pt},
    subcontainer/.style={rectangle, draw=blue!50, fill=blue!10, inner sep=5pt}
]

\node[container, minimum width=5cm, minimum height=4cm] (solutionset) at (0,0) {};
\node[above] at (solutionset.north) {Solution Set $U$};

\node[subcontainer] (base) at (-1, 0.8) {
    \begin{tabular}{c}
    $\alpha_{Base_U}$ \\
    $\varepsilon_{Base_U}$
    \end{tabular}
};
\node[above, font=\small] at (base.north) {$Base_U$};

\node[subcontainer] (decision) at (-1, -1) {
    \begin{tabular}{c}
    $\alpha_{Decision_U}$ \\
    $\chi_{Decision_U}$
    \end{tabular}
};
\node[above, font=\small] at (decision.north) {$Decision_U$};

\node[subcontainer] (candidate) at (1.2, 0.8) {
    \begin{tabular}{c}
    $\alpha_{I_U}$ \\
    $\varepsilon_{I_U}$
    \end{tabular}
};
\node[above, font=\small] at (candidate.north) {$I_U \in U$};

\node[subcontainer] (chi) at (1.2, -1) {$\chi_U$};
\node[above, font=\small] at (chi.north) {Char. Func.};

\draw[->, thick, font=\Large] (3, 0) -- node[above] {$\Phi$} node[below] {$\mathfrak{D}, \mathfrak{C}$} (5, 0);

\node[container, minimum width=5cm, minimum height=4cm] (phiset) at (8,0) {};
\node[above] at (phiset.north) {$\Phi(U)$};

\node[subcontainer] (flat) at (8, 0.8) {
    \begin{tabular}{c}
    $\alpha_{flat_U}$ \\
    $\chi_{flat_U}$
    \end{tabular}
};
\node[above, font=\small] at (flat.north) {Flattened Domain $flat_U$};

\node[subcontainer] (symi) at (8, -1) {
    \begin{tabular}{c}
    $\alpha_{symI_U}$ \\
    $\varepsilon_{symI_U}$
    \end{tabular}
};
\node[above, font=\small] at (symi.north) {Symbolic Candidate $symI_U$};

\end{tikzpicture}
\caption{Translation from solution sets to relational algebra via $\Phi$ in the context of $\mathfrak{D}, \mathfrak{C}$}
\label{fig:translation-overview}
\end{figure}
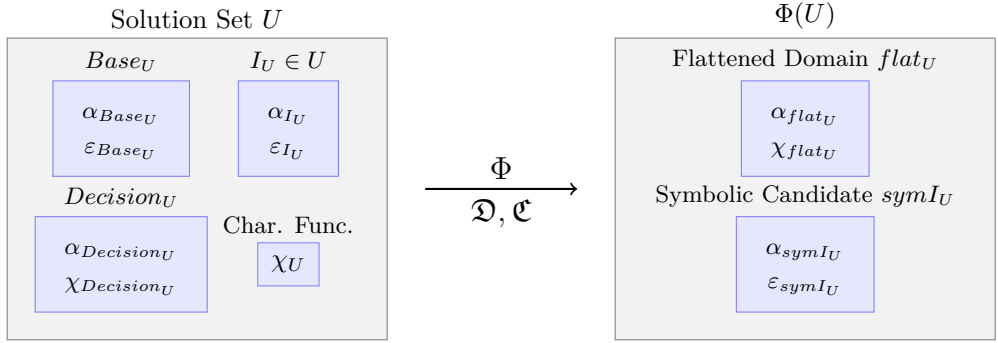

Having defined an algebra over these higher-order solution sets, our third contribution grounds its semantics by defining $\Phi$ as a homomorphic translation function back to standard $RA$. Figure~\ref{fig:translation-overview} illustrates this translation, noting that we are translating the results of $RA_{sol}$ operators—solution sets, rather than translating higher-order operators themselves.

Given solution set $U = \langle Base_U, Decision_U, \chi_U \rangle$, and its components as shown in the figure, the translation $\Phi(U)$ proceeds in four steps:

\subparagraph*{Step 1: Flattened Domain ($flat_U$).} A complete domain relation capturing the exponential search space as a repeated cross product. Representing all $|Decision_U|^{|Base_U|}$ possible extensions of the $Base$ relation with attribute set $\alpha_{flat_U} = \{a_i \mid a \in \alpha_{Decision_U}, i \in \{1,2,...,|Base_U|\}\}$. In this step we create the first conjunct of the translated characteristic function by replicating $Decision_U$'s characteristic function across each of the $|Base_U|$ decision replications in $flat_U$.

\subparagraph*{Step 2: Symbolic Candidate ($symI_U$).} An active domain relation preserving problem structure. It contains $|Base_U|$ tuples with attributes $\alpha_{Base_U} \cup \alpha_{Decision_U}$. Base attributes contain actual values while decision attributes contain symbolic $\langle references \rangle$ pointing to corresponding attributes in $flat_U$.

\subparagraph*{Step 3: Characteristic Function Join Semantics ($symI_U \bowtie R$).} Within the $\gamma$ expressions in $\chi_U$ joins on base attributes translate directly, while joins on decision attributes require encoding relations as functional dependencies. When $\alpha_R \cap \alpha_{Decision_U} \neq \varnothing$, we transform $R$ into a symbolic relation $R'$ with lookup expressions for dependent attributes.

\subparagraph*{Step 4: Characteristic Function Translation.} The translation $\Phi(\chi_U)$ is homomorphic: for any operator $\oplus$ and expressions $A$ and $B$, $\Phi(A \oplus B) = \Phi[\oplus](\Phi(A), \Phi(B))$. Aggregations over symbolic expressions preserve structure by aggregating the symbolic references for later evaluation under $\pi_{sol}$. (Complete translation details are deferred to Appendix~\ref{app:translation-solution-sets-relational-algebra}.)

To complete the example, \autoref{fig:latin-square-Phi} shows the fully translated Latin square. \codett{flatLatin} has the problem domain and the restrictions. \codett{symILatin} has the structure as a symbolic candidate.

    \begin{figure}[!htbp]
\centering
\begin{minipage}[c]{0.58\textwidth}
    \centering
    \begin{lstlisting}[mathescape=true, language=RA]
flatLatin := $\omega$[value1:1..2, value2:1..2, value3:1..2, value4:1..2](
  AllDifferent({value1,value3}) AND AllDifferent({value2, value4})
  AND AllDifferent({value1,value2}) AND AllDifferent({value3, value4})
  AND value1 = 1)
solutionLatin := $\pi$[value1, value2, value3, value4]
    (flatLatin)
    \end{lstlisting}
\end{minipage}
\hfill 
\begin{minipage}[c]{0.38\textwidth}
    \centering
    \begin{tabularx}{\textwidth}{|X|X|X|}
        \multicolumn{3}{l}{\codett{symILatin}} \\
        \hline
        \textbf{row:1..2}& \textbf{col:1..2} & \textbf{value:sym}  \\
        \hline
        1 & 1 & $\langle$value1$\rangle$  \\ \hline
        1 & 2 & $\langle$value2$\rangle$  \\ \hline
        2 & 1 & $\langle$value3$\rangle$  \\ \hline
        2 & 2 & $\langle$value4$\rangle$  \\ \hline
    \end{tabularx}
\end{minipage}
\caption{Latin square problem translated to \codett{flatLatin} and \codett{symILatin}}
\label{fig:latin-square-Phi}
\end{figure}

This translation bridges high-level solution sets to complete domain $RA$, enabling systematic transformation for query evaluation. We demonstrate translation to MiniZinc, a solver-independent constraint modelling language that preserves the declarative semantics of our algebras while providing access to diverse query evaluation algorithms. (See Appendix~\ref{app:why-minizinc} for MiniZinc's role as an intermediate language and Appendix~\ref{subapp:outer-operators-and-evaluation-backends} for the translation process.)

\section{Validation Through a Representative Problem} \label{sec:validation-through-representative-problems}

\subparagraph*{Cakes Production.} Taken from the MiniZinc documentation~\cite{cakes20241110} the Cake Production problem shows how data can parameterise an optimisation problem. \autoref{fig:cakes-on-a-page} traces the complete example. We are looking for a batch of \codett{Cakes} that we can make according to the \codett{Recipes} with the available \codett{Inventory} that maximises expected profit. \codett{Cakes} relation is our $Base$ relation and the $Decision$ relation is \codett{Qty}. The problem is fully stated in $RA_{sol}$, and translated into the $symI$ and $flat$ relations, thence to an evaluation algorithm (via MiniZinc in this case). The result \codett{LetsMakeBatch} relation is shown last. The solution in polymorphic SQL is in \autoref{fig:sql-cakes-intro} in the introduction.  (Due to space restrictions, we present just one example here; for others, including a Pareto-optimal subset selection example, see Appendix~\ref{app:additional-examples}.)

\newsavebox{\exCAKEBaseboxSavebox}
\begin{lrbox}{\exCAKEBaseboxSavebox}
\begin{tabular}{|l|r|}
\hline
cake & price \\
\hline
Banana & 400 \\
Chocolate & 450 \\
\hline
\end{tabular}
\end{lrbox}

\newsavebox{\exCAKEDecisionboxSavebox}
\begin{lrbox}{\exCAKEDecisionboxSavebox}
\begin{minipage}{6.5cm}
\begin{lstlisting}[mathescape=true, language=RA, backgroundcolor={}]
Qty = $\omega$[qty: 0..100](True)
\end{lstlisting}
\end{minipage}
\end{lrbox}

\newsavebox{\exCAKEContextboxSavebox}
\begin{lrbox}{\exCAKEContextboxSavebox}
\scalebox{0.85}{
\begin{tabular}{c}
\begin{tabular}{|l|r|}
\hline
\multicolumn{2}{|c|}{\textbf{Inventory}} \\
\hline
ingredient & avail \\
\hline
Banana & 6 \\
Cocoa & 500 \\
Flour & 4000 \\
Butter & 500 \\
Sugar & 2000 \\
\hline
\end{tabular}
\quad
\begin{tabular}{|l|l|r|}
\hline
\multicolumn{3}{|c|}{\textbf{Recipes}} \\
\hline
cake & ingr. & amt \\
\hline
Banana & Banana & 2 \\
Banana & Sugar & 75 \\
Banana & Flour & 250 \\
Banana & Butter & 100 \\
Choc. & Cocoa & 75 \\
Choc. & Sugar & 150 \\
Choc. & Flour & 200 \\
Choc. & Butter & 150 \\
\hline
\end{tabular}
\end{tabular}
}
\end{lrbox}

\newsavebox{\exCAKERasolboxSavebox}
\begin{lrbox}{\exCAKERasolboxSavebox}
\begin{minipage}{6.5cm}
\begin{lstlisting}[language=RA, basicstyle=\footnotesize\ttfamily, mathescape=true, backgroundcolor={}, frame=none]
B := $\omega_{sol}$(Cakes,Qty)
MakableBatches := $\sigma_{sol}$[
  $\gamma[\varnothing]$[bool_and(ret)$\rightarrow$ret](
    $\gamma$[ingredient][
      sum(qty*amount)<=min(avail)$\rightarrow$ret]
      (B $\bowtie$ Recipe $\bowtie$ Inventory))](B)
ProfitableBatch := $\lambda_{sol}$[1](
  $\tau_{sol}$[DESC][$\gamma[\varnothing]$[sum(qty*price)$\rightarrow$profit]
    (MakableBatches)](MakableBatches))
LetsMakeBatch := $\pi_{sol}$[$\varnothing$][cake,qty,profit](ProfitableBatch)
\end{lstlisting}
\end{minipage}
\end{lrbox}

\newsavebox{\exCAKESymIboxSavebox}
\begin{lrbox}{\exCAKESymIboxSavebox}
\begin{tabular}{|l|l|}
\hline
cake & qty \\
\hline
Banana & $\langle$qty1$\rangle$ \\
Chocolate & $\langle$qty2$\rangle$ \\
\hline
\end{tabular}
\end{lrbox}

\newsavebox{\exCAKEFlatboxSavebox}
\begin{lrbox}{\exCAKEFlatboxSavebox}
\begin{minipage}{6.5cm}
\begin{lstlisting}[mathescape=true, language=RA, backgroundcolor={}]
FlatCakes = $\omega$[qty1:0..100, qty2:0..100](
    250*qty1+200*qty2 <= 4000
    AND 2*qty1 <= 6
    AND 75*qty1+150*qty2 <= 2000
    AND 100*qty1+150*qty2 <= 500
    AND 75*qty2 <= 500
)
Results = $\pi$[$\varnothing$][*]($\lambda$[1]($\tau$[DESC][400*qty1+450*qty2](FlatCakes)))
\end{lstlisting}
\end{minipage}
\end{lrbox}

\newsavebox{\exCAKEIntermediateboxSavebox}
\begin{lrbox}{\exCAKEIntermediateboxSavebox}
\begin{minipage}{8cm}
\begin{lstlisting}[mathescape=true, language=MiniZinc,  backgroundcolor={}]
var 0..100: qty1; 
var 0..100: qty2; 
constraint 250*qty1+200*qty2 <= 4000; 
constraint 2*qty1 <= 6; 
constraint 75*qty1+150*qty2 <= 2000; 
constraint 100*qty1+150*qty2 <= 500; 
constraint 75*qty2 <= 500; 
solve maximize 400*qty1+450*qty2; 
\end{lstlisting}
\end{minipage}
\end{lrbox}

\newsavebox{\exCAKEOutputboxSavebox}
\begin{lrbox}{\exCAKEOutputboxSavebox}
\begin{tabular}{|l|r|r|}
\hline
cake & qty & profit \\
\hline
Banana & 2 & 800 \\
Chocolate & 2 & 900 \\
\hline
\end{tabular}
\end{lrbox}

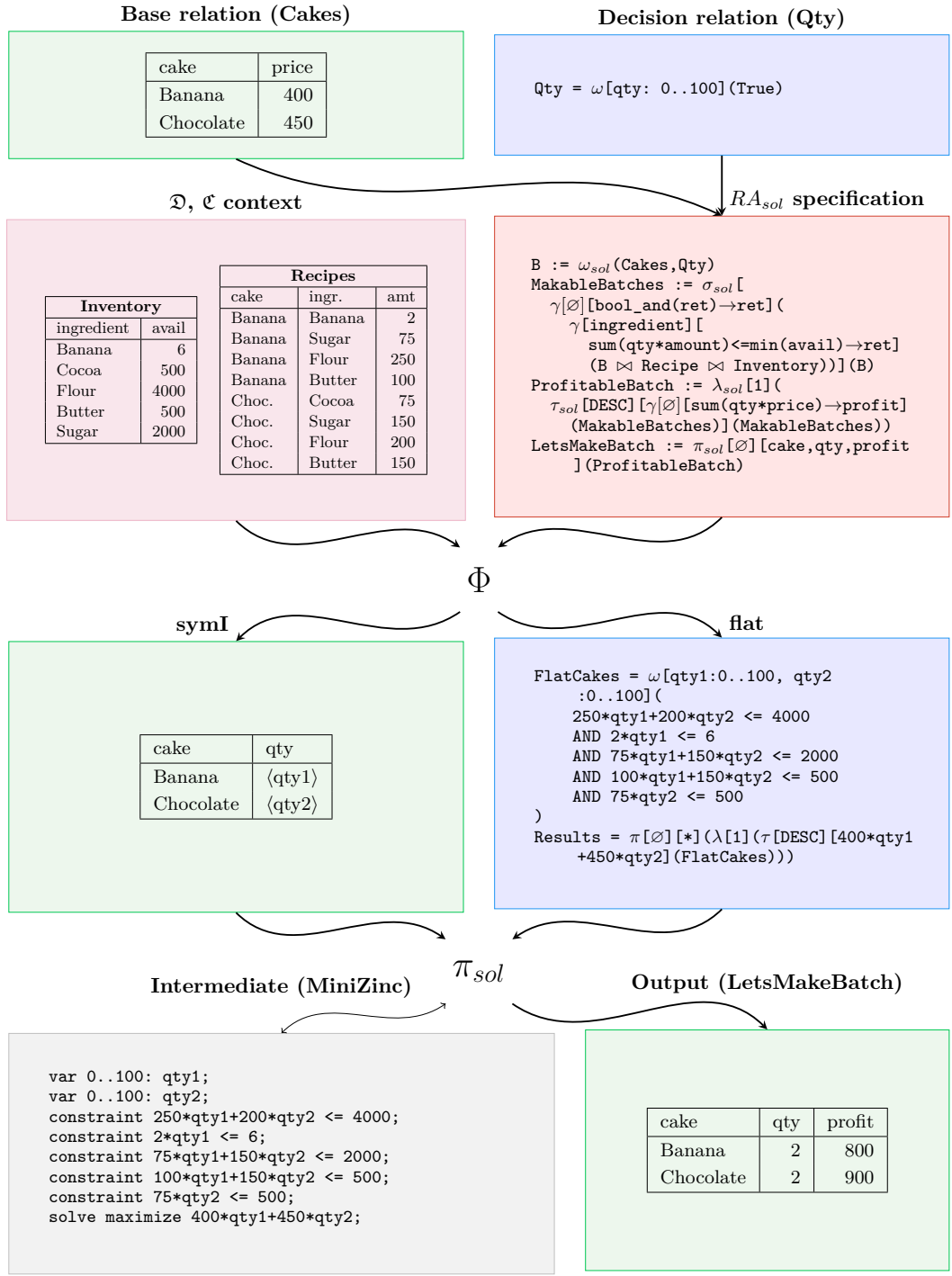
\begin{figure}[!htbp]
    \centering
    \resizebox{\textwidth}{!}{
    \begin{tikzpicture}[
        node distance=3.5cm,
        every node/.style={minimum height=1.5cm, font=\normalsize},
        inputbox/.style={rectangle, draw={rgb,255:red,0;green,200;blue,83}, fill={rgb,255:red,237;green,247;blue,237}, inner sep=10pt, minimum width=7.5cm, minimum height=2cm},
        decisionbox/.style={rectangle, draw={rgb,255:red,33;green,150;blue,243}, fill={rgb,255:red,232;green,232;blue,255}, inner sep=10pt, minimum width=7.5cm, minimum height=2cm},
        contextbox/.style={rectangle, draw=purple!30, fill=purple!10, inner sep=10pt, minimum width=7.5cm, minimum height=5cm},
        rasolbox/.style={rectangle, draw={rgb,255:red,210;green,64;blue,45}, fill={rgb,255:red,255;green,229;blue,229}, inner sep=10pt, minimum width=7.5cm, minimum height=5cm, text width=6.5cm, align=left, font=\footnotesize\ttfamily},
        transformbox/.style={ellipse, draw=black!0, fill=gray!0, inner sep=3pt, font=\LARGE\bfseries},
        flatbox/.style={rectangle, draw={rgb,255:red,33;green,150;blue,243}, fill={rgb,255:red,232;green,232;blue,255}, inner sep=10pt, minimum width=7.5cm, minimum height=4.5cm},
        symIbox/.style={rectangle, draw={rgb,255:red,0;green,200;blue,83}, fill={rgb,255:red,237;green,247;blue,237}, inner sep=10pt, minimum width=7.5cm, minimum height=4.5cm},
        intermediatebox/.style={rectangle, draw=gray!50, fill=gray!10, inner sep=10pt, minimum width=9cm, minimum height = 4cm},
        outputbox/.style={rectangle, draw={rgb,255:red,0;green,200;blue,83}, fill={rgb,255:red,237;green,247;blue,237}, inner sep=10pt, minimum width=6cm, minimum height = 4cm},
        labelstyleleft/.style={font=\normalsize\bfseries, anchor=south west},
        labelstyleright/.style={font=\normalsize\bfseries, anchor=south east},
        arrow/.style={->, thick, >=stealth}
    ]

    \node[inputbox] (base) {
        \usebox{\exCAKEBaseboxSavebox}
    };
    \node[labelstyleleft, above=-0.5cm and 0cm of base.north] {Base relation (Cakes)};

    \node[decisionbox, right=.5cm of base] (decision) {
        \usebox{\exCAKEDecisionboxSavebox}
    };
    \node[labelstyleright, above=-0.5cm and 0cm of decision.north] {Decision relation (Qty)};

    \node[contextbox, below=1cm of base] (context) {
        \usebox{\exCAKEContextboxSavebox}
    };
    \node[labelstyleleft, above =-0.5cm and 0cm of context.north] {$\mathfrak{D}$, $\mathfrak{C}$ context};

    \node[rasolbox, below=1cm of decision] (rasol) {
        \usebox{\exCAKERasolboxSavebox}
    };
    \node[labelstyleright, above right=-0.5cm and 0cm of rasol.north] {$RA_{sol}$ specification};

    \node[transformbox, below=0.25cm of $(context.south)!0.5!(rasol.south)$] (phi) {$\Phi$};

    \node[symIbox, below=2cm of context] (symi) {
        \usebox{\exCAKESymIboxSavebox}
    };
    \node[labelstyleleft, above left=-0.5cm and 0cm of symi.north] {symI};

    \node[flatbox, below=2cm of rasol] (flat) {
        \usebox{\exCAKEFlatboxSavebox}
    };
    \node[labelstyleright, above right=-0.5cm and 0cm of flat.north] {flat};

    \node[transformbox, below=0.25cm of $(symi.south)!0.5!(flat.south)$] (pisol) {$\pi_{sol}$};

    \node[intermediatebox, below=2cm of symi.south west, anchor=north west] (minizinc) {
        \usebox{\exCAKEIntermediateboxSavebox}
    };
    \node[labelstyleleft, above =0cm and 0cm of minizinc.north] {Intermediate (MiniZinc)};

    \node[outputbox, below=2cm of flat.south east, anchor=north east] (output) {
        \usebox{\exCAKEOutputboxSavebox}
    };
    \node[labelstyleright, above =0cm and 0cm of output.north] {Output (LetsMakeBatch)};

    \draw[arrow] (base.south) to[out=-25, in=155] (rasol.north);
    \draw[arrow] (decision.south) to[out=-90, in=90] (rasol.north);
    \draw[arrow] (context.south) to[out=-45, in=145] (phi.north west);
    \draw[arrow] (rasol.south) to[out=-135, in=35] (phi.north east);

    \draw[arrow] (phi.south west) to[out=-145, in=45] (symi.north);
    \draw[arrow] (phi.south east) to[out=-35, in=135] (flat.north);

    \draw[arrow] (symi.south) to[out=-45, in=145] (pisol.north west);
    \draw[arrow] (flat.south) to[out=-135, in=35] (pisol.north east);

    \draw[<->] (pisol.south west) to[out=-145, in=45] (minizinc.north);
    \draw[arrow] (pisol.south east) to[out=-35, in=135] (output.north);

    \end{tikzpicture}
    }
    \caption{Cakes production optimisation: Complete transformation from problem specification through $RA_{sol}$, homomorphic translation $\Phi$ to relational algebra, and evaluation via MiniZinc to final solution.}
    \label{fig:cakes-on-a-page}
\end{figure}

\section{Related Work} \label{sec:related-work}
We consider related work in two passes over the extensive history underlying our contributions. First, we examine antecedents to CDRs and second, we survey the evolution of subset selection and optimisation approaches in database contexts.

\subparagraph*{The Quest for Expressivity: Relational Database.} CDRs build on Hall's algorithmic relations~\cite{hall_1975_an.algebra.relations.machine.computation},
Maier and Warren's computed relations~\cite{maier_1981_incorporating.computed.relations.relational.databases, maier_1983_the.theory.of.relational.databases}, and we are following recent research on safe queries with external predicates~\cite{guagliardo_2025_queries.with.external.predicates} with interest. The focus of these works has been to model external functions as relations. The constraint database
tradition~\cite{revesz_1998_safe.query.languages.constraint.databases} sought to lift expressivity, especially for spatio-temporal queries.
Within the constraint database paradigm, Goldin's constraint query algebra~\cite{goldin_2004_constraint.query.algebras} comes closest in spirit to our proposal. It is an algebraic treatment of the paradigm's constraint relation.  The gap is that \say{Each constraint relation is a quantifier-free first-order DNF formula}~\cite{revesz_1998_safe.query.languages.constraint.databases}, and the consequence of this is that while constraint database recognises the need for
relations beyond active domains, the tuple is privileged. Our contribution of generalising relations via their characteristic
functions constructable with their own algebra appears novel. (See Appendix \ref{appsubsec:the-quest-for-expressivity-relational-database} for additional annotated bibliography.)

\subparagraph*{The Quest for Expressivity: Subset Selection and Optimisation.}
Our second contribution both inherits key ideas from, and may be distinguished from, decades of research interest in subset selection and optimisation by the relational database community. As we review the theory underlying these contributions, we acknowledge that all the works below include results that have no parallel in this paper, such as: evaluation algorithms e.g. SketchRefine~\cite{brucato_2016_scalable.package.queries.relational.database}, prototypes, optimisers, and user studies.

Research contributions from the early 1990s to the present form a Pareto front of capabilities, each optimising different aspects of the problem. Key innovations include: Boolean aggregation functions for filtering (SQLMP~\cite{choobineh_1991_sqlmp.data.sublanguage.representation.formulation}), algebraic foundations~\cite{valluri_2004_subset.queries.in.relational.databases, goldin_2004_constraint.query.algebras}, search space creation via functional dependency specifications (SCL~\cite{siva_2011_enabling.relational.databases.effective.csp}), recognition of sets of relations as the core abstraction and formal translation semantics (CombSQL+~\cite{sakanashi_2018_transformation.combinatorial.optimization.written.sql}), reuse of existing SQL clauses (DivDB~\cite{vieira_2011_divdb}), and introducing search spaces into SQL(package queries~\cite{brucato_2016_scalable.package.queries.relational.database}). The most expressive prior work, including over infinite domains, is SolveDB~\cite{siksnys_2016_solvedb.optimization.problem.solvers.sql}. Our solution sets integrate these insights within a unified algebraic framework with formal semantics.
While these contributions advanced important capabilities, a unified theoretical foundation remained elusive.

Two prior works developed relational algebras specifically for subset selection and optimisation: Valluri and Karlapalem's subset algebra~\cite{valluri_2004_subset.queries.in.relational.databases} achieves algebraic completeness but remains constrained by tuple-by-tuple semantics. Cadoli and Mancini's NP-Alg~\cite{cadoli_2007_relational.algebra.sql.constraint.modelling} is a constraint language treating relations as variables with nondeterministic semantics. This departure from the functional compositional semantics of standard $RA$ makes it unsuitable as the foundation we seek.

We have found many prior attempts to express subset selection and optimisation search spaces in SQL. Some depend on SQL's Three-Valued Logic(3VL) and overload \codett{NULL} to mean \say{this value is a variable in this problem}~\cite{choobineh_1991_sqlmp.data.sublanguage.representation.formulation, siksnys_2016_solvedb.optimization.problem.solvers.sql}. SQL's Data Definition Language DDL is extended in SCL~\cite{siva_2011_enabling.relational.databases.effective.csp}. Other authors have expressed the search space as guessed, non-deterministic relations~\cite{cadoli_2007_relational.algebra.sql.constraint.modelling, mancini_2010_local.search.over.relational.databases, sakanashi_2018_transformation.combinatorial.optimization.written.sql}. Still others specify the search space as a package (power multiset of a relation)~\cite{brucato_2016_scalable.package.queries.relational.database}. Finally, some authors generate the space by introducing both \say{variables} and \say{constraints} as attribute types in ordinary database tables~\cite{valdron_2020_data.driven.relational.constraint.programming}. Along with the variety we have just canvassed, each proposal introduces new clauses into SQL to accommodate the specification and filtering of the search space.

Additionally, some research threads e.g. sampling and clustering queries~\cite{agarwal_2024_computing.well.representative.summary.conjunctive,arenas_2024_tractability.diversity.query.answers.ultrametrics, gan_2024_optimal.dynamic.parameterized.subset.sampling, galhotra_2024_k.clustering.comparison.distance.oracles} appear to lack a way of specifying their problems in $RA$ or SQL despite early attempts~\cite{vieira_2011_divdb}.

Recently, we were encouraged to find that query optimisation research independently described patterns that align with our framework: Koch's quantifier elimination technique~\cite{koch_2024_query.optimization.by.quantifier.elimination} effectively treats relations as solution sets (with $Decision$ as identity), when replacing multi-way joins with aggregation, though without seeming to make this theoretical connection.

(For a contribution by contribution survey of prior work see Appendix \ref{appsec:the-quest-for-expressivity-subset-selection-and-optimisation}.)

We now confirm that solution sets do, in fact, match SolveDB, the Pareto front exemplar for prior expressivity.

\section{Expressiveness of Solution Sets} \label{sec:expressiveness-of-solution-sets}

We analyse the computational complexity and expressive power of solution sets, demonstrating that it matches the most expressive prior approach.

From our survey, SolveDB achieves the highest expressiveness among previous systems, with search spaces encompassing all functions $f: R \rightarrow C$ where $R$ is finite and $C$ may be infinite (containing domains like \codett{INT} and \codett{FLOAT}).
Solution sets achieve equivalent expressiveness through the same functional structure. For a solution set $U = \langle Base_U, Decision_U, \chi_U \rangle$, the domain corresponds exactly to the set of all functions from $Base_U$ to $Decision_U$: $\{f: Base_U \rightarrow Decision_U\}$. When $Decision_U$ includes infinite domains, the expressiveness is equivalent to that of SolveDB. (Appendix \ref{subapp:energy} shows the SolveDB Energy Balance problem as a worked example.)

Our contribution is not increased expressiveness, but instead providing this power through a principled $RA$ that maintains semantic clarity and compositional properties, which we have not found in prior approaches.

\section{Conclusions and Future Work} \label{sec:conclusion-and-future-work}

We have presented a unified algebraic foundation for subset selection and combinatorial optimisation queries through three complementary contributions: complete domain $RA$ and the higher-order algebra $RA_{sol}$ over solution sets with translation to $RA$. By elevating characteristic functions to first-class status, we have shown that a strict requirement for query safety does not preclude access to higher levels of expressivity. By introducing the relational exponentiation operator, we lifted the expressivity to consider all functions $f: Base \rightarrow Decision$. We are inspired by progress on GQL, and offer the algebras as foundational theoretical work on the path to unifying our fragmented approaches in subset selection and optimisation.

\subparagraph*{Research Direction.} Key Challenges include:
\begin{itemize}
    \item \textbf{Detailed Proofs} of algebraic properties and characterisation of computability, decidability, and termination issues.
\item \textbf{Formal connections to constraint semantics.} With a focus on approaches to missing values, partial functions, and relations as first-class constraint variables, mapping between the relational algebraic framework and established constraint programming formalisms.
    \item \textbf{Exploring SQL Language Design.} For data management, subset selection and optimisation, should SQL be polymorphic? How do we build on SCL's pioneering work in specifying solution sets via DDL integrity constraints, ergonomically support partial functions $f:Base \rightarrow Decision$ and leverage SQL's multisets and three-valued logic?
    \item \textbf{Query optimiser architecture.} Given that query evaluation can draw from the diverse algorithmic traditions we canvased
in the introduction, what principles should guide the choice between federation (where specialised evaluators handle subproblems in their domains of expertise) versus integration (where techniques from multiple paradigms are combined within a single evaluation framework)?
\item \textbf{Cost-based optimisation for solution sets.} Develop cost models and heuristics to guide evaluation algorithm selection.
    \item \textbf{Edge-case-free language target for AI.} We believe that these algebras can provide an attractive language target for LLM prompt-to-query applications covering data management, and prescriptive analytics.
\end{itemize}

\subparagraph*{Implementation Path.} A query compiler translating our three algebras to be evaluated by appropriate algorithms (e.g. via databases, constraint solvers) would enable experimental validation and future research.

This work provides the first unified theoretical foundation that spans data management, subset selection and optimisation, enabling fluid navigation between \say{what is} and what \say{might be}.

\bibliography{references}

\begin{thebibliography}{10}

\bibitem{abiteboul_1994_foundations.of.databases}
S.~Abiteboul, R.~Hull, and V.~Vianu.
\newblock {\em {Foundations of Databases}}.
\newblock Addison Wesley, 1994.
\newblock \href {https://doi.org/10.5860/choice.33-0359}
  {\path{doi:10.5860/choice.33-0359}}.

\bibitem{agarwal_2024_computing.well.representative.summary.conjunctive}
Pankaj~K. Agarwal, Aryan Esmailpour, Xiao Hu, Stavros Sintos, and Jun Yang.
\newblock {Computing A Well-Representative Summary of Conjunctive Query
  Results}.
\newblock {\em Proc. ACM Manag. Data}, 2024.
\newblock \href {https://doi.org/10.1145/3695835} {\path{doi:10.1145/3695835}}.

\bibitem{arenas_2024_tractability.diversity.query.answers.ultrametrics}
Marcelo Arenas, Timo~Camillo Merkl, Reinhard Pichler, and Cristian Riveros.
\newblock {Towards Tractability of the Diversity of Query Answers: Ultrametrics
  to the Rescue}.
\newblock {\em Proc. ACM Manag. Data}, 2024.
\newblock \href {https://doi.org/10.1145/3695833} {\path{doi:10.1145/3695833}}.

\bibitem{beldiceanu_2010_global.constraint.catalog}
Nicolas Beldiceanu, Mats Carlsson, and Jean-Xavier Rampon.
\newblock {Global Constraint Catalog}.
\newblock Technical report, Swedish Institute of Computer Science, Kista,
  Sweden, 2010.

\bibitem{brucato_2017_scalable.execution.engine.package.queries}
Matteo Brucato, Azza Abouzied, and Alexandra Meliou.
\newblock {A scalable execution engine for package queries}.
\newblock {\em SIGMOD Record}, 46(1), 2017.
\newblock \href {https://doi.org/10.1145/3093754.3093761}
  {\path{doi:10.1145/3093754.3093761}}.

\bibitem{brucato_2016_scalable.package.queries.relational.database}
Matteo Brucato, Juan~Felipe Beltran, Azza Abouzied, and Alexandra Meliou.
\newblock {Scalable package queries in relational database systems}.
\newblock {\em Proceedings of the VLDB Endowment}, 9(7), 2016.
\newblock \href {https://doi.org/10.14778/2904483.2904489}
  {\path{doi:10.14778/2904483.2904489}}.

\bibitem{brucato_2020_spaqltools.stochastic.package.query.interface}
Matteo Brucato, M.~Mannino, A.~Abouzied, P.~Haas, and A.~Meliou.
\newblock {sPaQLTooLs: A Stochastic Package Query Interface for Scalable
  Constrained Optimization}.
\newblock {\em Proceedings of the VLDB Endowment}, 2020.
\newblock \href {https://doi.org/10.14778/3415478.3415499}
  {\path{doi:10.14778/3415478.3415499}}.

\bibitem{brucato_2020_stochastic.packagequeries.in.probabilistic.databases}
Matteo Brucato, Nishant Yadav, A.~Abouzied, P.~Haas, and A.~Meliou.
\newblock {Stochastic Package Queries in Probabilistic Databases}.
\newblock In {\em Proceedings of the 2020 ACM SIGMOD International Conference
  on Management of Data}, 2020.

\bibitem{cadoli_2007_relational.algebra.sql.constraint.modelling}
Marco Cadoli and Toni Mancini.
\newblock {Combining relational algebra, SQL, constraint modelling, and local
  search}.
\newblock {\em Theory and Practice of Logic Programming}, 7(1-2), 2007.
\newblock \href {https://doi.org/10.1017/S1471068406002857}
  {\path{doi:10.1017/S1471068406002857}}.

\bibitem{choobineh_1991_sqlmp.data.sublanguage.representation.formulation}
J.~Choobineh.
\newblock {SQLMP: A Data Sublanguage for Representation and Formulation of
  Linear Mathematical Models}.
\newblock {\em INFORMS journal on computing}, 1991.
\newblock \href {https://doi.org/10.1287/ijoc.3.4.358}
  {\path{doi:10.1287/ijoc.3.4.358}}.

\bibitem{codd_1970_relational.model.data.large.shared}
E~F Codd.
\newblock {A Relational Model of Data for Large Shared Data Banks}.
\newblock {\em Communications of the ACM}, 13(6), 1970.
\newblock \href {https://doi.org/10.1145/362384.362685}
  {\path{doi:10.1145/362384.362685}}.

\bibitem{fernandes_2014_packagebuilder.querying.for.packages.tuples}
Kevin Fernandes, Matteo Brucato, R.~Ramakrishna, A.~Abouzeid, and A.~Meliou.
\newblock {PackageBuilder: querying for packages of tuples}.
\newblock In {\em SIGMOD Conference}, 2014.
\newblock \href {https://doi.org/10.1145/2588555.2612667}
  {\path{doi:10.1145/2588555.2612667}}.

\bibitem{flener_2003_introducing.esra.relational.language.modelling}
P.~Flener, J.~Pearson, and Magnus {\AA}gren.
\newblock {Introducing esra, a Relational Language for Modelling Combinatorial
  Problems}.
\newblock {\em International Workshop/Symposium on Logic-based Program
  Synthesis and Transformation}, 2003.
\newblock \href {https://doi.org/10.1007/978-3-540-45193-8{\_}95}
  {\path{doi:10.1007/978-3-540-45193-8{\_}95}}.

\bibitem{flener_2001_towards.relational.modelling.combinatorial.optimisation}
Pierre Flener.
\newblock {Towards Relational Modelling of Combinatorial Optimisation
  Problems}.
\newblock In {\em In Proceedings of IJCAI-2001 Workshop on Modelling and
  Solving Problems with Constraints. International Joint Conference on
  Artificial Intelligence}, 2001.

\bibitem{floyd_1967_nondeterministic.algorithms}
R.~W. Floyd.
\newblock {Nondeterministic Algorithms}.
\newblock {\em JACM}, 1967.
\newblock \href {https://doi.org/10.1145/321420.321422}
  {\path{doi:10.1145/321420.321422}}.

\bibitem{frisch_2009_proper.treatment.undefinedness.constraint.languages}
Alan~M Frisch and Peter~J Stuckey.
\newblock {The proper treatment of undefinedness in constraint languages}.
\newblock In {\em Lecture Notes in Computer Science (including subseries
  Lecture Notes in Artificial Intelligence and Lecture Notes in
  Bioinformatics)}, volume 5732 LNCS, 2009.
\newblock \href {https://doi.org/10.1007/978-3-642-04244-7{\_}30}
  {\path{doi:10.1007/978-3-642-04244-7{\_}30}}.

\bibitem{galhotra_2024_k.clustering.comparison.distance.oracles}
Sainyam Galhotra, Rahul Raychaudhury, and Stavros Sintos.
\newblock {k-Clustering with Comparison and Distance Oracles}.
\newblock {\em Proc. ACM Manag. Data}, 2024.
\newblock \href {https://doi.org/10.1145/3695830} {\path{doi:10.1145/3695830}}.

\bibitem{gan_2024_optimal.dynamic.parameterized.subset.sampling}
Junhao Gan, S.~Umboh, Hanzhi Wang, Anthony Wirth, and Zhuo Zhang.
\newblock {Optimal Dynamic Parameterized Subset Sampling}.
\newblock {\em Proc. ACM Manag. Data}, 2024.
\newblock \href {https://doi.org/10.1145/3695827} {\path{doi:10.1145/3695827}}.

\bibitem{garey_1979_computers.intractability.guide.np.completeness}
Michael~R Garey and David~S Johnson.
\newblock {\em {Computers and Intractability: A Guide to the Theory of
  NP-Completeness.}}
\newblock W H Freeman, San Francisco, CA, USA, 1979.

\bibitem{gelder_1991_safety.and.translation.relational.calculus}
A.~V. Gelder and R.~Topor.
\newblock {Safety and translation of relational calculus}.
\newblock {\em TODS}, 1991.
\newblock \href {https://doi.org/10.1145/114325.103712}
  {\path{doi:10.1145/114325.103712}}.

\bibitem{goldin_2004_constraint.query.algebras}
Dina~Q. Goldin and P.~Kanellakis.
\newblock {Constraint query algebras}.
\newblock {\em Constraints}, 2004.
\newblock \href {https://doi.org/10.1007/BF00143878}
  {\path{doi:10.1007/BF00143878}}.

\bibitem{guagliardo_2025_queries.with.external.predicates}
P.~Guagliardo, Leonid Libkin, Victor Marsault, Wim Martens, Filip Murlak,
  L.~Peterfreund, and Cristina Sirangelo.
\newblock {Queries with External Predicates}.
\newblock {\em International Conference on Database Theory}, 2025.
\newblock \href {https://doi.org/10.4230/LIPIcs.ICDT.2025.22}
  {\path{doi:10.4230/LIPIcs.ICDT.2025.22}}.

\bibitem{gyssens_1992_powerset.algebra.natural.tool.handle}
M.~Gyssens and D.~V. Gucht.
\newblock {The Powerset Algebra as a Natural Tool to Handle Nested Database
  Relations}.
\newblock {\em Journal of computer and system sciences (Print)}, 1992.
\newblock \href {https://doi.org/10.1016/0022-0000(92)90041-G}
  {\path{doi:10.1016/0022-0000(92)90041-G}}.

\bibitem{hall_1975_an.algebra.relations.machine.computation}
Patrick A~V Hall, Peter Hitchcock, and Stephen Todd.
\newblock {An algebra of relations for machine computation}.
\newblock In {\em POPL '75}, 1975.

\bibitem{hansen_1989_integrating.relational.databases.constraint.languages}
M.~R. Hansen, B.~S. Hansen, P.~Lucas, and P.~E. Boas.
\newblock {Integrating Relational Databases and Constraint Languages}.
\newblock {\em Computer languages}, 1989.
\newblock \href {https://doi.org/10.1016/0096-0551(89)90014-3}
  {\path{doi:10.1016/0096-0551(89)90014-3}}.

\bibitem{hirst_1993_completeness.results.recursive.data.bases}
T.~Hirst and D.~Harel.
\newblock {Completeness results for recursive data bases}.
\newblock {\em Journal of computer and system sciences (Print)}, 1993.
\newblock \href {https://doi.org/10.1145/153850.153905}
  {\path{doi:10.1145/153850.153905}}.

\bibitem{hooker_2011_integrated.methods.for.optimization}
J.~Hooker.
\newblock {Integrated methods for optimization}.
\newblock In {\em International Series in Operations Research and Management
  Science}, 2011.
\newblock \href {https://doi.org/10.1007/978-1-4614-1900-6}
  {\path{doi:10.1007/978-1-4614-1900-6}}.

\bibitem{hrbacek_1978_introduction.to.set.theory}
K.~Hrbacek and T.~Jech.
\newblock {\em {Introduction to Set Theory}}.
\newblock New York : M. Dekker, 1978.
\newblock \href {https://doi.org/10.2307/3621546} {\path{doi:10.2307/3621546}}.

\bibitem{jaffar_1987_constraint.logic.programming}
J~Jaffar and J.-L. Lassez.
\newblock {Constraint Logic Programming}.
\newblock In {\em Proceedings of the 14th ACM SIGACT-SIGPLAN Symposium on
  Principles of Programming Languages}, POPL '87, pages 111--119, New York, NY,
  USA, 1987. Association for Computing Machinery.
\newblock \href {https://doi.org/10.1145/41625.41635}
  {\path{doi:10.1145/41625.41635}}.

\bibitem{kanellakis_1995_constraint.query.languages}
Paris~C Kanellakis, Gabriel~M Kuper, and Peter~Z Revesz.
\newblock {Constraint Query Languages}.
\newblock {\em Journal of Computer and System Sciences}, 51(1):26--52, 1995.
\newblock URL: \url{https://doi.org/10.1145/298514.298582}, \href
  {https://doi.org/10.1006/jcss.1995.1051} {\path{doi:10.1006/jcss.1995.1051}}.

\bibitem{klug_1982_equivalence.relational.algebra.relational.calculus}
Anthony~C. Klug.
\newblock {Equivalence of Relational Algebra and Relational Calculus Query
  Languages Having Aggregate Functions}.
\newblock {\em JACM}, 1982.
\newblock \href {https://doi.org/10.1145/322326.322332}
  {\path{doi:10.1145/322326.322332}}.

\bibitem{koch_2024_query.optimization.by.quantifier.elimination}
Christoph Koch and Peter Lindner.
\newblock {Query Optimization by Quantifier Elimination}.
\newblock {\em Proc. ACM Manag. Data}, 2024.
\newblock \href {https://doi.org/10.1145/3651607} {\path{doi:10.1145/3651607}}.

\bibitem{kursawe_1990_variant.evolution.strategies.vector.optimization}
F.~Kursawe.
\newblock {A Variant of Evolution Strategies for Vector Optimization}.
\newblock {\em Parallel Problem Solving from Nature}, 1990.
\newblock \href {https://doi.org/10.1007/BFb0029752}
  {\path{doi:10.1007/BFb0029752}}.

\bibitem{mai_2023_scaling.package.queries.billion.tuples}
Anh Mai, Matteo Brucato, A.~Abouzeid, Peter~J. Haas, and A.~Meliou.
\newblock {Scaling Package Queries to a Billion Tuples via Hierarchical
  Partitioning and Customized Optimization}.
\newblock {\em Proceedings of the VLDB Endowment}, 2023.
\newblock \href {https://doi.org/10.48550/arXiv.2307.02860}
  {\path{doi:10.48550/arXiv.2307.02860}}.

\bibitem{maier_1983_the.theory.of.relational.databases}
David Maier.
\newblock {\em {The theory of relational databases}}.
\newblock Computer Science Press, Rockville, 1983.

\bibitem{maier_1981_incorporating.computed.relations.relational.databases}
David Maier and David~S Warren.
\newblock {Incorporating computed relations in relational databases}.
\newblock In {\em Proceedings of the ACM SIGMOD International Conference on
  Management of Data}, 1981.
\newblock \href {https://doi.org/10.1145/582318.582345}
  {\path{doi:10.1145/582318.582345}}.

\bibitem{mancini_2010_local.search.over.relational.databases}
Toni Mancini, P.~Flener, and J.~Pearson.
\newblock {Local search over relational databases}.
\newblock Technical report, Uppsala University, 2010.

\bibitem{mancini_2012_combinatorial.problem.relational.databases.view}
Toni Mancini, P.~Flener, and J.~Pearson.
\newblock {Combinatorial problem solving over relational databases: view
  synthesis through constraint-based local search}.
\newblock {\em ACM Symposium on Applied Computing}, 2012.
\newblock \href {https://doi.org/10.1145/2245276.2245295}
  {\path{doi:10.1145/2245276.2245295}}.

\bibitem{marques.silva_2021_conflict.driven.clause.learning.sat}
Joao Marques-Silva, I.~Lynce, and S.~Malik.
\newblock {Conflict-Driven Clause Learning SAT Solvers}.
\newblock In {\em Handbook of Satisfiability}, 2021.
\newblock \href {https://doi.org/10.3233/978-1-58603-929-5-131}
  {\path{doi:10.3233/978-1-58603-929-5-131}}.

\bibitem{marriott_1998_programming.with.constraints.an.introduction}
Kim Marriott and Peter~James Stuckey.
\newblock {\em {Programming with Constraints: An Introduction}}.
\newblock MIT Press, 1998.

\bibitem{meliou_2025_data.management.perspectives.prescriptive.analytics}
A.~Meliou, A.~Abouzeid, Peter~J. Haas, R.~R. Haque, Anh~L. Mai, and Vasileios
  Vittis.
\newblock {Data Management Perspectives on Prescriptive Analytics (Invited
  Talk)}.
\newblock {\em International Conference on Database Theory}, 2025.
\newblock \href {https://doi.org/10.4230/LIPIcs.ICDT.2025.2}
  {\path{doi:10.4230/LIPIcs.ICDT.2025.2}}.

\bibitem{cakes20241110}
{MiniZinc}.
\newblock {An Arithmetic Optimisation Example}, 11 2024.
\newblock URL:
  \url{https://docs.minizinc.dev/en/stable/modelling.html#an-arithmetic-optimisation-example}.

\bibitem{moesmann_2024_data.driven.prescriptive.analytics.applications}
Martin Moesmann and T.~Pedersen.
\newblock {Data-Driven Prescriptive Analytics Applications: A Comprehensive
  Survey}.
\newblock {\em Information Systems}, 2024.
\newblock \href {https://doi.org/10.48550/arXiv.2412.00034}
  {\path{doi:10.48550/arXiv.2412.00034}}.

\bibitem{nethercote_2007_minizinc.standard.cp.modelling.language}
N.~Nethercote, Peter~James Stuckey, Ralph Becket, S.~Brand, Gregory~J. Duck,
  and Guido Tack.
\newblock {MiniZinc: Towards a Standard CP Modelling Language}.
\newblock {\em International Conference on Principles and Practice of
  Constraint Programming}, 2007.
\newblock \href {https://doi.org/10.1007/978-3-540-74970-7{\_}38}
  {\path{doi:10.1007/978-3-540-74970-7{\_}38}}.

\bibitem{paredaens_1992_converting.nested.algebra.expressions.flat}
J.~Paredaens and D.~V. Gucht.
\newblock {Converting nested algebra expressions into flat algebra
  expressions}.
\newblock {\em ACM Transactions on Database Systems}, 1992.
\newblock \href {https://doi.org/10.1145/128765.128768}
  {\path{doi:10.1145/128765.128768}}.

\bibitem{pratten_2024_relational.expressions.data.transformation.computation}
David~Robert Pratten and Luke Mathieson.
\newblock {Relational Expressions for Data Transformation and Computation}.
\newblock In {\em LNCS,volume 14386}, pages 241--255, 2024.
\newblock URL: \url{https://link.springer.com/10.1007/978-3-031-47843-7_17},
  \href {https://doi.org/10.1007/978-3-031-47843-7{\_}17}
  {\path{doi:10.1007/978-3-031-47843-7{\_}17}}.

\bibitem{revesz_1998_safe.query.languages.constraint.databases}
P.~Revesz.
\newblock {Safe query languages for constraint databases}.
\newblock {\em TODS}, 1998.
\newblock \href {https://doi.org/10.1145/288086.288088}
  {\path{doi:10.1145/288086.288088}}.

\bibitem{rogova_2023_a.researchers.digest.of.gql}
Alexandra Rogova, D.~Vrgo{\v{c}}, Nadime Francis, Amélie Gheerbrant,
  P.~Guagliardo, L.~Libkin, Victor Marsault, Wim Martens, Filip Murlak,
  L.~Peterfreund, F.~Geerts, and Brecht Vandevoort.
\newblock {A Researcher’s Digest of GQL}.
\newblock In {\em 26th International Conference on Database Theory (ICDT
  2023)}, 2023.

\bibitem{gutierrez_2013_implementation.relation.domain.constraint.programming}
G.~Sabogal, P.~V. Roy, and Sascha~Van Cauwelaert.
\newblock {Implementation of the relation domain for constraint programming}.
\newblock In {\em International Conference on Principles and Practice of
  Constraint Programming}, 2013.

\bibitem{sakanashi_2018_transformation.combinatorial.optimization.written.sql}
Genki Sakanashi and Masahiko Sakai.
\newblock {Transformation of Combinatorial Optimization Problems Written in
  Extended SQL into Constraint Problems}.
\newblock {\em ACM-SIGPLAN International Conference on Principles and Practice
  of Declarative Programming}, 2018.
\newblock \href {https://doi.org/10.1145/3236950.3236963}
  {\path{doi:10.1145/3236950.3236963}}.

\bibitem{sakanashi_2020_transformation.sql.combinatorial.optimization.constraint}
Genki Sakanashi and Masahiko Sakai.
\newblock {Transformation of SQL-based combinatorial optimization problems into
  Constraint problems}.
\newblock {\em The Japanese Society for Artificial intelligence}, 112:12--17, 3
  2020.
\newblock \href {https://doi.org/10.11517/jsaifpai.112.0{\_}03}
  {\path{doi:10.11517/jsaifpai.112.0{\_}03}}.

\bibitem{siksnys_2017_demonstrating.solvedb.sql.dbms.optimization}
Laurynas Siksnys and T.~Pedersen.
\newblock {Demonstrating SolveDB: An SQL-Based DBMS for Optimization
  Applications}.
\newblock {\em IEEE International Conference on Data Engineering}, 2017.
\newblock \href {https://doi.org/10.1109/ICDE.2017.180}
  {\path{doi:10.1109/ICDE.2017.180}}.

\bibitem{siksnys_2021_solvedb.sql.based.prescriptive.analytics}
Laurynas Siksnys, T.~Pedersen, T.~D. Nielsen, and Davide Frazzetto.
\newblock {SolveDB+: SQL-Based Prescriptive Analytics}.
\newblock {\em International Conference on Extending Database Technology},
  2021.
\newblock \href {https://doi.org/10.5441/002/edbt.2021.13}
  {\path{doi:10.5441/002/edbt.2021.13}}.

\bibitem{siksnys_2016_solvedb.optimization.problem.solvers.sql}
Laurynas {\v{S}}ik{\v{s}}nys and Torben~Bach Pedersen.
\newblock {SolveDB: Integrating optimization problem solvers into SQL
  databases}.
\newblock In {\em ACM International Conference Proceeding Series}, volume
  18-20-July-2016, 2016.
\newblock \href {https://doi.org/10.1145/2949689.2949693}
  {\path{doi:10.1145/2949689.2949693}}.

\bibitem{siva_2011_enabling.relational.databases.effective.csp}
Sebastien Siva.
\newblock {\em {Enabling Relational Databases for Effective CSP Solving}}.
\newblock PhD thesis, Emory University, Atlanda, GA, 2011.

\bibitem{valdron_2020_data.driven.relational.constraint.programming}
Michael Valdron and Ken~Q Pu.
\newblock {Data Driven Relational Constraint Programming}.
\newblock In {\em Proceedings - 2020 IEEE 21st International Conference on
  Information Reuse and Integration for Data Science, IRI 2020}, 2020.
\newblock \href {https://doi.org/10.1109/IRI49571.2020.00030}
  {\path{doi:10.1109/IRI49571.2020.00030}}.

\bibitem{valluri_2004_subset.queries.in.relational.databases}
Satyanarayana~R. Valluri and K.~Karlapalem.
\newblock {Subset Queries in Relational Databases}.
\newblock {\em arXiv.org}, 2004.

\bibitem{vieira_2011_divdb}
Marcos~R. Vieira, H.~Razente, M.~Barioni, Marios Hadjieleftheriou,
  D.~Srivastava, C.~Traina, and V.~Tsotras.
\newblock {DivDB}.
\newblock {\em Proceedings of the VLDB Endowment}, 2011.
\newblock \href {https://doi.org/10.14778/3402755.3402779}
  {\path{doi:10.14778/3402755.3402779}}.

\end{thebibliography}

\begin{appendix}

\section{Guide to Notation}

\begin{tikzpicture}
\node[anchor=north west] at (0,0) {
\begin{minipage}{0.45\textwidth}
\textbf{Core Structures}\\[0.5ex]
\begin{tabularx}{\linewidth}{p{1.1cm}>{\raggedright\arraybackslash}X}
$\mathfrak{D}$ & Database (set of ADRs) \\
$\mathfrak{C}$ & Set of complete domain relations \\
$\mathfrak{U}$ & Set of solution sets \\
$\mathfrak{A}$ & Set of aggregation functions
\end{tabularx}

\vspace{1.5ex}
\textbf{Active Domain Relations (ADRs)}\\[0.5ex]
\begin{tabularx}{\linewidth}{p{1.1cm}>{\raggedright\arraybackslash}X}
$R, S$ & ADRs \\
$\langle \alpha_R, \varepsilon_R \rangle$ & Active domain relation structure (attributes, extension) \\
$\alpha_R$ & Attribute set of relation $R$ \\
$\text{dom}(a)$ & Domain of attribute $a$ \\
$\varepsilon_R$ & Extension (set of tuples) of relation $R$ \\
$\text{dom}(R)$ & Domain of relation $R$ (Cartesian product of attribute domains) \\
$t$ & Tuple \\
$c_{R,t,a}$ & Constant for attribute $a$ in tuple $t$ of relation $R$. Also $c_{t,a}$, $c_{a}$, or just $c$ in context. \\

\end{tabularx}

\vspace{1.5ex}
\textbf{Complete Domain Relations (CDRs)}\\[0.5ex]
\begin{tabularx}{\linewidth}{p{1.1cm}>{\raggedright\arraybackslash}X}
$C, D$ & CDRs \\
$\langle \alpha_C, \chi_C \rangle$ & Complete domain relation structure (attributes, characteristic function) \\
$\alpha_C$ & Attribute set of relation $C$ \\
$\chi_C$ & Characteristic function of $C$ \\
$\text{dom}(C)$ & Domain of complete domain relation $C$ (Cartesian product of attribute domains) \\

\end{tabularx}

\end{minipage}
};

\draw[gray] (7,0) -- (7,-15);

\node[anchor=north west] at (7.5,0) {
\begin{minipage}{0.45\textwidth}
\textbf{Solution Sets}\\[0.5ex]
\begin{tabularx}{1\linewidth}{p{3.1cm}>{\raggedright\arraybackslash}X}
$f: Base \rightarrow Decision$ & Space of all functions from $Base$ to $Decision$. \\
$|Decision|^{|Base|}$ & Solution set cardinality \\
$U, V$ & Solution sets \\
$\langle Base_U, Decision_U, \chi_U \rangle$ & Solution set structure \\
$Base_U$ & Base relation of solution set $U$ \\
$Decision_U$ & Decision relation of solution set $U$ \\
$\chi_U$ & Characteristic function of solution set $U$ \\
$I_U$ & Candidate relations in $U$
\end{tabularx}

\vspace{1.5ex}
\textbf{Translation}\\[0.5ex]
\begin{tabularx}{1\linewidth}{p{1.2cm}>{\raggedright\arraybackslash}X}
$\Phi$ & Translation function \\
$flat_U$ & Flattened complete domain for $U$ \\
$symI_U$ & Symbolic candidate for $U$ \\
$\langle a_i \rangle$ & Symbolic reference to $a_i$ \\
$I\text{Expr}_U$ & Candidate expression \\
\end{tabularx}

\vspace{1.5ex}
\textbf{Attribute Domains}\\[0.5ex]
\begin{tabularx}{\linewidth}{p{1.2cm}>{\raggedright\arraybackslash}X}
$INT$ & Integer \\
$FLOAT$ & Float \\
$BOOL$ & Boolean with values $True$ and $False$\\
$1..5$ & INT Range \\
$1.1..5.8$ & FLOAT Range \\
$\text{IN }\pi[a](R)$ & Analogous to SQL's \codett{REFERENCES a(R)} \\
\end{tabularx}

\end{minipage}
};
\end{tikzpicture}
\newpage

\section{Relational Operator Closure Summary}

\autoref{apptab:summary of operators} shows the relational operators across three algebras with their closure properties.

\begin{table}[!htbp]
\centering
\caption{Operator Closure Properties Across the Three Algebras}
\begin{tabular}{|l|p{3cm}|c|c|c|}
\hline
\textbf{Operator} & \textbf{Description} & \textbf{Active} & \textbf{Complete} & \textbf{Solution} \\
$\oplus$ & & \textbf{Domain} $\oplus$ & \textbf{Domain} $\oplus$ & \textbf{sol} $\oplus_{sol}$ \\
\hline
$\omega$ & Constructor & Closed & Closed & Closed \\ \hline
$\bowtie$ & Natural join & Closed & Closed & Closed \\ \hline
$\times$ & Cross product & Closed & Closed & Closed \\ \hline
$\cap$ & Intersection & Closed & Closed & Closed \\ \hline
$-$ & Difference & Closed & Closed & Closed \\ \hline
$\sigma$ & Selection & Closed & Closed & Closed \\ \hline
$\cup$ & Union & Closed & Closed & Closed \\ \hline
$\rho$ & Rename & Closed & Closed & Closed \\ \hline
$\gamma$ & Group-by-aggregate & Closed & N/A & N/A \\ \hline
$\pi$ & Projection & Closed & \multicolumn{2}{c|}{$\rightarrow$ Active Domain} \\ \hline
$\lambda$ & Limit & \multicolumn{3}{c|}{Outer (guides evaluation)} \\ \hline
$\tau$ & Order by & \multicolumn{3}{c|}{Outer (guides evaluation)} \\ \hline
\end{tabular}

\label{apptab:summary of operators}
\end{table}

\section{Additional Details for Active Domain Relational Algebra} \label{app:additional-details-for-active-domain-relational-algebra}
This appendix provides the formal notation and additional technical details for active domain $RA$ deferred from Section~\ref{sec:preliminaries-active-domain-relations}. While the main text assumes familiarity with standard relational operators, we provide here their precise notation as used throughout this paper, which would have interrupted the flow of establishing our three main contributions.

\subsection{The \texorpdfstring{$\omega$}{omega} Operator}
We introduce a constructor for ADRs, the $\omega$ operator, analogous to SQL's \codett{CREATE TABLE} with immediate data insertion via \codett{VALUES}. \autoref{fig:example-table-definition} shows an example with commonly occurring domains. The idiom \codett{IN $\pi$[id](Categories)} constrains the domain to values from an existing relation, analogous to SQL's \codett{REFERENCES Categories(id)}.
\begin{lstlisting}[mathescape=true, language=RA,caption={Example of relation construction via relational algebraic $\omega$ operator},label=fig:example-table-definition,float=ht,
  abovecaptionskip=-\medskipamount]
R := $\omega$[id: INT,name: VARCHAR, weight: FLOAT, status: ENUM(active, inactive), size: 1..99, category: IN $\pi$[id](Categories)
     ]({$\langle$1, 'Fred',67.0, active, 3, 234$\rangle$, ...})
\end{lstlisting}

\subsection{Active Domain Relational Operators}
Given ADRs $R$ and $S$, and recalling that they are defined respectively as an attribute set and extension pair $\langle \alpha_R, \varepsilon_R \rangle$ and $\langle \alpha_S, \varepsilon_S \rangle$, and given $\theta_R$ an arbitrary Boolean expression over $\alpha_R$, and $\alpha_\oplus$ a non-empty subset of attributes in $\alpha_R$ for operator $\oplus$: \autoref{tab:relational-operators-for-active-domain-relations} shows our notation for the algebra's operators.
\begin{table}[!htbp]
    \centering
        \caption{Relational Operators for ADRs}
    \begin{tabularx}{\textwidth}{|>{\raggedright\arraybackslash}X|>{\raggedright\arraybackslash}X|} \hline 
        \textbf{Operator $\oplus$ Notation} \\ \hline
        Create Active Domain Relation $\omega[\text{a}_1{:}\text{d}_1, $ $ \text{a}_2{:}\text{d}_2, ..., $ $ \text{a}_i{:}\text{d}_i](\varepsilon) $ where $a_i$ are attribute identifiers and $d_i$ are domain specifications, and $\varepsilon$ is a set of tuples of constants $\{\langle c_1, c_2, ... c_i \rangle, ...\}$ which defaults to $\varnothing$. \\ \hline
        Natural-Join $R \bowtie S$     \\ \hline
        Cross-Product $R \times S : \alpha_R \cap \alpha_S = \varnothing  $  \\ \hline
        Intersection $R \cap S : \alpha_R = \alpha_S $   \\ \hline
        Difference $R - S : \alpha_R = \alpha_S $  \\ \hline
        Selection $\sigma[\theta_R](R)$  \\[1pt] \hline
        Projection $\pi [\alpha_\pi](R) : \alpha_\pi \subseteq \alpha_R $  \\[3pt] \hline  
        Union $R \cup S : \alpha_R = \alpha_S $    \\ \hline

        Rename $\rho[renamespec](R)$   \\ \hline
        Group-By-Aggregate $\gamma\left [\alpha_\gamma\right]\left[agg() \rightarrow b, ...\right](R):$ $b$ is the attribute name assigned to the aggregated value  \\ \hline
        Order $\tau [\alpha_\tau] (R) : \alpha_\tau \subseteq \alpha_R$     \\ \hline
        Limit $\lambda[n](R)$     \\ \hline
        \end{tabularx}

    \label{tab:relational-operators-for-active-domain-relations}
\end{table}

\section{Additional Details for Complete Domain Relational Algebra} \label{app:additional-details-for-complete-domain-relational-algebra}

This appendix collects the technical details for complete domain $RA$ deferred from Section~\ref{sec:an-algebra-for-complete-domain-relations}. These subsections expand on specific aspects—from the theoretical basis in optimisation problems to query safety proofs and operational details—that would have interrupted the main narrative flow of establishing our first contribution.

\subsection{Notation} The constant value for a given active domain relation $R$, tuple $t$ and attribute $a$ may be denoted as $c_{R,t,a} \in \text{dom}(a)$ --- when the relation or tuple under discussion is clear from the context, the $R,t$ may be elided giving $c_{t,a}$ or $c_{a}$. 

\subsection{Relational algebra is a Native Language for Expressing Optimisation Problems} \label{subapp:constraint-solving-basis-in-relations}

This subsection expands on the claim from Section~\ref{sec:an-algebra-for-complete-domain-relations} that CDRs naturally express the same optimisation problems addressed by constraint-solving technologies, showing the direct correspondence between standard optimisation formulations and relational algebraic expressions.

If we shift the terminology in the definition of CDRs: attribute $\rightarrow$ variable, characteristic function $\rightarrow$ constraint, we have the domain of constraint programming(CP), linear programming (LP), Boolean satisfiability (SAT), and Satisfiability Modulo Theories (SMT)~\cite{marriott_1998_programming.with.constraints.an.introduction}. If we add the ability to specify an order of candidate solutions based on some objective function, we have the domain of optimisation problems. In fact, according to Hooker in his magisterial text~\cite{hooker_2011_integrated.methods.for.optimization}, the theoretical basis of optimisation is relations. The following quote [p20] is included here to show the exact parallel with the CDRs described above.
\begin{quote}
.. an optimization problem can be written
\begin{align*}
\text{min (or max)} f(x) \\
C(x) \\
x \in D
\end{align*}

where $f(x)$ is a real-valued function of variable $x$ and $\mathcal{D}$ is the $domain$ of $x$. The function $f(x)$ is to be minimised (or maximised) subject to a set $C$ of constraints, each of which is either satisfied or violated by any given $x \in D$. Generally, $x$ is a tuple $(x_1, ..., x_n)$ and D is a Cartesian product $D_1 \times ... \times D_n$, where each $x_j \in D_j$. The notation $C(x)$ means that $x$ satisfies all the constraints in $C$.
\end{quote}

To make the connection with relational algebra explicit, here is an informal translation of Hooker's prototypical optimisation problem above into a familiar form: $$\lambda[1](\tau[f(x)](\sigma[C(x)](D)))$$

This translation reveals that relational algebra is a natural specification language for optimisation problems. The outer operators provide guidance for the constraint-solving: The ordering by $\tau$[f(x)] specifies the objective function. At the same time, the limit $\lambda$ might be relaxed to request more than one result or omitted to specify a satisfaction problem.

\subsection{Complete Domain Relational Algebra Example} \label{subapp:complete-domain-relational-algebra-example}
Here we demonstrate the complete domain operators in action through a practical example: an Australian GST calculator that illustrates how CDRs support multi-directional computation without explicit variable declaration and may be constructed from simpler relations using natural join.

\autoref{lst:gst-example} shows a calculator for the Australian Goods and Services Tax (GST) that we will use as a pedagogical example:
\begin{lstlisting}[mathescape=true, language=RA,caption={Australian GST Example},label=lst:gst-example,float=ht,
  abovecaptionskip=-\medskipamount]
GST := $\omega$[price: FLOAT, gst: FLOAT, exgst: FLOAT](
    gst = price / 11 AND exgst = price - gst
)
\end{lstlisting}
This defines an infinite relation containing all valid combinations of prices and GST amounts. Unlike an active domain relation that would need to enumerate specific values, this complete domain relation captures the entire infinite set of valid GST calculations.

The power of this approach becomes clear when we join with actual data. \autoref{fig:gst-application-example} shows relation \codett{Prices} and the result of natural join $\bowtie$ with $\text{GST}$.

\begin{figure}[!htbp]
    \begin{subfigure}[t]{0.45\textwidth}
        \centering
        \begin{tabular}{c}
        \multicolumn{1}{c}{\textbf{Prices}} \\
        \hline
        price \\
        \hline
        110.00 \\
        55.00 \\
        \hline
        \end{tabular}
        \caption{Prices table}
    \end{subfigure}
    \hfill
    \begin{subfigure}[t]{0.45\textwidth}
        \centering
        \begin{tabular}{|c|c|c|}
        \multicolumn{3}{c}{\textbf{Prices $\bowtie$ GST}} \\
        \hline
        price & gst & exgst \\
        \hline
        110.00 & 10.00 & 100.00 \\
        55.00 & 5.00 & 50.00 \\
        \hline
        \end{tabular}
        \caption{Prices with GST}
    \end{subfigure}
    \caption{Applying the GST relation using natural join}
    \label{fig:gst-application-example}
\end{figure}

Moreover, this relation works multi-directionally. We could equally start with ex-GST amounts and calculate prices, or with GST amounts and derive both prices and ex-GST values—all using the same complete domain relation.

As an illustration of the algebra at work, \autoref{lst:gst-by-bowtie-example} shows how we can rebuild the \codett{GST} complete relation out of two simpler CDRs.

\begin{lstlisting}[mathescape=true, language=RA,caption={Natural join of two CDRs to recreate GST},label=lst:gst-by-bowtie-example,float=ht,
  abovecaptionskip=-\medskipamount]
PriceGST := $\omega$[price:FLOAT, gst:FLOAT](price/11 = gst)
PriceExGST := $\omega$[price:FLOAT, gst:FLOAT, exgst:FLOAT](exgst = price - gst)
GST2 := PriceGST $\bowtie$ PriceExGST
GST == GST2 -- True
\end{lstlisting}

\subsection{Query Safety in the Complete Domain Algebra} \label{appsubsec:query-safety-in-the-complete-domain-algebra}
We now demonstrate that query safety is not compromised by extending to CDRs. This claim will be established in two steps. Firstly, we will demonstrate that ADRs occupy a well-defined region in the space of all CDRs, characterised by the property of domain independence. Then, we will demonstrate that this safe region is reachable in polynomial time after applying any of the relational algebraic operators.

\subparagraph*{Domain-independent Subset of Complete Domain Relations.} Every active domain relation may be considered the union of singleton relations, each singleton relation being the natural join of unary and singleton relations, one for each attribute. If we build a characteristic function in this fashion, we end up with a disjunction of conjunctions in what is commonly called Disjunctive Normal Form (DNF). Specifically, every active domain relation $R$ has a characteristic function of the form
$$\chi_R = \bigvee_{t \in \varepsilon_R} \left( \bigwedge_{a \in \alpha_R} a=c_{t,a}\right)$$
where $|\varepsilon_R| < k$ for some natural number $k$.
Looking at our earlier example, we can see that the extension provided to the active domain constructor can be considered syntactic sugar for a characteristic function: R and S are equal-valued (\autoref{lst:ext-is-cf}).
\begin{lstlisting}[mathescape=true, language=RA,caption={Extensions are syntactic sugar for characteristic functions},label=lst:ext-is-cf,float=ht,
  abovecaptionskip=-\medskipamount]
R := $\omega$[id: INT, name: VARCHAR, weight: FLOAT, status: ENUM(active, inactive), size: 1..99, category: IN $\pi$[id](Categories)]
    ({$\langle$1, 'Fred',67.0, active, 3, 234$\rangle$, ...})
S := $\omega$[id: INT, name: VARCHAR, weight: FLOAT, status: ENUM(active, inactive), size: 1..99, category: IN $\pi$[id](Categories)](
    (id = 1 AND name = 'Fred' AND weight = 67.0 AND status = active AND size = 3 AND category = 234)
    OR
    ...
)
R == S -- True
\end{lstlisting}

So here is the domain-independent subset. Relations with their characteristic function in DNF are domain independent because generalising any domain does not affect the value of the relation. Put another way, the equality conjuncts reference only the explicit constants $c_{t,a}$, making the relation's extension independent of the choice of domain—we could, for example, change an attribute's domain from \codett{0..1} to \codett{-100..100} without affecting the relation's value.

\subparagraph*{Reachability under Operations.} When CDRs with DNF characteristic functions undergo relational operations, the safe region (DNF) is reachable in polynomial time with domain independence preserved, even with an implementation that ignores optimisations like hash joins.  We need to consider only natural join, union and difference as they are the basis of the algebra. Given two ADRs $R$ and $S$ with DNF characteristic functions, for natural join we may distribute conjunction over $\chi_R$ and $\chi_S$, resulting in $|R|\cdot|S|$ disjuncts, of which those with conflicting values for any common attributes are eliminated, leaving a formula in DNF in $O(|R|\cdot|S|)$. Union directly generates a formula in DNF in linear time. For Difference, since $\alpha_R = \alpha_S$ is required, each disjunct in both DNF formulas specifies values for all attributes. The difference operation removes from R's disjuncts those that appear in S, computable in $O(|R| + |S|)$ time. If input relations $R$ and $S$ are in DNF the relational algebraic operators preserve DNF and domain independence.

Thus, we have demonstrated that all relational algebraic queries over relations in the active domain relation (DNF) region of CDRs are safe queries.

\begin{remark}
The preservation of query safety under CDR operations relies fundamentally on the algebraic nature of our approach. While first-order logic (FOL) and relational calculus (RC) suffer from undecidable query safety—making it impossible to determine algorithmically whether an arbitrary formula returns finite results—relational algebra provides constructive guarantees. Every operation in our complete domain algebra preserves the domain-independent DNF region, ensuring finite results for finite inputs. This distinction is crucial: whereas FOL/RC require syntactic restrictions whose satisfaction is undecidable in general, CDR operators inherently maintain safety through their definitions. For a comprehensive treatment of safety in logical versus algebraic query languages, see Van Gelder and Topor~\cite{gelder_1991_safety.and.translation.relational.calculus}.
\end{remark}

\subsection{Projection as a Transition Operator} \label{appsubsec:projection-as-a-transition-operator}
Here, we provide additional explanation of the role for projection, ubiquitous in relational database queries and yet practically absent in constraint-solving technologies apart from formatting output.

At its core, projection is about eliminating dimensions from a cross product. When we project $\pi[\{a\}](R)$ where $R$ has attributes $\{a, b\}$, we're asking: given $R \subseteq \text{dom}(a) \times \text{dom}(b)$, what is the set of $a$ values that participate? For ADRs, this is trivial because each tuple explicitly provides both components of the cross product. The tuple $\langle a{:}1, b{:}2 \rangle$ tells us directly that $a{:}1$ participates. We collect these $a$ values across all tuples. However, outside the DNF region occupied by ADRs, things are not as simple.

In the constraint-solving community, it has long been recognised that \say{Projection corresponds to quantifier elimination [variable elimination] and is the nontrivial operation}~\cite{kanellakis_1995_constraint.query.languages}. In the general case, this requires solving the problem at hand to find the values in some dimensions that participate in the solution. In solving technologies, projection is used, generally as it is in relational databases, where we have a set of solutions (tuples) to project.

Given these two converging traditions, we make a pragmatic choice: we define projection applied with Limit $\lambda$ as always closed over ADRs, and of course, Limit is optional for finite domains. Projection is thus the operator that signals the transition from complete domain to active domain semantics—the boundary where evaluation must occur.

\subsection{Safely Joining Active Domain and Complete Domain Relations} \label{appsubsec:safely-joining-active-domain-and-complete-domain-relations}
While solution sets \textbf{don't} depend on such joins, we take up this topic here to answer a question likely to be of interest to readers. When CDRs are joined with active domain data, they function analogously to SQL's GENERATED ALWAYS AS columns, applying computational rules to concrete data. However, unlike SQL's schema-bound computed columns, CDRs are first-class entities that can be composed, reused, and applied dynamically to different base relations. Joins between active and complete domain relations are well defined since we are computing the conjunction of two well-defined characteristic functions. If the complete domain relation is finite, or if every attribute in the complete relation becomes functionally dependent on the active domain relation, well and good.  We are not attempting to answer the computability implications of joining with CDRs containing infinite attributes, but do refer the reader to active research in this domain, e.g.~\cite{guagliardo_2025_queries.with.external.predicates}.

\section{Additional Details for Solution Set Algebra \texorpdfstring{$RA_{sol}$}{RA\_sol}} \label{app:additional-details-solution-set-algebra}

This appendix provides additional technical details for the solution set algebra $RA_{sol}$ introduced in Section~\ref{sec:proposed-higher-order-algebra-for-solution-sets}. These subsections expand on specific aspects—equivalent definitions, operator properties, and compositional details—that are deferred from the main text to maintain narrative flow.

\subsection{Solution Set Equivalent Definitions} \label{subapp:solution-set-equivalent-definitions}

Solution sets have additional and applicable definitions that can inform our discussion from two different perspectives.

The first perspective will assist us in fixing the semantics of solution sets through translation to ordinary $RA$, and is where the domain solution set $U$ is seen as the repeated cross product of the $Decision$ relation once for each tuple in the $Base$ relation, as follows. Let  $\{t_1, t_2, ..., t_{|Base_U|}\}$ be the tuples of $Base$,  $(d_1, d_2, ..., d_{|Base_U|})$ be a possible extension from $Decision$, $Decision_U^{|Base_U|}$ means the $|Base_U|$-fold Cartesian product of $Decision_U$ and  $t_i\cup\; d_i$ represents the union of attribute-value pairs from tuples $t_i$ and $d_i$ then:
\begin{equation}
\text{dom}(U) = \left\{ \left\{ t_i\cup\; d_i \mid i=1,2,\ldots,|Base_U| \right\} \mid (d_1, d_2, ..., d_{|Base_U|}) \in Decision_U^{|Base_U|} \right\}
\label{eq:solution-set-domain-2}
\end{equation}
The second perspective will help us understand the role of the $\gamma$ operator in characteristic functions and how we may use relational algebraic expressions with ADRs in $\mathfrak{D}$, and CDRs $\mathfrak{C}$ to assist with filtering candidate solutions. We adopt a characteristic function notation that makes the possible participation of other relations explicit. We define solution set $U$ as $U = \{I_U \in \text{dom}(U) | \chi_U(I\text{Expr}_U)\}$. An $I\text{Expr}_U$ includes: (1) the candidate relation $I_U$ itself, (2) relational algebraic expressions including $I_U$ and relations from $\mathfrak{D}$ (e.g., $I_U \bowtie R$ for $R \in \mathfrak{D}$), (3) algebraic expressions with CDRs from $\mathfrak{C}$ (e.g., $I_U \bowtie C$ for $C \in \mathfrak{C}$), and (4) compositions thereof. (Further details on restrictions that apply to these expressions can be found in \autoref{subsubsec:iexprU-details}.) These candidate expressions are reduced to Boolean values via $\gamma$ operators with Boolean aggregation functions. So for solution set $U$ interpreted as a set of $I_U$'s we can say that $\chi_U$ is of the general form $\gamma[][boolAgg()](I\text{Expr}_U)$ where $boolAgg \in boolAgg$ e.g., \codett{AllDifferent()}, or \codett{Bool\_And()}\footnote{We assume that the unary singleton relation returned by a such a $\gamma$ operator can be interpreted as a scalar Boolean value.}. In the Cakes example~\autoref{sec:validation-through-representative-problems}, we ensured that we have the ingredients in stock by joining candidate cake batches with recipes and inventories $Batch \bowtie Recipe \bowtie Inventory$.

\subsection{\texorpdfstring{$RA_{sol}$}{RA\_sol} Operators and Properties} \label{subapp:relational-operators-for-solution-sets}

This section expands on the solution set operators, with particular attention to how characteristic functions must be adapted when joining solution sets.

\subparagraph*{Natural-Join $\bowtie_{sol}$ and Cross-Product $\times_{sol}$.} These operators are defined in terms of the natural join of their components, and a lifting function that adapts the source characteristic functions to the higher-dimensional search space formed by the natural join.

When natural joining two solution sets $U$ and $V$, a fundamental challenge arises because each characteristic function must be evaluated over the joined solution candidates' attribute set, which contains attributes from both $U$ and $V$. For a characteristic function originally defined over $\alpha_{I_U}$, the $lift(\chi_U, \alpha_{Base_V})$ function transforms the restriction to hold universally across all possible partitions defined by $V$'s base attributes, applying $\chi_U$ to each partition and then universally quantifying over those results. This ensures that $U$'s restrictions are applied regardless of how the solution candidates are partitioned according to $V$'s structure. Natural Join is a symmetrical operation, and the same procedure is applied for $\chi_V$ using $Base_U$ to partition the joined solution candidate.

Let's define the $lift()$ functions. Recalling that solution set characteristic functions are relational algebraic $\gamma$ operators, we can formalise the definition as follows: $lift(\chi_{U},\alpha_{Base_V}) \wedge lift(\chi_{V}, \alpha_{Base_U})$ converts the original characteristic functions, combining:
\begin{equation}
\begin{split}
& \chi_U = \gamma[\varnothing][boolAgg_U() \rightarrow res](I_U) \text{, and}  \chi_V = \gamma[\varnothing][boolAgg_V() \rightarrow res](I_V) ~\text{into } \chi_{U \bowtie_{sol} V} \text{ as} \\
&\quad\gamma[\varnothing][Bool\_And(res) \rightarrow res](\gamma[\alpha_{Base_V}][boolAgg_U() \rightarrow res](I_{U\bowtie_{sol} V})) \\
&\wedge \gamma[\varnothing][Bool\_And(res) \rightarrow res](\gamma[\alpha_{Base_U}][boolAgg_V() \rightarrow res](I_{U\bowtie_{sol} V}))
\end{split}
\label{eq:lift-function}
\end{equation}
The partitioning is implemented through the inner $\gamma[\alpha_{Base_V}]$ and $\gamma[\alpha_{Base_U}]$ operations, which group the joined candidate relation by the base attributes from the opposite solution set. Universal quantification is achieved through the outer $\gamma[\varnothing][Bool\_And(res) \rightarrow res]$ operations, ensuring that the original restrictions hold across all partitions.

With this definition in place, we can illustrate natural join by recreating the Latin square problem's effective search space using a natural join of two more straightforward solution sets. In \autoref{fig:latin-Square-Solution-raset-via-natural-join} we start by defining a \codett{BoardRow} and a solution space with unique row values. Then a \codett{BoardCol} and a solution set with unique values.  \codett{EffectiveSearchSpace2} is constructed out of the two one-dimensional solution spaces: \codett{UniqueValuesInRow} and \codett{UniqueValuesInCol}.

\begin{lstlisting}[mathescape=true, language=RA,caption={Latin square Solution via $\bowtie_{sol}$},label=fig:latin-Square-Solution-raset-via-natural-join,float=ht,
  abovecaptionskip=-\medskipamount]
-- A row search space with unique values
BoardRow := $\pi$[row]($\omega$[row:IN $\pi$[value](Values)](True))
SearchRowSpace := $\omega_{sol}$(BoardRow,Values)
UniqueValuesInRow := $\sigma_{sol}$[
$\gamma$[$\varnothing$][AllDifferent(value) $\rightarrow$ ret](SearchRowSpace)
](SearchRowSpace)
-- A column search space with unique values
BoardCol := $\pi$[col]($\omega$[col:IN $\pi$[value](Values)](True))
SearchColSpace := $\omega_{sol}$(BoardCol,Values)
UniqueValuesInCol := $\sigma_{sol}$[
$\gamma$[$\varnothing$][AllDifferent(value) $\rightarrow$ ret](SearchColSpace)
](SearchColSpace)
-- Construct the same Latin square board and compare
EffectiveSearchSpace2 := UniqueValuesInRow $\bowtie_{sol}$ UniqueValuesInCol
EffectiveSearchSpace2 == EffectiveSearchSpace -- True
\end{lstlisting}

\subparagraph*{Selection $\sigma_{sol}$.} Analogously to CDRs, selection is defined in terms of natural join rather than as a separate operator. In this case, the lifting function is identity since there is no change of solution space dimensionality. By construction, the only way to filter or restrict a solution set is via a relational algebraic $\gamma$ expression with a Boolean aggregation function.

\subparagraph*{Set Operators.} Intersection and Difference require identical base relations ($Base_U = Base_V$) and identical decision attribute sets ($\alpha_{Decision_U} = \alpha_{Decision_V}$), ensuring that both characteristic functions $\chi_U$ and $\chi_V$ are already defined over the same candidate schema. Since both solution sets operate over identical domains, their characteristic functions can be directly combined through Boolean operations without dimensional transformation (lifting). Union operates under the same compatibility constraints, allowing for direct disjunction of characteristic functions: $\chi_{U \cup_{sol} V} = \chi_U \vee \chi_V$.

\subparagraph*{Outer Operators Order $\tau_{sol}$ and Limit $\lambda_{sol}$.} The outer operators Order $\tau_{sol}$ and Limit $\lambda_{sol}$ yield sequences of solution candidates rather than a solution set. Once these operators are applied, further core $RA_{sol}$ operators may not be applied, as the result is no longer a solution set. These operators provide guidance to the evaluation algorithm(s). This is analogous to the situation in SQL where the \codett{ORDER BY} clause is permissible only as part of the outer-most query and provides guidance to the database optimiser.

\subparagraph*{Projection $\pi_{sol}$.} Following the same reasoning that we applied to complete domain algebra projection, $\pi_{sol}$ is the operator that bridges from solution sets back to standard active-domain relations. $\pi_{sol}[candRankAttr][\alpha_\pi](U)$ materialises the solution set by triggering evaluation of the higher-order relational algebraic expression and returns actual solutions as tuples in a standard relation. The optional $candRankAttr$ parameter distinguishes multiple solutions when they exist, preventing ambiguity in the result relation.

\section{Why MiniZinc?} \label{app:why-minizinc}

While nothing in these algebras depends on MiniZinc, it was chosen as an exemplar intermediate language due to its algebraic and relational approach, and its role as a gateway to diverse query evaluation strategies.

\subparagraph*{Universality.} MiniZinc is a solver-independent constraint modelling language designed to preserve declarative specifications while enabling query evaluation through dozens of different approaches~\cite{nethercote_2007_minizinc.standard.cp.modelling.language}.

\subparagraph*{Algebraic and Relational Connections} The constraint-solving communities developed an early appreciation for the convenience and expressivity of algebraic specification~\cite{jaffar_1987_constraint.logic.programming} complementing logic-based formulations. We are not able to provide a full tracing of these developments; however MiniZinc stands in this lineage.  In addition, MiniZinc has been explicitly influenced by the relational model, especially in its treatment of undefinedness~\cite{frisch_2009_proper.treatment.undefinedness.constraint.languages}.

\subparagraph*{MiniZinc's PREDICATE is a Complete Domain Relation}
While we are unaware of an implementation of a complete domain relation in a database context, Minizinc's \codett{PREDICATE} construct is a faithful implementation of the construct, without an associated algebra. \autoref{lst:minizinc-cdr} is the GST example from Appendix \ref{subapp:complete-domain-relational-algebra-example} implemented in MiniZinc.
\begin{lstlisting}[language=Minizinc,caption={A faithful CDR implementation in MinZinc},label=lst:minizinc-cdr,float=ht,
  abovecaptionskip=-\medskipamount]
predicate gst(var float: price, var float: exgst, var float: gst) =
let {
    constraint price/11 = gst;
    constraint exgst = price-gst;
    }
in true;
\end{lstlisting}

The close correspondence between CDRs and MiniZinc predicates demonstrates that our framework naturally maps to existing declarative paradigms, which provide access to diverse query evaluation strategies through solvers such as Gecode, Chuffed, and OR-Tools, without our algebras committing to any particular approach.

\section{Translation from Solution Sets to Relational Algebra} \label{app:translation-solution-sets-relational-algebra}
This appendix collects the technical details for the translation from solution sets to relational algebra deferred from Section~\ref{sec:semantics-translating-solution-sets-to-relations}. We introduce $\Phi$ in four steps.  We translate the domain to $flat_U$, then translate the structure to $symI_U$. We then clarify the candidate expression join semantics, and finally demonstrate how $\Phi$ translates the characteristic function $\chi_U$.  We will show the translation of the Latin square example as we proceed. Note that the characteristic function $\chi_{flat_U}$ of the $flat_U$ relation has two conjuncts: the first is translated domain restrictions of $Decision_U$ (first step), the second conjunct is the translation of this solution set's characteristic function(last step).

\begin{figure}[!htbp]
\centering
\begin{tikzpicture}[
    box/.style={rectangle, draw=black, fill=white, minimum width=1.5cm, minimum height=0.8cm},
    container/.style={rectangle, draw=black!50, fill=gray!10, inner sep=8pt},
    subcontainer/.style={rectangle, draw=blue!50, fill=blue!10, inner sep=5pt}
]

\node[container, minimum width=5cm, minimum height=4cm] (solutionset) at (0,0) {};
\node[above] at (solutionset.north) {Solution Set $U$};

\node[subcontainer] (base) at (-1, 0.8) {
    \begin{tabular}{c}
    $\alpha_{Base_U}$ \\
    $\varepsilon_{Base_U}$
    \end{tabular}
};
\node[above, font=\small] at (base.north) {$Base_U$};

\node[subcontainer] (decision) at (-1, -1) {
    \begin{tabular}{c}
    $\alpha_{Decision_U}$ \\
    $\chi_{Decision_U}$
    \end{tabular}
};
\node[above, font=\small] at (decision.north) {$Decision_U$};

\node[subcontainer] (candidate) at (1.2, 0.8) {
    \begin{tabular}{c}
    $\alpha_{I_U}$ \\
    $\varepsilon_{I_U}$
    \end{tabular}
};
\node[above, font=\small] at (candidate.north) {$I_U \in U$};

\node[subcontainer] (chi) at (1.2, -1) {$\chi_U$};
\node[above, font=\small] at (chi.north) {Char. Func.};

\draw[->, thick, font=\Large] (3, 0) -- node[above] {$\Phi$} node[below] {$\mathfrak{D}, \mathfrak{C}$} (5, 0);

\node[container, minimum width=5cm, minimum height=4cm] (phiset) at (8,0) {};
\node[above] at (phiset.north) {$\Phi(U)$};

\node[subcontainer] (flat) at (8, 0.8) {
    \begin{tabular}{c}
    $\alpha_{flat_U}$ \\
    $\chi_{flat_U}$
    \end{tabular}
};
\node[above, font=\small] at (flat.north) {Flattened Domain $flat_U$};

\node[subcontainer] (symi) at (8, -1) {
    \begin{tabular}{c}
    $\alpha_{symI_U}$ \\
    $\varepsilon_{symI_U}$
    \end{tabular}
};
\node[above, font=\small] at (symi.north) {Symbolic Candidate $symI_U$};

\end{tikzpicture}
\caption{Translation from solution sets to relational algebra via $\Phi$ in the context of $\mathfrak{D}, \mathfrak{C}$}
\label{fig:translation-overview-appendix}
\end{figure}
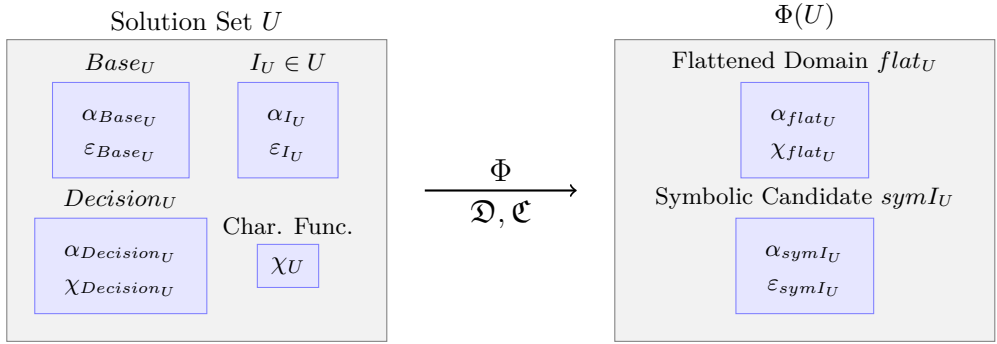

Figure~\ref{fig:translation-overview-appendix} recalls the translation structure from Section~\ref{sec:semantics-translating-solution-sets-to-relations}.

\subsection{Translating the Domain: The Flattened \texorpdfstring{$flat_U$}{flat\_U}.}
Recalling \autoref{eq:solution-set-domain-2}, which highlights the repeated cross product in the construction of a solution set, the domain of $flat_U$ will be the Cartesian product of $Decision_U$ taken $|Base_U|$ times, constrained by the characteristic function of $Decision_U$. $flat_U$ will have an attribute set $\alpha_{flat_U} = \{a_i \mid a \in \alpha_{Decision_U}, i \in \{1,2,...,|Base_U|\}\}$, where each attribute in $Decision_U$ is replicated with arbitrary attribute renaming. The restrictions that are part of $Decision_U$ are applied as follows. The first conjunct of the characteristic function $\chi_{flat_U}$ is constructed by repeated application of the characteristic function $\chi_{Decision_U}$ once for each tuple in the base relation. $\chi_{flat_U} = \bigwedge_{i=1}^{|Base_U|} \chi_{Decision_U}^i$. Where $\chi_{Decision_U}^i$ is the restriction $\chi_{Decision_U}$ with each attribute $a \in \alpha_{Decision_U}$ replaced by its corresponding renamed attribute $a \in flat_U$. The translation ensures that the constraints on valid decision values are satisfied simultaneously across all tuples of the $Base$ relation.

The domain of the Latin square problem translates to a $flat$ relation as shown in \autoref{fig:latin-square-flat}.

\begin{lstlisting}[mathescape=true, language=RA,caption={Domain of the Latin square problem translated to $flat$},label=fig:latin-square-flat,float=ht,
  abovecaptionskip=-\medskipamount]
flatLatin := $\omega$[value1:1..2, value2:1..2, value3:1..2, value4:1..2](True)
\end{lstlisting}

\subsection{Translating the Structure: The Symbolic Candidate \texorpdfstring{$symI_U$}{symI\_U}.} The role of the symbolic candidate $symI_U$ is to connect the combinatorial possibilities in $flat_U$ with the structure of the solution set and the problem domain and to enable joins with relations in $\mathfrak{D}$ and $\mathfrak{C}$.

The symbolic candidate is in the same shape as $I_U$. Its attribute set $\alpha_{symI_U} = \alpha_{Base_U} \cup \alpha_{Decision_U}$ with $|Base_U|$ tuples. The extension $\varepsilon_{symI_U}$ is formed as follows:

\hangindent=3em \hangafter=0
\noindent For each tuple $t_i \in \varepsilon_{Base_U}$ (where $1 \leq i \leq |Base_U|$), we create a corresponding tuple in $\varepsilon_{symI_U}$:

\hangindent=5em \hangafter=0
\noindent For each attribute $a \in \alpha_{Base_U}$: $c_{symI,t_i,a} = c_{Base_U,t_i,a}$ (the actual value from the base relation), and

\hangindent=5em \hangafter=0
\noindent for each attribute $a \in \alpha_{Decision_U}$: $c_{symI,t_i,a} = \langle a_i\rangle$ (a symbolic reference to the attribute in $flat_U$ that corresponds to the $i$th replication of this attribute).

In this fashion, the structure of the problem is preserved,  the symbolic candidate maps the constant values from the $Base$ relation tuple to the $Decision$ values via symbolic references to the attributes in the $flat$ relation.

The structure of the Latin square problem translates to a symbolic candidate \codett{symILatin} relation as shown in \autoref{fig:latin-square-symI}.
\begin{table}[!htbp]
\centering
\caption{Structure of the Latin square problem translated to $symI$}
\begin{tabularx}{0.4\textwidth}{|X|X|X|}
    \multicolumn{3}{l}{\codett{symILatin}} \\
    \hline
    \textbf{row:1..2}& \textbf{col:1..2} & \textbf{value:sym}  \\            \hline
    1 & 1 & $\langle$value1$\rangle$  \\ \hline
    1 & 2 & $\langle$value2$\rangle$  \\ \hline
    2 & 1 & $\langle$value3$\rangle$  \\ \hline
    2 & 2 & $\langle$value4$\rangle$  \\ \hline
\end{tabularx}

\label{fig:latin-square-symI}
\end{table}
We now turn to clarifying the semantics of relational algebraic operators in solution set characteristic function.

\subsection{Relational Algebraic Expressions in \texorpdfstring{$I\text{Expr}_U$}{IExpr\_U}} \label{subsubsec:iexprU-details}
Before showing the translation of characteristic functions, we clarify the restrictions on relational algebraic expressions in $I\text{Expr}_U$ with $R \in \mathfrak{D}$ and $C \in \mathfrak{C}$.

As we have defined above, the outermost operator for the characteristic function will be a $\gamma[\varnothing][boolAgg() \rightarrow res](I\text{Expr}_U)$ operator. The restrictions on constructing $I\text{Expr}_U$ are as follows:

\subparagraph*{Join Semantics for Candidate Expressions.} When joining candidate relations with external relations (whether from $\mathfrak{D}$ or $\mathfrak{C}$), the fundamental invariant is that each $I_U$ represents a function $f: Base_U \to Decision_U$. Preserving this leads to two cases:

Joins on base attributes only: $I_U \bowtie R$ where $\alpha_R \cap \alpha_{Decision_U} = \varnothing$. These joins filter or extend candidates based on relationships in the data. The functional dependency $Base_U \to Decision_U$ is preserved since only base attributes participate in the join. For joins with CDRs $C$, care must be taken to ensure sufficient base attributes are matched in the join to guarantee finiteness (as discussed in \autoref{appsubsec:safely-joining-active-domain-and-complete-domain-relations}).

Joins involving decision attributes: $I_U \bowtie R$ where $\alpha_R \cap \alpha_{Decision_U} \neq \varnothing$. These must preserve the function $f: Base_U \to Decision_U$ by treating the join as creating a functional dependency—the decision attributes determine values in $R$, creating derived dependencies from $Base_U$ through $Decision_U$ to the other attributes in $R$. This enables constraints that reference values conditioned on decisions (e.g., looking up weights for selected items). This requirement applies equally to CDRs $I_U \bowtie C$ where $\alpha_C \cap \alpha_{Decision_U} \neq \varnothing$, like a join to the example GST relation.

These join semantics ensure that characteristic functions can express rich constraints while maintaining the solution set's fundamental structure as a set of functions $f: Base_U \to Decision_U$.

\subparagraph*{Selection Semantics}: When selecting against $I_U$ the available attributes for restriction are those in $\alpha_{Base_U}$.  Selections over attributes in $\alpha_{Decision_U}$ are done by the $\gamma$ operator at the candidate relation level, rather than the tuple level.

\subparagraph*{Set Operation Semantics}: The set operations, Union ($\cup$), Intersection ($\cap$), and Difference ($-$) may not be applied to $I_U$ since there are no other candidate relations in scope to apply them to. Such operations are across solution sets and are defined at the level of the algebra over solution sets $RA_{sol}$, introduced above.

\subparagraph*{Outer Operators}: The outer operators---Ordering $\tau$ and Limit $\lambda$ may not be applied to $I_U$.

Given the above restrictions on the relational algebraic expressions in $I\text{Expr}_U$, here is how we translate the solution set characteristic function $\chi_U$ to $RA$ over $flat_U$ and $symI_U$:

\subsection{Translating the Characteristic Function \texorpdfstring{$\chi_U$}{chi\_U}} \label{subsubsec:translating-the-characteristic-function}
Our task is to map $\chi_U(I\text{Expr}_U)$ to an additional conjunct of $\chi_{flat_U}$. The characteristic function $\chi_U(I\text{Expr}_U)$ has a recursive structure, and we adopt a homomorphic translation approach that preserves it.  This approach follows the principle that for any operator $\oplus$ and expressions $A$ and $B$, the translation satisfies $\Phi(A \oplus B) = \Phi[\oplus](\Phi(A), \Phi(B))$, where $\Phi$ is our translation function.  At the end of this section, there is a worked example using the Latin square problem, an example of the join on decision attributes is deferred until \autoref{app:additional-examples}. Given $R \in \mathfrak{D}$ and $C \in \mathfrak{C}$, and subject to the restrictions above, we specify $\Phi(\chi_U(I\text{Expr}_U))$ by cases.

\subparagraph*{Base Cases.}
$\Phi(I_U) = symI_U$ —-- the candidate relation is replaced by its symbolic representation. $\Phi(R) = R$, and $\Phi(C) = C$ --- external relations remain unchanged.

\subparagraph*{Recursive Cases}
The join logic for both active domain and complete domain relations is the same. What is shown here for $R$ also applies to $C$.
\begin{itemize}

\item \textbf{Joins on base attributes only}: If $\alpha_R \cap \alpha_{Decision_U} = \varnothing$:
  $\Phi(I\text{Expr}_U \bowtie R) = \Phi(I\text{Expr}_U) \bowtie R$.
  By commutativity: $\Phi(R \bowtie I\text{Expr}_U) = R \bowtie \Phi(I\text{Expr}_U)$. The result will be an $I\text{Expr}_U$ with attribute set $\alpha_R \cup \alpha_{I\text{Expr}_U}$. The cardinality of the result will vary depending on the common attributes between $I\text{Expr}_U$ and $R$. Where there are no common attributes, the natural join becomes a cross product with cardinality $|I\text{Expr}_U| \times |R|$. Where there are common attributes (from the base relation part of $I\text{Expr}_U$), the join will filter to matching tuples only. The symbolic references to decision variables are replicated in \textbf{all} matching tuples, just like ordinary attribute values.
\item \textbf{Joins on decision attributes}: If $\alpha_R \cap \alpha_{Decision_U} \ne \varnothing$, then we need to encode R as a set of functionally dependent lookups, and then join it. We create a singleton symbolic relation $R'$ with the same attribute set as $R$ that is functionally dependent on $\alpha_{Base_U}$.
\begin{enumerate}
    \item We separate the attributes of $R$ into two groups: join matched attributes $\alpha_{matched}$ with individual attributes $a_m \in \alpha_{matched}$ , and the unmatched (dependent) attributes $\alpha_{dependent}$ with individual attributes $a_d \in \alpha_{dependent}$.
    \item The single tuple in $R'$ is constructed as follows: The value of attributes $a_m \in \alpha_{matched}$ will be a symbolic reference to themselves: $\langle a_m \rangle$.  The value of $a_d \in \alpha_{dependent}$ will be a relational algebraic expression retrieving the value given the matched attributes: $\pi[a_d](\sigma[ \bigwedge (a_m=\langle a_m\rangle) \mid a_m \in \alpha_{matched}](R)$. The resultant $R'$ is $|\alpha_{dependent}|$ total functions over $\alpha_{matched}$.
    \item $\Phi(I\text{Expr}_U \bowtie R) = \Phi(I\text{Expr}_U) \bowtie R'$. By commutativity: $\Phi(R \bowtie I\text{Expr}_U) = R' \bowtie \Phi(I\text{Expr}_U)$
\end{enumerate}
The result will be an $I\text{Expr}_U$ with attribute set $\alpha_R \cup \alpha_{I\text{Expr}_U}$. The cardinality of the result is unchanged, the same as the input $|I\text{Expr}_U|$. Each tuple in the input $I\text{Expr}_U$ will be extended by the unmatched attributes in $R'$, and the $\langle a_m\rangle$ values will be substituted in each tuple with the matching values in the input $I\text{Expr}_U$.
\item \textbf{Selection.} $\Phi(\sigma[\theta](I\text{Expr}_U)) = \sigma[\theta](\Phi(I\text{Expr}_U))$. The result will be an $I\text{Expr}_U$ restricted by $\theta$.
\item \textbf{Projection.} $\Phi(\pi[\alpha_\pi](I\text{Expr}_U)) = \pi[\alpha_\pi](\Phi(I\text{Expr}_U))$. The result will be an $I\text{Expr}_U$ projected over $\alpha_\pi$.
\item \textbf{Rename.} $\Phi(\rho[a \rightarrow b](I\text{Expr}_U)) = \rho[a \rightarrow b](\Phi(I\text{Expr}_U))$. The result will be an $I\text{Expr}_U$ with attribute $a$ renamed to $b$.
\item \textbf{Aggregation.} Aggregation over symbolic expressions must preserve the symbolic structure rather than computing actual values.
$$\Phi(\gamma[\alpha_\gamma][agg(args) \rightarrow res](I\text{Expr}_U)) = \gamma[\alpha_\gamma][agg(args) \rightarrow res](\Phi(I\text{Expr}_U))$$

When $\Phi(I\text{Expr}_U)$ contains symbolic references, the aggregation operation collects arguments rather than evaluating the aggregation. For example, instead of trying to evaluate $SUM(value)$, the translation will collect the symbolic values to be summed, e.g. $SUM(\langle value1 \rangle, \langle value2 \rangle, ...)$ as required. These symbolic aggregations will be evaluated under the control of the $RA_{sol}$ algebra, as part of evaluating a $\pi_{sol}$ operation.
\end{itemize}

With these definitions, we have fully specified the homomorphic translation function $\Phi$ that we introduced in \autoref{fig:translation-overview-appendix}. This translation preserves the structure of the solution set and establishes its semantics in terms of $RA$. We have mapped the domain to $flat_U$, the structure to $symI_U$, and given the homomorphic translation of the characteristic function $\chi_U$, which is the second conjunct of $\chi_{{flat}_U}$.

For the Latin square problem, given the translated domain $flat$ relation (\autoref{fig:latin-square-flat}), and the translated structure $symI$ relation (\autoref{fig:latin-square-symI}), we may complete the translation of the characteristic function starting with the \codett{UniqueValuesInRows} restriction (\autoref{fig:latinSquareSolutionraset}):
\begin{enumerate}
    \item Base Case. $\Phi$(SearchSpace) $\rightarrow$ \codett{symILatin} (\autoref{fig:latin-square-symI})
    \item Inner $\gamma$. $\Phi$($\gamma$[row][AllDifferent(value) $\rightarrow$ ret](\codett{symILatin}) $\rightarrow$ \codett{symIExpr2} (\autoref{fig:latinSquaretranslatestep2})
    \item Outer $\gamma$. $\Phi$($\gamma$[][Bool\_And(ret) $\rightarrow$ ret](symIExpr1) $\rightarrow$ \codett{symIExpr3} (\autoref{fig:latinSquaretranslatestep3})
    \item After this restriction, \codett{flatLatin}$ = True \wedge $ AllDifferent(\{$value1$,
    $value3$\})  $\wedge$ \newline
    AllDifferent(\{$value2$,
    $value4\}$)
    \item Repeating steps 1-3 for \codett{UniqueValuesInCols} restriction yields \codett{flatLatin}$ = True \wedge $ AllDifferent(\{$value1$,
    $value3$\})  $\wedge$
    AllDifferent(\{$value2$,
    $value4$\}) $\wedge $ AllDifferent(\{$value1$,
    $value2$\})  $\wedge$
    AllDifferent(\{$value3$,
    $value4$\})
    \item The subset restriction for \codett{singularSolution} translates to a restriction on the value of the top left cell yielding: \codett{flatLatin}$ = True \wedge $ AllDifferent(\{$value1$,
    $value3$\})  $\wedge$ \newline
    AllDifferent(\{$value2$,
    $value4$\}) $\wedge $ AllDifferent(\{$value1$,
    $value2$\})  $\wedge$
    AllDifferent(\{$value3$,
    $value4$\}) $\wedge$ value1 = 1
\end{enumerate}

\autoref{fig:latinSquareFullChi_flat} shows the fully translated $flat$ relation for the Latin square example. It is but a short step from here to translate this $RA$ to an evaluation algorithm which can evaluate it efficiently, and that is what we turn our attention to now.

\begin{table}
\centering
\caption{Latin square, \codett{symIExpr2} after step 2.}
\begin{tabularx}{0.4\textwidth}{|X|}
    \multicolumn{1}{l}{\codett{symIExpr2}} \\
    \hline
    \textbf{ret}  \\            \hline
    AllDifferent(\{$\langle$value1$\rangle$,
    $\langle$value3$\rangle\})$  \\ \hline
    AllDifferent(\{$\langle$value2$\rangle$,
    $\langle$value4$\rangle\})$  \\ \hline
\end{tabularx}

\label{fig:latinSquaretranslatestep2}
\end{table}
\begin{table}
\centering
\caption{Latin square, \codett{symIExpr3} after step 3.}
\begin{tabularx}{0.4\textwidth}{|>{\raggedright\arraybackslash}X|}
    \multicolumn{1}{l}{\codett{symIExpr3}} \\
    \hline
    \textbf{ret}  \\            \hline
     AllDifferent(\{$\langle$value1$\rangle$,
    $\langle$value3$\rangle\})$  $\wedge$
    AllDifferent(\{$\langle$value2$\rangle$,
    $\langle$value4$\rangle\})$  \\ \hline
\end{tabularx}

\label{fig:latinSquaretranslatestep3}
\end{table}

\begin{lstlisting}[mathescape=true, language=RA,caption={Latin square, solution set translation to $RA$},label=fig:latinSquareFullChi_flat,float=ht,
  abovecaptionskip=-\medskipamount]
flatLatin := $\omega$[value1:1..2, value2:1..2, value3:1..2, value4:1..2](
    AllDifferent({value1,value3}) AND AllDifferent({value2, value4})
    AND AllDifferent({value1,value2}) AND AllDifferent({value3, value4})
    AND value1 = 1
    solutionLatin := $\pi$[value1, value2, value3, value4](flatLatin)
)
\end{lstlisting}

\subsection{Outer Operators Guide Query Evaluation} \label{subapp:outer-operators-and-evaluation-backends}

The outer operators—$\tau_{sol}$, $\lambda_{sol}$, and $\pi_{sol}$—not only break closure over solution sets but also guide the query optimiser concerning the problem class to be evaluated. Here are the instructions provided by various combinations of outer operators ranging from decision, through satisfaction, to optimisation:

\begin{itemize}
\item $\pi_{sol}[\text{\tiny ...}][\text{\tiny ...}](\lambda_{sol}[1](U)) \rightarrow$ decision query; is there one?
\item $\pi_{sol}[\text{\tiny ...}][\text{\tiny ...}](\lambda_{sol}[k](U)) \rightarrow$ satisfaction query; with cardinality limit $k$
\item $\pi_{sol}[\text{\tiny ...}][\text{\tiny ...}](U) \rightarrow$ satisfaction query; find all
\item $\pi_{sol}[\text{\tiny ...}][\text{\tiny ...}](\lambda_{sol}[1](\tau_{sol}[\mu](U))) \rightarrow$ optimisation query; with objective $\mu$
\end{itemize}

Our examples are translated to an intermediate form in MiniZinc, and we claim that a semantically accurate translation to such intermediate forms is mechanical. Witness the nearly 1:1 correspondence between the translation of the Latin square solution (\autoref{fig:latinSquareFullChi_flat}) and the MiniZinc language, as shown in \autoref{fig:latinSquareSolutionMZ}. \autoref{fig:latinSquareSolutionMZResults} shows the results returned by MiniZinc after evaluation.

\begin{lstlisting}[mathescape=true, language=MiniZinc,caption={Latin square, Translation of relational algebra to MiniZinc},label=fig:latinSquareSolutionMZ,float=ht,
  abovecaptionskip=-\medskipamount]
include "globals.mzn";

var 1..2: value1;
var 1..2: value2;
var 1..2: value3;
var 1..2: value4;

constraint all_different([value1, value3]);
constraint all_different([value2, value4]);
constraint all_different([value1, value2]);
constraint all_different([value3, value4]);
constraint value1 = 1;

solve satisfy;
\end{lstlisting}

\begin{lstlisting}[mathescape=true, language=JSON,caption={Latin square, singleton candidate returned by MiniZinc},label=fig:latinSquareSolutionMZResults,float=ht,
  abovecaptionskip=-\medskipamount]
{
  "candidates": [{"value1": 1, "value2": 2, "value3": 2, "value4": 1}],
  "status": "SATISFIED"
}
\end{lstlisting}

The instantiation of the results as an active domain relation for further processing by operators surrounding the $\pi_{sol}$ is also mechanical. Returned values are substituted back into $symI$, reuniting the $Base$ relation values with their corresponding $Decision$ values. In the case of the Latin square problem, the relation returned by $\pi_{sol}$ is shown in \autoref{fig:latinSquareSolutionpiResults}.
\begin{table}[!htbp]
\centering
\caption{Latin square, solution returned by $\pi_{sol}$}
\begin{tabularx}{0.4\textwidth}{|X|X|X|}
    \multicolumn{3}{l}{LatinSolution} \\
    \hline
    \textbf{row:1..2}& \textbf{col:1..2} & \textbf{value:1..2}  \\            \hline
    1 & 1 & 1  \\ \hline
    1 & 2 & 2  \\ \hline
    2 & 1 & 2  \\ \hline
    2 & 2 & 1  \\ \hline
    \end{tabularx}

\label{fig:latinSquareSolutionpiResults}
\end{table}

Under the control of $\pi_{sol}$ and the other outer operators, solution sets may be translated to ordinary $RA$, to further intermediate representations as required by evaluation algorithms, and the results returned as ADRs for further processing.

\section{Survey of Prior Approaches to Increasing RA/SQL Expressivity} \label{app:survey-prior-approaches-increasing-ra-sql-expressivity}

\subsection{The Quest for Expressivity: Relational Database} \label{appsubsec:the-quest-for-expressivity-relational-database}

This section provides additional bibliographic notes on the evolution of attempts to extend $RA$ and SQL beyond active domains, supplementing the discussion in Section~\ref{sec:related-work}.

While not formalising them in an algebra, Hansen et al.~\cite{hansen_1989_integrating.relational.databases.constraint.languages} refer to CDRs as \say{rules} and illustrate how they can be joined with a data relation under a variety of access patterns (constituting a functional dependency) to calculate Ohm's law. Similarly, not pursuing an algebraic analysis, Hirst and Harel~\cite{hirst_1993_completeness.results.recursive.data.bases} describe recursive databases which might, for example, include infinite-valued trigonometric functions as relations. For a more detailed account of these developments, see~\cite{pratten_2024_relational.expressions.data.transformation.computation}.

\subsection{The Quest for Expressivity: Subset Selection and Optimisation } \label{appsec:the-quest-for-expressivity-subset-selection-and-optimisation}

This section provides detailed bibliographic notes on the evolution of subset selection and optimisation approaches in databases from the 1990s to present, supplementing the overview in Section~\ref{sec:related-work}.

\subsubsection{SQLMP: early 1990s}
Choobineh described the first synthesis of the relational model and mathematical programming SQLMP in the early 1990s~\cite{choobineh_1991_sqlmp.data.sublanguage.representation.formulation}. To create a search space SQLMP takes an ordinary database table and interprets it as all functions from the provided data columns to the columns that are empty/\codett{NULL}. The empty values are interpreted as \say{this missing constant is a variable}. SQLMP searches over both finite and infinite domains. SQLMP supports reuse through packaging but not the composition of queries over search spaces. 

We follow Choobineh in using Boolean aggregation functions as the heart of filtering search spaces. \autoref{lst:sqlmp-constraint} shows the SQLMP Boolean aggregation constraint that will be \codett{True} only for candidates where subsets of the records for each value of \codett{t} have a total of attribute \codett{x} less than 123.7. In this, SQLMP elides quantification; universal in this case: $\forall t$.

\begin{lstlisting}[language=SQL, mathescape=true, caption={SQLMP Boolean aggregation constraint},label=lst:sqlmp-constraint, float=ht,
  abovecaptionskip=-\medskipamount]
CONSTRAINT capacity_cons
   SUM(x) < 123.7
   FROM jtcx
   GROUP BY t
\end{lstlisting}

SQLMP is designed for optimisation problems with single solutions rather than satisfaction problems that may have multiple solutions. Choobineh noted that \say{No equivalent expressions exist in the relational model of data for expressing ... objective functions} and accordingly introduced objective functions through explicit \codett{MINIMIZE} and \codett{MAXIMIZE} clauses, drawing these constructs from mathematical programming.

\subsubsection{Nested Algebra and Powerset Algebra: 1990s}
Investigations into nested relational algebra~\cite{paredaens_1992_converting.nested.algebra.expressions.flat} and powerset relational algebra~\cite{gyssens_1992_powerset.algebra.natural.tool.handle} concluded that they did not provide a lift in expressivity over the \say{flat} relational algebra.

\subsubsection{ESRA: early 2000s}
In the early 2000s, Flener~\cite{flener_2001_towards.relational.modelling.combinatorial.optimisation, flener_2003_introducing.esra.relational.language.modelling} introduced the \textit{Executable Symbolism for Relational Algebra} ESRA language with the goal of advancing \say{solver-independent, high-level relational constraint modelling.} ESRA applies relational concepts to constraint programming, particularly introducing database Entity-Relationship design principles to structure constraint models. While ESRA uses the term \say{relational algebra}, it develops its own formalism for constraint modelling rather than extending Codd's relational algebra~\cite{codd_1970_relational.model.data.large.shared}. ESRA's contribution lies in demonstrating how relational thinking—particularly entity-relationship modelling—can organize and structure constraint programming problems.

\subsubsection{Subset Relational Algebra and SQL: 2004}
Valluri and Karlapalem's 2004~\cite{valluri_2004_subset.queries.in.relational.databases} contribution framed a class of optimisation problems as a search over subsets of database relations. They propose a complete $RA$ extended to address subsets and corresponding extensions to SQL. Given relation $R$, their powerful algebra searches over all functions $f: R \rightarrow \{True, False\}$ and demonstrates how it can be utilised (in principle) to solve a set of combinatorial optimisation problems within the context of databases. \say{In principle} because, as they acknowledge, their subset algebra is not efficiently evaluable. The gap lies in their approach: the algebra's semantics are defined in terms of set operations over tuples instead of set characteristic functions, which would have provided a pathway to efficient evaluation using solving technologies.

\subsubsection{NP-Alg and ConSQL+ : 2007-2012}
Next, we consider a research contribution spanning from 2007 to 2012 by Cadoli, Mancini, Flener, and Pearson~\cite{cadoli_2007_relational.algebra.sql.constraint.modelling, mancini_2010_local.search.over.relational.databases, mancini_2012_combinatorial.problem.relational.databases.view}. This research partnership yielded an exploration of $RA$ as a basis for integrating relational databases and combinatorial optimisation. Called NP-Alg, the extension can express NP-complete decision problems such as k-colouring, independent sets and clique. In the spirit of SQLMP the group also presented a strict superset of SQL  ConSQL+ for evaluating combinatorial optimisation problems. We will consider these two contributions in turn.

NP-Alg creates a search space derived from data in the database using non-determinism as a first-class language feature~\cite{floyd_1967_nondeterministic.algorithms}. Given $i$ as an index, \codett{GUESS} $Q_i$ denotes relations $Q_i$ with an arbitrary extension, and given $j$ as an index, we can denote ordinary ADRs as $R_j$. Then expressions in NP-Alg are success-on-empty constraints using ordinary relational algebraic operators between the $Q_i$'s as variables and $R_j$'s as constants.  NP-Alg is confined to handling materialisable discrete domains, and doesn't support optimisation problems. NP-Alg does not address how multiple solution candidates to a combinatorial problem may be introduced into the \say{ordinary} processing world of relational operators.

The intuition underlying NP-Alg, that a more powerful $RA$ should underpin a more powerful SQL, is well taken. However, later work by Gutierrez, Van Roy, and Cauwelaert~\cite{gutierrez_2013_implementation.relation.domain.constraint.programming} characterises this pattern of using $RA$ operators as creating a \say{relation constraint domain with multiple decision variables}.  Combined with explicit nondeterminism in the programming model to create variables, NP-Alg can be understood as embedding $RA$ operators within a constraint specification language, using relations as constraint variables with nondeterministic semantics rather than extending the functional composition semantics of standard $RA$.

ConSQL+ shares with NP-Alg an explicit nondeterminism to express combinatorial problems as an extension of SQL. Search spaces were created using a second-order \codett{VIEW} within a new \codett{SPECIFICATION} context. These views have attributes from ordinary ADRs along with nondeterministic \codett{CHOOSE()} attributes which correspond to NP-Alg's \codett{GUESS}. Each second-order view represents all functions $f : D \rightarrow C$, where $D$ is data from the database and $C$ are guessed values. Within the \codett{SPECIFICATION} context, then, constraints are expressed using SQL \codett{CHECK} constraints similar Choobineh's SQLMP.  Typically, ConSQL+ frames constraints as \codett{NOT EXISTS (SQL query)}. In addition, like SQLMP optimisation objective functions are introduced through explicit \codett{MAXIMIZE} or \codett{MINIMIZE} clauses.

At the same time, the achievements of ConSQL+ come with some limitations. ConSQL+ is not fully translatable to NP-Alg as the latter can't express optimisation objective functions.

\subsubsection{SCL: 2008-2011}
Drawing on learnings from ConSQL, Siva~\cite{siva_2011_enabling.relational.databases.effective.csp} in his 2011 doctoral thesis introduced SCL, which pioneered the use of functional dependencies to construct search spaces explicitly. SCL is the only prior work we have found that creates search spaces through functional dependency specifications—an approach central to our solution sets. The following excerpt (\autoref{lst:scl-problem-specification}) from the solution to a round-robin tournament problem demonstrates how this can be achieved using Siva's SQL Constraint Data Engine Command Language SCL, a deft extension of SQL DDL.

\begin{lstlisting}[language=SQL, mathescape=true, caption={SCL Round-Robin Problem Specification},label=lst:scl-problem-specification, float=ht,
  abovecaptionskip=-\medskipamount]
CREATE CONSTRAINT TABLE schedule (
    week INT FOREIGN KEY REFERENCES weeks (id),
    period INT FOREIGN KEY REFERENCES periods (id),
    home INT FOREIGN KEY REFERENCES teams (id),
    away INT FOREIGN KEY REFERENCES teams (id),

    KEY (home, away),

-- No team can play itself.
CONSTRAINT C0 CHECK (NOT EXISTS
    (SELECT * FROM schedule s
        WHERE s.home <= s.away ))
...
\end{lstlisting}
Here, we have a \codett{CONSTRAINT TABLE} defining a search space. The domains of the search space attributes are specified via foreign key references to the four domains of interest. The search space is all subsets of the powerset of $week \times period \times home \times away$. Crucially, the \codett{KEY} clause imposes a functional dependency from $home \times away$ to $week \times period$. Thus, the search space consists of all functions $f: home \times away \rightarrow week \times period$. This explicit use of functional dependencies to define search spaces as sets of functions anticipates our solution set formulation, though SCL expresses this through SQL's DDL rather than algebraically.

SCL demonstrates that functional dependencies can naturally express combinatorial search spaces in relational terms. While SCL is limited to discrete domains and decision problems, its functional dependency insight represents a significant conceptual advance that we build upon in our solution sets.

\subsubsection{Sampling and Clustering Queries: 2011-2024}
DivDB~\cite{vieira_2011_divdb} proposed SQL extensions for diversity queries, notably reusing the ordinary \codett{ORDER BY} clause to order candidate subsets rather than creating non-standard clauses. DivDB specified subset search spaces by overloading the GROUP BY clause to generate candidate subsets, though this approach raises challenges for reconciling with standard SQL semantics. While DivDB provided a user-facing language, recent contributions in sampling and clustering~\cite{agarwal_2024_computing.well.representative.summary.conjunctive,arenas_2024_tractability.diversity.query.answers.ultrametrics, gan_2024_optimal.dynamic.parameterized.subset.sampling, galhotra_2024_k.clustering.comparison.distance.oracles} focus on optimised algorithms without a relational language for problem specification.

\subsubsection{Package Queries: 2014-2025}
Under the rubric of package queries, Fernandes, Brucato, Ramakrishna, Abouzied, Meliou, Beltran, Mai, Wang, and Haas made research contributions concerning SQL-based combinatorial optimisation from 2014-2025~\cite{fernandes_2014_packagebuilder.querying.for.packages.tuples, brucato_2016_scalable.package.queries.relational.database, brucato_2020_spaqltools.stochastic.package.query.interface, mai_2023_scaling.package.queries.billion.tuples}.  A package is a \say{collection of tuples with certain global properties deﬁned on the collection as a whole.}~\cite{fernandes_2014_packagebuilder.querying.for.packages.tuples}. Recent work on stochastic optimisation has adopted the \say{missing values are variables} approach to expand the search space to include functions to infinite attributes~\cite{brucato_2020_spaqltools.stochastic.package.query.interface}.  Package queries may be expressed in a dialect of SQL called PaQL. The expression \codett{PACKAGE(R) [REPEAT M]} denotes a search space as a powermultiset of possible collections of tuples from relation $R$ with multiplicity $M$. It expresses a search space of all functions $f: R \rightarrow \{0..M\}$: for some natural number $M$. Solution sets in this paper are a generalisation of the package construct.

A key contribution of this thread of work is the advances made through the SketchRefine algorithm, which enabled scaling of such package queries to millions of tuples and stochastic optimisation. The 2024 contribution by Mai et al. extended this algorithm's scaling to billions of tuples~\cite{mai_2023_scaling.package.queries.billion.tuples}. The interested reader is referred to a recent survey article covering this and related Prescriptive Analytics research~\cite{meliou_2025_data.management.perspectives.prescriptive.analytics}. PaQL's filtering constraints (SUCH THAT clause) and objective functions (\codett{MINIMIZE} and \codett{MAXIMIZE} clauses) follow the pattern established in SQLMP, ConSQL+ and SCL. PaQL is designed primarily for single solution optimisation. The package query framework focuses on algorithmic advances and SQL extensions rather than developing a corresponding $RA$.

\subsubsection{SolveDB(+): 2016-2021}
Between 2016 and 2021, researchers from Aalborg University, Denmark, including Siksnys, Pedersen, Nielsen and Frazzetto, extended SQL as SolveDB to express and solve combinatorial problems with an implementation in PostgreSQL~\cite{siksnys_2016_solvedb.optimization.problem.solvers.sql, siksnys_2017_demonstrating.solvedb.sql.dbms.optimization, siksnys_2021_solvedb.sql.based.prescriptive.analytics}. To create a search space, SolveDB uses a \codett{var IN query} clause, which, like SQLMP above, overloads \codett{NULL} values with the meaning of \say{variable}. SolveDB has the most expressive search space that we have encountered so far in our review of prior work. Given a finite relation $R$ and potentially infinite relation $S$ with, for example, \codett{INT} and \codett{FLOAT} attributes, the search space considered by SolveDB is all functions $f: R \rightarrow S$ and contains problems in NP-complete/hard, including decision and optimisation problems. Concerning the composition of queries, enhancements in 2021~\cite{siksnys_2021_solvedb.sql.based.prescriptive.analytics} increased the reusability of queries through common decision table expressions and model inlining.

SolveDB adds a separate query context to SQL called \codett{SOLVESELECT}.  Within this query context, \codett{SUBJECTO} clauses use the filtering aggregated queries familiar to us from the earlier work to express constraints. As in earlier contributions, \codett{MINIMIZE} and \codett{MAXIMIZE} clauses introduce an aggregated query as an objective function. Finally, the \codett{WITH solver} clause provides a way of giving hints to the query evaluator on which of the available solvers to use. Like other SQL extensions surveyed, SolveDB focuses on language design and implementation rather than developing an underlying $RA$ for these operations.

\subsubsection{CombSQL+: 2018-2020}
Between 2018 and 2020, Sakanashi and Sakai from Nagoya University, Japan, contributed CombSQL+ as an extension of SQL for combinatorial optimisation problems~\cite{sakanashi_2018_transformation.combinatorial.optimization.written.sql, sakanashi_2020_transformation.sql.combinatorial.optimization.constraint}. CombSQL+ has roots in earlier work by their research group and in Cadoli and Manicini's ConSQL. Like the earlier ConSQL, CombSQL+ defines search spaces for combinatorial optimisation problems by incorporating explicit nondeterminism via a \codett{CHOOSE()} function. Given a database relation $R$ and a relation $S$ formed by the cross product of non-deterministically chosen values from finite domains, the solution space expressed by CombSQL+ is all functions $f: R \rightarrow S$. Similar to earlier research, filtering of candidate tables is via aggregation queries in a \codett{SUCH THAT} clause and objective functions by \codett{MINIMIZE} and \codett{MAXIMIZE} clauses.

CombSQL+ represents a step forward in the programming model offered for combinatorial optimisation. CombSQL+ is the first contribution in the literature to explicitly inform the programmer that transitioning from ordinary SQL to combinatorial optimisation involves a shift from tables/relations to sets of tables/relations.

The second significant contribution of COMBSQL+ is a translation algorithm that converts SQL queries over sets of relations into input for a Constraint Programming/SMT solver. This pioneering algorithm shows that a disciplined translation is possible. However, due to its multi-paradigm approach, it is not directly applicable to the three algebras introduced here. Specifically, the algorithm is based on search spaces as \say{tuples of sets of relations} rather than just sets of relations, it synthesises set-theoretic concepts with constraint programming constructs (variables and evaluations), and directly targets CP/SMT solvers rather than evaluation strategy neutral $RA$.

CombSQL+ can specify finite domain optimisation problems.

\subsubsection{Unified Relational Query Language (URQL): 2020}
In 2020, Valdron and Pu~\cite{valdron_2020_data.driven.relational.constraint.programming} introduced an unnamed extension to SQL for unifying relational databases and constraint satisfaction, which they described as a \say{unified relational query language}.  Accordingly, let's call it URQL for this discussion. URQL defines the search space for combinatorial satisfaction problems by introducing decision variables and constraints into the database as column types and values. In contrast to SQLMP and SolveDB, instead of signalling variables by a missing value, in URQL there is a special constructor \codett{new\_var()} used in \codett{INSERT} statements for this purpose. The goals for satisfaction are provided through a \codett{CREATE GOAL[S]} statement, and referenced variables are captured ready for translating related queries to a constraint satisfaction problem (CSP).

URQL's approach of treating both variables and constraints as first-class database values represents a distinct design choice, storing the problem specification itself as data rather than extending the query language or algebra.

This multi-decade survey reveals a clear pattern: despite significant algorithmic advances and creative SQL extensions, each approach remains isolated in its own silo—using incompatible semantics (\codett{NULL}s as variables, nondeterministic relations, DDL specifications, or first-class variable types) and supporting different problem classes (decision-only, discrete-only, or single-solution)—underscoring the need for a unified algebraic foundation that can express all these capabilities within a single, compositional framework.

\section{Additional Examples} \label{app:additional-examples}

The Cakes Production example in \autoref{sec:validation-through-representative-problems} demonstrated how data parameterises an optimisation problem within our unified framework. This appendix presents four additional examples that showcase capabilities beyond basic optimisation:
\begin{itemize}
    \item \textbf{Market Selection}, a multi-stage problem where solutions cascade through sequential decisions (\autoref{subapp:multi}),
    \item \textbf{Meal Planning}, showing joins on decision attributes that enable decision value-dependent lookups during solving (\autoref{subapp:eg-meal-planner}),
    \item \textbf{Energy Balancing}, showing optimisation over infinite domains~(\autoref{subapp:energy}), and finally,
    \item \textbf{Pareto-optimal subset selection}~(\autoref{subapp:kursawe}).
\end{itemize}
Together with the earlier Latin Square example from the main text, these demonstrate that our algebraic framework achieves the expressiveness of the most powerful prior approaches while maintaining compositional clarity.

\subsection{Multi-Stage Optimisation through Composition.} \label{subapp:multi}
In addition to the combination of optimisation problems through natural join into a single higher-dimensional optimisation problem, sequential composition is also naturally expressed through the $RA_{sol}$ algebra. For example, a business expansion decision might involve selecting a new market based on state-level data. Once a market is chosen, locations for stores may be determined based on more detailed population and geographic data.  In $RA_{sol}$, this combined problem may be expressed naturally with the selected market of the first problem parameterising the base relation of the second.  \autoref{fig:multi-stage} sketches the idea with each stage's solutions becoming the structural foundation for the subsequent stage, declaratively expressing sequential decision-making problems.

\begin{lstlisting}[mathescape=true, language=RA,caption={Multi-stage optimisation pseudocode},label=fig:multi-stage,float=ht,
  abovecaptionskip=-\medskipamount]
BestMarket := $\lambda[1](\tau[\mu](\omega_{sol}$(MarketsToConsider,DecisionM)))
-- possible locations in this market
MarketLocations :=  $\sigma[market](\pi_{sol}$(BestMarket)) $\bowtie$ LocationsToConsider
-- up to five locations
Locations := $\lambda[5](\tau[\mu](\omega_{sol}$(MarketLocations,DecisionML)))
\end{lstlisting}

\subsection{Meal Planner} \label{subapp:eg-meal-planner}
The Meal Planner problem is adapted from Brucato et. al.'s introduction to PaQL~\cite{brucato_2017_scalable.execution.engine.package.queries} to demonstrate joins on $Decision$ attributes and how relations in the evaluation context may be made available at evaluation time. A set of recipes should be combined to create a day's meals, ensuring that the meals are different, gluten-free, and the daily kCal total is between 2.0 and 2.5. The three meals are named, and recipes are assigned to a specific meal rather than just included in an undifferentiated daily plan. Figures \ref{fig:meal-planner-specification} and \ref{fig:meal-planner-translation} trace the complete example.

\begin{figure}[!htbp]
\centering
\includegraphics[width=0.8\linewidth]{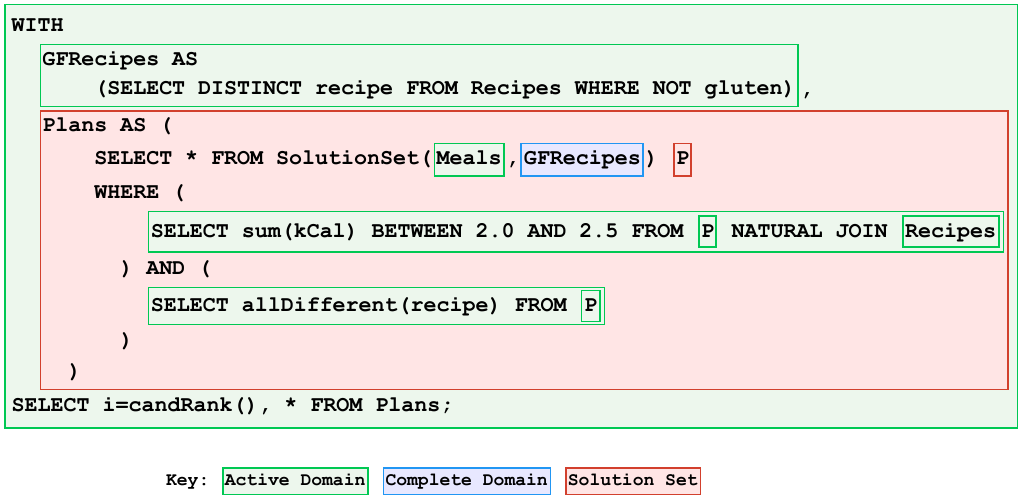}
\caption{Meal Planner Illustration SQL}
\label{fig:meal-planner-sql-sol}
\end{figure}

Starting with \autoref{fig:meal-planner-specification}, we are assigning recipes to meals, and so the \codett{Meals} relation is an appropriate $Base$ relation. We capture the choice of a gluten-free recipe for each meal in \codett{GFRecipes} as our $Decision$ relation. This construction takes advantage of the fact that any unary active domain relation may be interpreted as a complete domain relation. The relevant context is in the \codett{Recipes} relation. The problem is fully stated in $RA_{sol}$. The solution is illustrated in SQL (\autoref{fig:meal-planner-sql-sol}).

Proceeding to the translation and evaluation, $\Phi$ follows the procedure for joining on decision attributes in \autoref{subsubsec:translating-the-characteristic-function}, and transforms \codett{Recipes} into \codett{Recipes'} as shown in \autoref{fig:meal-planner-Recipes-prime}. Then this is joined with \codett{symIP}, resulting in the relation shown in \autoref{fig:meal-planner-symIExprMP1}. $\Phi$ then generates the $symI$ and $flat$ relations as shown in \autoref{fig:meal-planner-translation}. Under the guidance of $\pi_{sol}$, the translation to the intermediate form required by MiniZinc idiomatically expresses the Recipe relation as a set of keys and three lookup arrays, one for each $\pi$ symbolic lookup. There are twelve solutions output, two sets of satisfying recipes that can be assigned to each of the three meals in six ways. To accommodate the multiple candidate solutions, the $\pi_{sol}$ accepts an attribute name \codett{i} as a candidate identifier.

With this, we complete the end-to-end translation and evaluation of an assignment combinatorial problem that is expressed via a join on a decision attribute.

\begin{table}[!htbp]
\centering
\caption{Meal Planner $Recipes'$ relation}
\begin{tabularx}{1\textwidth}{|p{2.9cm}|>{\raggedright\arraybackslash}X|>{\raggedright\arraybackslash}X|>{\raggedright\arraybackslash}X|}
    \multicolumn{4}{l}{\codett{Recipes'}} \\
    \hline
    \textbf{recipe:\newline decisionAttr}& \textbf{satFat: sym} & \textbf{kCal: sym} & \textbf{gluten: sym}  \\            \hline
    $\langle recipe\rangle$ & $\pi[satFat](\sigma[recipe=\langle recipe\rangle](Recipes)$ & $\pi[kCal](\sigma[recipe=\langle recipe\rangle](Recipes)$ & $\pi[gluten](\sigma[recipe=\langle recipe\rangle](Recipes)$ \\ \hline
\end{tabularx}

\label{fig:meal-planner-Recipes-prime}
\end{table}

\begin{table}[!htbp]
\centering
\caption{Meal Planner:  symIP $\bowtie$ Recipes' }
\begin{tabularx}{1\textwidth}{|p{1.5cm}|p{2cm}|>{\raggedright\arraybackslash}X|>{\raggedright\arraybackslash}X|>{\raggedright\arraybackslash}X|}
    \multicolumn{5}{l}{\codett{symIP $\bowtie$ Recipes'}} \\ \hline
    \textbf{meal:\newline STRING} & \textbf{recipe:sym} & \textbf{satFat: sym} & \textbf{kCal: sym} & \textbf{gluten: sym} \\            \hline
    Breakfast & $\langle recipe1 \rangle$ & $\pi[satFat](\sigma[recipe=\langle recipe1\rangle](Recipes)$ & $\pi[kCal](\sigma[recipe=\langle recipe1\rangle](Recipes)$ & $\pi[gluten](\sigma[recipe=\langle recipe1\rangle](Recipes)$ \\ \hline
    Lunch  & $\langle recipe2 \rangle$ &$\pi[satFat](\sigma[recipe=\langle recipe2\rangle](Recipes)$ &$\pi[kCal](\sigma[recipe=\langle recipe2\rangle](Recipes)$ & $\pi[gluten](\sigma[recipe=\langle recipe2\rangle](Recipes)$ \\ \hline
    Dinner & $\langle recipe3 \rangle$ &$\pi[satFat](\sigma[recipe=\langle recipe3\rangle](Recipes)$ &$\pi[kCal](\sigma[recipe=\langle recipe3\rangle](Recipes)$ &$\pi[gluten](\sigma[recipe=\langle recipe3\rangle](Recipes)$ \\ \hline
    \end{tabularx}

\label{fig:meal-planner-symIExprMP1}
\end{table}

\newsavebox{\exMPBaseboxSavebox}
\begin{lrbox}{\exMPBaseboxSavebox}
\begin{tabular}{|l|}
\hline
\textbf{meal} \\
\hline
Breakfast \\
Lunch \\
Dinner \\
\hline
\end{tabular}
\end{lrbox}

\newsavebox{\exMPDecisionboxSavebox}
\begin{lrbox}{\exMPDecisionboxSavebox}
\begin{lstlisting}[mathescape=true, language=RA, backgroundcolor={}]
NonGluten := $\pi[recipe](\sigma$[NOT gluten](Recipes))
GFRecipes := $\omega$[recipe: IN NonGluten](True)
\end{lstlisting}
\end{lrbox}

\newsavebox{\exMPContextboxSavebox}
\begin{lrbox}{\exMPContextboxSavebox}
\begin{tabular}{|l|r|r|r|}
\hline
\multicolumn{4}{|c|}{\textbf{Recipes}} \\
\hline
recipe & satFat & kCal & gluten \\
\hline
1 & 7.1 & 0.45 & False \\
2 & 5.2 & 0.55 & False \\
3 & 1.0 & 0.20 & True \\
4 & 3.2 & 0.25 & False \\
5 & 6.5 & 0.15 & False \\
6 & 2.0 & 1.20 & False \\
7 & 4.0 & 0.90 & True \\
\hline
\end{tabular}

\end{lrbox}

\newsavebox{\exMPRasolboxSavebox}
\begin{lrbox}{\exMPRasolboxSavebox}
\begin{minipage}{7.5cm}
\begin{lstlisting}[language=RA, basicstyle=\normalsize\ttfamily, mathescape=true, backgroundcolor={}, frame=none]
P := $\omega_{sol}$(Meals,GFRecipes)
Plans := $\sigma_{sol}$[
  $\gamma[\varnothing]$[sum(kCal) BETWEEN 2.0 AND 2.5]
    (P $\bowtie$ Recipes)
  AND $\gamma[\varnothing]$[AllDifferent(recipe)](P)
  ](P)
LetsUsePlans := $\pi_{sol}$[i][meal, recipe](Plans)
\end{lstlisting}
\end{minipage}
\end{lrbox}

\newsavebox{\exMPSymIboxSavebox}
\begin{lrbox}{\exMPSymIboxSavebox}
\begin{tabular}{|l|l|}
\hline
\textbf{meal} & \textbf{recipe} \\
\hline
Breakfast & $\langle$recipe1$\rangle$ \\
Lunch & $\langle$recipe2$\rangle$ \\
Dinner & $\langle$recipe3$\rangle$ \\
\hline
\end{tabular}
\end{lrbox}

\newsavebox{\exMPFlatboxSavebox}
\begin{lrbox}{\exMPFlatboxSavebox}
\begin{minipage}{7.5cm}
\begin{lstlisting}[mathescape=true, language=RA, backgroundcolor={}]
NonGluten := $\pi[recipe](\sigma$[NOT gluten](Recipes))
FlatP2 := $\omega$[recipe1: IN NonGluten, recipe2: IN NonGluten, recipe3: IN NonGluten](
    AllDifferent(recipe1, recipe2, recipe3) AND
    sum($\pi[kCal](\sigma[recipe=\langle recipe1\rangle](Recipes))$,
        $\pi[kCal](\sigma[recipe=\langle recipe2\rangle](Recipes))$,
        $\pi[kCal](\sigma[recipe=\langle recipe3\rangle](Recipes))$
    ) BETWEEN 2.0 AND 2.5
)
Results := $\pi[i][recipe1, recipe2, recipe3](FlatP2)$
\end{lstlisting}
\end{minipage}
\end{lrbox}

\newsavebox{\exMPIntermediateboxSavebox}
\begin{lrbox}{\exMPIntermediateboxSavebox}
\begin{minipage}{6.5cm}
\begin{lstlisting}[language=MiniZinc, backgroundcolor={}]
include "globals.mzn";

predicate between(var float: x, float: lower, float: upper) =
    lower <= x /\ x <= upper;

set of int: recipe = 1..7;
array[recipe] of float: satFat = [7.1,5.2,1.0,3.2,6.5,2.0,4.0];
array[recipe] of float: kCal = [0.45, 0.55, 0.20, 0.25, 0.15, 1.2, 0.9];
array[recipe] of bool: gluten = [false, false, true, false, false, false, true];

set of int: NonGluten = {1,2,4,5,6};

var NonGluten: recipe1;
var NonGluten: recipe2;
var NonGluten: recipe3;

constraint all_different([recipe1, recipe2, recipe3]);
constraint between(sum([kCal[recipe1],kCal[recipe2],kCal[recipe3]]),2.0,2.5);

solve satisfy;
\end{lstlisting}
\end{minipage}
\end{lrbox}

\newsavebox{\exMPOutputboxSavebox}
\begin{lrbox}{\exMPOutputboxSavebox}
\begin{tabular}{|r|l|r|}
\hline
\textbf{i} & \textbf{meal} & \textbf{recipe} \\
\hline
1 & Breakfast & 2 \\
1 & Lunch & 4 \\
1 & Dinner & 6 \\
... & ... & ... \\
6 & Breakfast & 1 \\
6 & Lunch & 2 \\
6 & Dinner & 6 \\
        ... & ... & ... \\
\hline
\end{tabular}
\end{lrbox}

\begin{figure}[!htbp]
    \centering
    \resizebox{\textwidth}{!}{
    \begin{tikzpicture}[
        node distance=5cm,
        every node/.style={minimum height=1.5cm, font=\normalsize},
        basebox/.style={rectangle, draw={rgb,255:red,0;green,200;blue,83}, fill={rgb,255:red,237;green,247;blue,237}, inner sep=10pt, minimum width=7.5cm, minimum height=3.5cm},
        decisionbox/.style={rectangle, draw={rgb,255:red,33;green,150;blue,243}, fill={rgb,255:red,232;green,232;blue,255}, inner sep=10pt, minimum width=7.5cm, minimum height=3.5cm},
        contextbox/.style={rectangle, draw=purple!30, fill=purple!10, inner sep=10pt, minimum width=7.5cm, minimum height=6cm},
        rasolbox/.style={rectangle, draw={rgb,255:red,210;green,64;blue,45}, fill={rgb,255:red,255;green,229;blue,229}, inner sep=10pt, minimum width=7.5cm, minimum height=6cm},
        transformbox/.style={ellipse, draw=black!0, fill=gray!0, inner sep=3pt, font=\LARGE\bfseries},
        labelstyleleft/.style={font=\normalsize\bfseries, anchor=south west},
        labelstyleright/.style={font=\normalsize\bfseries, anchor=south east},
        arrow/.style={->, thick, >=stealth}
    ]

    \node[basebox] (base) {
        \scalebox{1.2}{\usebox{\exMPBaseboxSavebox}}
    };
    \node[labelstyleleft, above=-0.5cm and 0cm of base.north] {Base relation (\codett{Meals})};

    \node[decisionbox, right=.5cm of base] (decision) {
        \scalebox{1.0}{\usebox{\exMPDecisionboxSavebox}}
    };
    \node[labelstyleright, above=-0.5cm and 0cm of decision.north] {Decision relation (\codett{GFRecipes})};

    \node[contextbox, below=1cm of base] (context) {
        \scalebox{1.0}{\usebox{\exMPContextboxSavebox}}
    };
    \node[labelstyleleft, above =-0.5cm and 0cm of context.north] {$\mathfrak{D}$, $\mathfrak{C}$ context};

    \node[rasolbox, below=1cm of decision] (rasol) {
        \scalebox{1.0}{\usebox{\exMPRasolboxSavebox}}
    };
    \node[labelstyleright, above right=-0.5cm and 0cm of rasol.north] {$RA_{sol}$ specification};

    \node[transformbox, below=0.5cm of $(context.south)!0.5!(rasol.south)$] (phi) {$\Phi$};

    \node[below=0.5cm of phi, font=\normalsize\itshape] {(Continued in Figure \ref{fig:meal-planner-translation})};

    \draw[arrow] (base.south) to[out=-25, in=155] (rasol.north);
    \draw[arrow] (decision.south) to[out=-90, in=90] (rasol.north);
    \draw[arrow] (context.south) to[out=-60, in=135] (phi.north west);
    \draw[arrow] (rasol.south) to[out=-135, in=45] (phi.north east);

    \end{tikzpicture}
    }
    \caption{Meal planner optimisation (Part A): Problem specification through $RA_{sol}$ leading to homomorphic translation $\Phi$.}
    \label{fig:meal-planner-specification}
\end{figure}
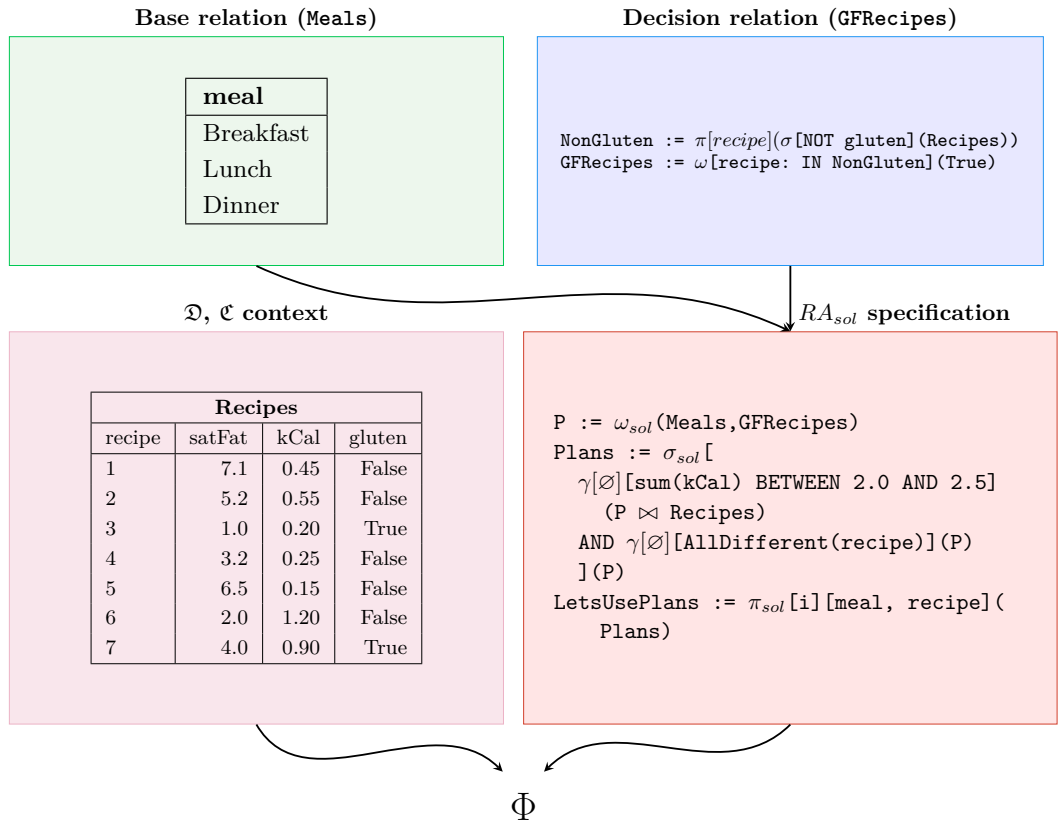

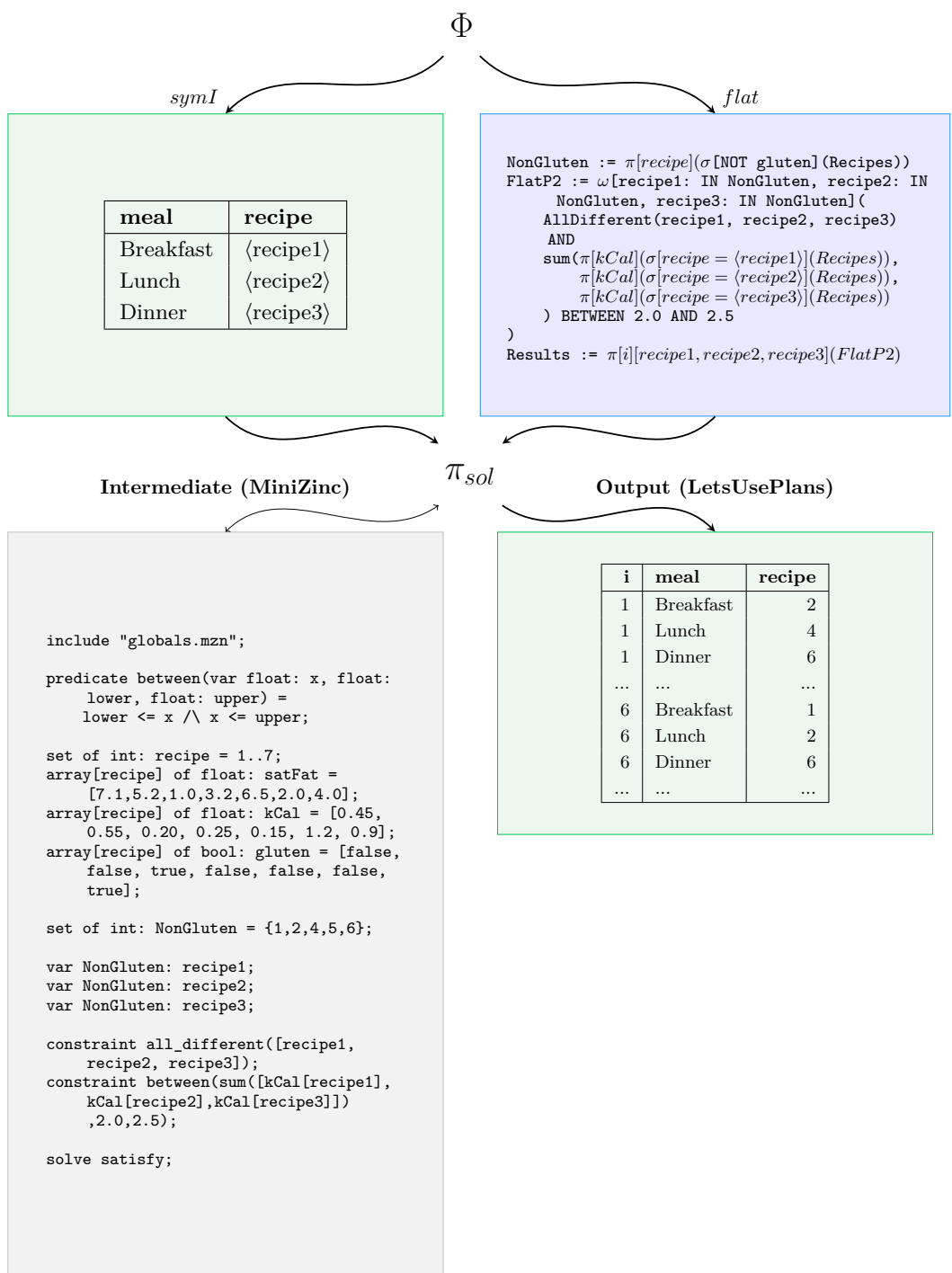
\begin{figure}[!htbp]
    \centering
    \resizebox{\textwidth}{!}{
    \begin{tikzpicture}[
        node distance=5cm,
        every node/.style={minimum height=1.5cm, font=\normalsize},
        transformbox/.style={ellipse, draw=black!0, fill=gray!0, inner sep=3pt, font=\LARGE\bfseries},
        flatbox/.style={rectangle, draw={rgb,255:red,33;green,150;blue,243}, fill={rgb,255:red,232;green,232;blue,255}, inner sep=10pt, minimum width=7.5cm, minimum height=5.25cm},
        symIbox/.style={rectangle, draw={rgb,255:red,0;green,200;blue,83}, fill={rgb,255:red,237;green,247;blue,237}, inner sep=10pt, minimum width=7.5cm, minimum height=5.25cm},
        intermediatebox/.style={rectangle, draw=gray!50, fill=gray!10, inner sep=10pt, minimum width=7.5cm, minimum height=13cm},
        outputbox/.style={rectangle, draw={rgb,255:red,0;green,200;blue,83}, fill={rgb,255:red,237;green,247;blue,237}, inner sep=10pt, minimum width=7.5cm, minimum height=5.25cm},
        labelstyleleft/.style={font=\normalsize\bfseries, anchor=south west},
        labelstyleright/.style={font=\normalsize\bfseries, anchor=south east},
        arrow/.style={->, thick, >=stealth}
    ]

    \node[transformbox] (phi) at (0,0) {$\Phi$};

    \node[above=0.5cm of phi, font=\normalsize\itshape] {(Continued from Figure \ref{fig:meal-planner-specification})};

    \node[symIbox, below left=1cm and 0cm of phi] (symi) {
        \scalebox{1.2}{\usebox{\exMPSymIboxSavebox}}
    };
    \node[labelstyleleft, above left=-0.5cm and 0cm of symi.north] {$symI$};

    \node[flatbox, below right=1cm and 0cm of phi] (flat) {
        \scalebox{1.0}{\usebox{\exMPFlatboxSavebox}}
    };
    \node[labelstyleright, above right=-0.5cm and 0cm of flat.north] {$flat$};

    \node[transformbox, below=.25cm of $(symi.south)!0.5!(flat.south)$] (pisol) {$\pi_{sol}$};

    \node[intermediatebox, below=2cm of symi] (minizinc) {
        \begin{minipage}[c][4.8cm][c]{6.8cm}
        \centering
        \scalebox{1}{\usebox{\exMPIntermediateboxSavebox}}
        \end{minipage}
    };
    \node[labelstyleleft, above =0cm and 0cm of minizinc.north] {Intermediate (MiniZinc)};

    \node[outputbox, below=2cm of flat] (output) {
        \scalebox{1}{\usebox{\exMPOutputboxSavebox}}
    };
    \node[labelstyleright, above =0cm and 0cm of output.north] {Output (LetsUsePlans)};

    \draw[arrow] (phi.south west) to[out=-135, in=45] (symi.north);
    \draw[arrow] (phi.south east) to[out=-45, in=135] (flat.north);
    \draw[arrow] (symi.south) to[out=-45, in=145] (pisol.north west);
    \draw[arrow] (flat.south) to[out=-135, in=35] (pisol.north east);
    \draw[<->] (pisol.south west) to[out=-145, in=45] (minizinc.north);
    \draw[arrow] (pisol.south east) to[out=-35, in=135] (output.north);

    \end{tikzpicture}
    }
    \caption{Meal planner optimisation (Part B): Translation through $\Phi$ to relational algebra, and evaluation via MiniZinc to final solution.}
    \label{fig:meal-planner-translation}
\end{figure}

\FloatBarrier
\subsection{Energy Balancing} \label{subapp:energy}
Siksynys and Pedersen's  SolveDB paper~\cite{siksnys_2016_solvedb.optimization.problem.solvers.sql} introduced an Energy Balancing problem, and it is included here to illustrate optimisation over continuous attributes. Individual high/low bands representing expected energy supply (+ive) and demand (-ive) over multiple periods for consumers are captured in so-called \textit{flexibility objects} (flexobject). The goal is to assign demand and supply to each flexobject in each period to minimise the energy transferred across all periods. This is represented by the linear programming (LP) problem in \autoref{fig:energy-balance-LP}, where $F$ is the total number of flexobjects and $T$ is the number of periods.
\begin{figure}[!htbp]
\centering
$$\min \sum_{t=1}^{T} \left|\sum_{f=1}^{F} e_{ft}\right| \quad \text{s.t. } eL_{ft} \leq e_{ft} \leq eH_{ft}, \; f = 1,\ldots,F, \; t = 1,\ldots,T$$
\caption{Energy Balance LP Problem}
\label{fig:energy-balance-LP}
\end{figure}
Figures \ref{fig:energy-balance-specification} and \ref{fig:energy-balance-translation} trace the complete example.

Starting with \autoref{fig:energy-balance-specification} our $Base$ relation \codett{FlexObjects} shows data for two flexobjects and four periods. We want to assign an energy assignment for each of these, and \codett{AssignedEnergy} as our $Decision$ relation can capture this. The problem doesn't require any context, and the problem is fully stated in $RA_{sol}$.

\begin{figure}[!htbp]
\centering
\includegraphics[width=0.8\linewidth]{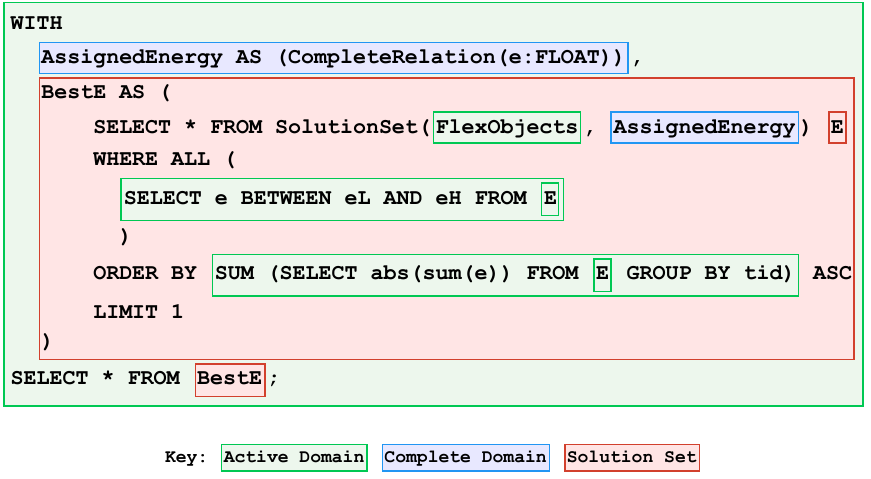}
\caption{Energy Balance Illustration SQL}
\label{fig:energy-balance-sql-sol}
\end{figure}

Proceeding with translation and evaluation, $\Phi$ generates the $symI$ and $flat$ relations as shown in \autoref{fig:energy-balance-translation}. Under the guidance of $\pi_{sol}$, the translation to an intermediate representation generates MiniZinc (in this case), and the output relation \codett{LetsUseE} has a net energy transfer of 2.5 units. The solution is illustrated in SQL (\autoref{fig:energy-balance-sql-sol}).

\newsavebox{\exEBBaseboxSavebox}
\begin{lrbox}{\exEBBaseboxSavebox}
\begin{tabularx}{.4\textwidth}{|>{\raggedright\arraybackslash}X|>{\raggedright\arraybackslash}X|>{\raggedright\arraybackslash}X|>{\raggedright\arraybackslash}X|}
    \hline
    \textbf{fid:\newline \codett{INT}}& \textbf{tid:\newline \codett{INT}} & \textbf{eL:\newline \smaller \codett{FLOAT}} & \textbf{eH:\newline \smaller \codett{FLOAT}}  \\
    1 & 1 & 1.1 & 3.4 \\
    1 & 2 & -4.3 & 1.2 \\
    1 & 3 & 3.2 & 4.7 \\
    2 & 2 & 2.5 & 5.1 \\
    2 & 3 & -3.3 & -1.0 \\
    2 & 4 & 1.4 & 2.2 \\ \hline
\end{tabularx}
\end{lrbox}

\newsavebox{\exEBDecisionboxSavebox}
\begin{lrbox}{\exEBDecisionboxSavebox}
\begin{lstlisting}[mathescape=true, language=RA, backgroundcolor={}]
AssignedEnergy = $\omega$[e: FLOAT](True)
\end{lstlisting}
\end{lrbox}

\newsavebox{\exEBContextboxSavebox}
\begin{lrbox}{\exEBContextboxSavebox}
---
\end{lrbox}

\newsavebox{\exEBRasolboxSavebox}
\begin{lrbox}{\exEBRasolboxSavebox}
\begin{minipage}{7.5cm}
\begin{lstlisting}[language=RA, basicstyle=\normalsize\ttfamily, mathescape=true, backgroundcolor={}, frame=none]
E := $\omega_{sol}$(FlexObjects,AssignedEnergy)
FeasibleE := $\sigma_{sol}$[
    $\gamma$[$\varnothing$][bool_and(eL<=e<=eH) $\rightarrow$ ret](E)
    ](E)
BestE := $\lambda_{sol}$[1](
    $\tau_{sol}$[ASC][
        $\gamma$[$\varnothing$][sum(t) $\rightarrow$ e](
            $\gamma$[tid][abs(sum(e)) $\rightarrow$ t]
              (FeasibleE))
          ](FeasibleE))
LetsUseE := $\pi_{sol}$[$\varnothing$][fid, tid, eL, eH, e](BestE)
\end{lstlisting}
\end{minipage}
\end{lrbox}

\newsavebox{\exEBSymIboxSavebox}
\begin{lrbox}{\exEBSymIboxSavebox}
\begin{tabularx}{0.45\textwidth}{|X|X|X|X|X|}
    \hline
    \textbf{fid:\newline \codett{INT}}& \textbf{tid:\newline \codett{INT}} & \textbf{eL:\newline \smaller \codett{FLOAT}} & \textbf{eH:\newline \smaller \codett{FLOAT}}& \textbf{e:\newline sym} \\            \hline
    1 & 1 & 1.1 & 3.4 & $\langle e1 \rangle$\\ \hline
    1 & 2 & -4.3 & 1.2 & $\langle e2 \rangle$\\ \hline
    1 & 3 & 3.2 & 4.7 & $\langle e3 \rangle$\\ \hline
    2 & 2 & 2.5 & 5.1 & $\langle e4 \rangle$\\ \hline
    2 & 3 & -3.3 & -1.0 & $\langle e5 \rangle$ \\ \hline
    2 & 4 & 1.4 & 2.2 & $\langle e6 \rangle$\\ \hline
\end{tabularx}

\end{lrbox}

\newsavebox{\exEBFlatboxSavebox}
\begin{lrbox}{\exEBFlatboxSavebox}
\begin{minipage}{7.5cm}
\begin{lstlisting}[mathescape=true, language=RA, backgroundcolor={}]
FlatE := $\omega$[e1 FLOAT, e2 FLOAT, e3 FLOAT, e4 FLOAT, e5 FLOAT, e6 FLOAT](
    between(e1, 1.1, 3.4) AND between(e2, -4.3, 1.2)
    AND between(e3, 3.2, 4.7) AND between(e4, 2.5, 5.1)
    AND between(e5, -3.3, -1.0) AND between(e6, 1.4, 2.2)
)
Results = $\pi$[$\varnothing$][*]($\lambda$[1](
    $\tau$[ASC][sum([abs(e1),abs(e2+e4),abs(e3+e5),abs(e6)])](FlatE))
)
\end{lstlisting}
\end{minipage}
\end{lrbox}

\newsavebox{\exEBIntermediateboxSavebox}
\begin{lrbox}{\exEBIntermediateboxSavebox}
\begin{minipage}{6.5cm}
\begin{lstlisting}[language=MiniZinc, backgroundcolor={}]
predicate between(var float: x, float: lower, float: upper) =
    lower <= x /\ x <= upper;

var float: e1;
var float: e2;
var float: e3;
var float: e4;
var float: e5;
var float: e6;

constraint between(e1, 1.1, 3.4);
constraint between(e2, -4.3, 1.2);
constraint between(e3, 3.2, 4.7);
constraint between(e4, 2.5, 5.1);
constraint between(e5, -3.3, -1.0);
constraint between(e6, 1.4, 2.2);

solve minimize sum([abs(e1),abs(e2+e4),abs(e3+e5),abs(e6)]);
\end{lstlisting}
\end{minipage}
\end{lrbox}

\newsavebox{\exEBOutputboxSavebox}
\begin{lrbox}{\exEBOutputboxSavebox}
\begin{tabularx}{0.5\textwidth}{|X|X|X|X|X|}

    \hline
    \textbf{fid:\newline \codett{INT}}& \textbf{tid:\newline \codett{INT}} & \textbf{eL:\newline \smaller \codett{FLOAT}} & \textbf{eH:\newline \smaller \codett{FLOAT}}& \textbf{e:\newline \smaller \codett{FLOAT}} \\            \hline
    1 & 1 & 1.1 & 3.4 & 1.1\\
    1 & 2 & -4.3 & 1.2 & -2.5\\
    1 & 3 & 3.2 & 4.7 & 3.2\\
    2 & 2 & 2.5 & 5.1 & 2.5\\
    2 & 3 & -3.3 & -1.0 & -3.2 \\
    2 & 4 & 1.4 & 2.2 & 1.4\\ \hline
\end{tabularx}
\end{lrbox}

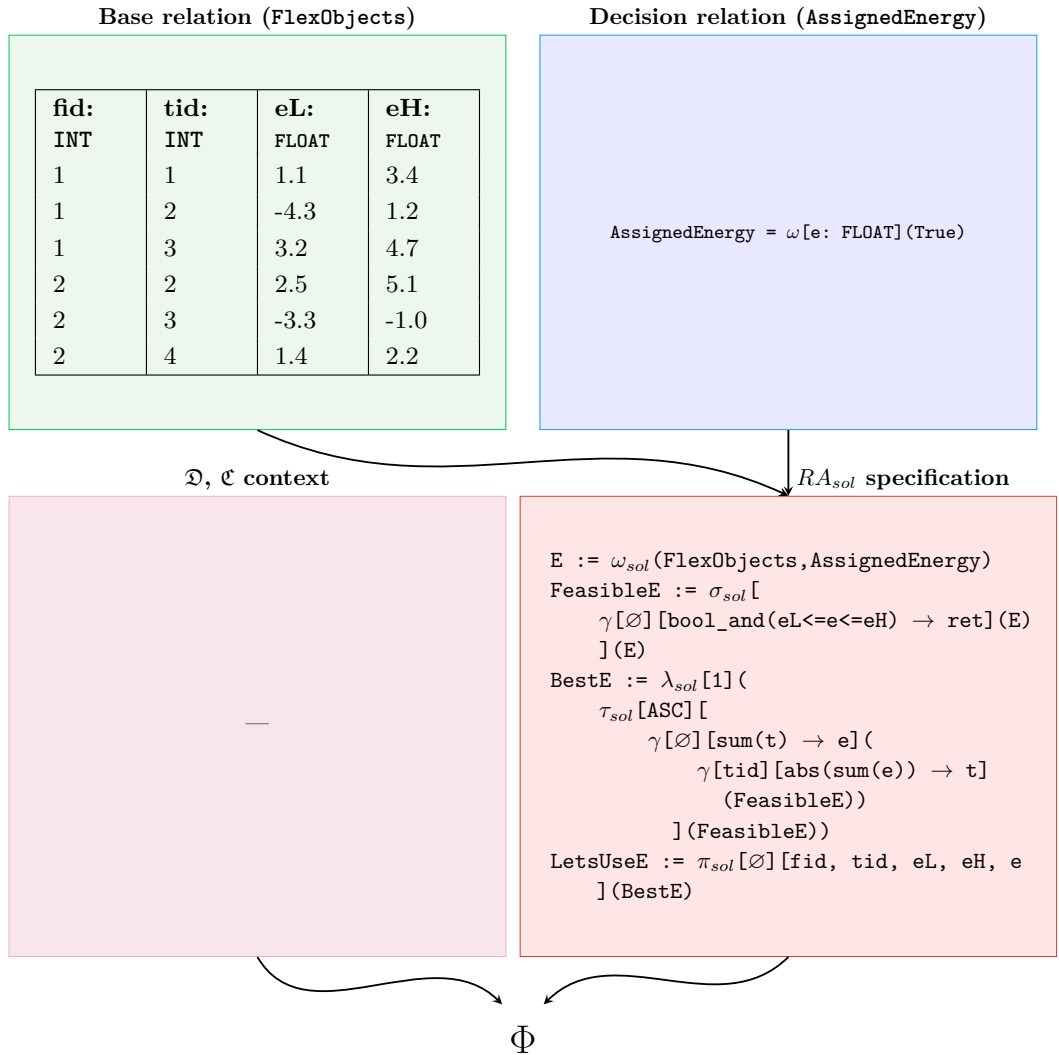
\begin{figure}[!htbp]
    \centering
    \resizebox{\textwidth}{!}{
    \begin{tikzpicture}[
        node distance=5cm,
        every node/.style={minimum height=1.5cm, font=\normalsize},
        basebox/.style={rectangle, draw={rgb,255:red,0;green,200;blue,83}, fill={rgb,255:red,237;green,247;blue,237}, inner sep=10pt, minimum width=7.5cm, minimum height=6cm},
        decisionbox/.style={rectangle, draw={rgb,255:red,33;green,150;blue,243}, fill={rgb,255:red,232;green,232;blue,255}, inner sep=10pt, minimum width=7.5cm, minimum height=6cm},
        contextbox/.style={rectangle, draw=purple!30, fill=purple!10, inner sep=10pt, minimum width=7.5cm, minimum height=7cm},
        rasolbox/.style={rectangle, draw={rgb,255:red,210;green,64;blue,45}, fill={rgb,255:red,255;green,229;blue,229}, inner sep=10pt, minimum width=7.5cm, minimum height=7cm},
        transformbox/.style={ellipse, draw=black!0, fill=gray!0, inner sep=3pt, font=\LARGE\bfseries},
        labelstyleleft/.style={font=\normalsize\bfseries, anchor=south west},
        labelstyleright/.style={font=\normalsize\bfseries, anchor=south east},
        arrow/.style={->, thick, >=stealth}
    ]

    \node[basebox] (base) {
        \scalebox{1.2}{\usebox{\exEBBaseboxSavebox}}
    };
    \node[labelstyleleft, above=-0.5cm and 0cm of base.north] {Base relation (\codett{FlexObjects})};

    \node[decisionbox, right=.5cm of base] (decision) {
        \scalebox{1.0}{\usebox{\exEBDecisionboxSavebox}}
    };
    \node[labelstyleright, above=-0.5cm and 0cm of decision.north] {Decision relation (\codett{AssignedEnergy})};

    \node[contextbox, below=1cm of base] (context) {
        \scalebox{1.0}{\usebox{\exEBContextboxSavebox}}
    };
    \node[labelstyleleft, above =-0.5cm and 0cm of context.north] {$\mathfrak{D}$, $\mathfrak{C}$ context};

    \node[rasolbox, below=1cm of decision] (rasol) {
        \scalebox{1.0}{\usebox{\exEBRasolboxSavebox}}
    };
    \node[labelstyleright, above right=-0.5cm and 0cm of rasol.north] {$RA_{sol}$ specification};

    \node[transformbox, below=0.5cm of $(context.south)!0.5!(rasol.south)$] (phi) {$\Phi$};

    \node[below=0.5cm of phi, font=\normalsize\itshape] {(Continued in Figure \ref{fig:energy-balance-translation})};

    \draw[arrow] (base.south) to[out=-25, in=155] (rasol.north);
    \draw[arrow] (decision.south) to[out=-90, in=90] (rasol.north);
    \draw[arrow] (context.south) to[out=-60, in=135] (phi.north west);
    \draw[arrow] (rasol.south) to[out=-135, in=45] (phi.north east);

    \end{tikzpicture}
    }
    \caption{Energy Balance optimisation (Part A): Problem specification through $RA_{sol}$ leading to homomorphic translation $\Phi$.}
    \label{fig:energy-balance-specification}
\end{figure}

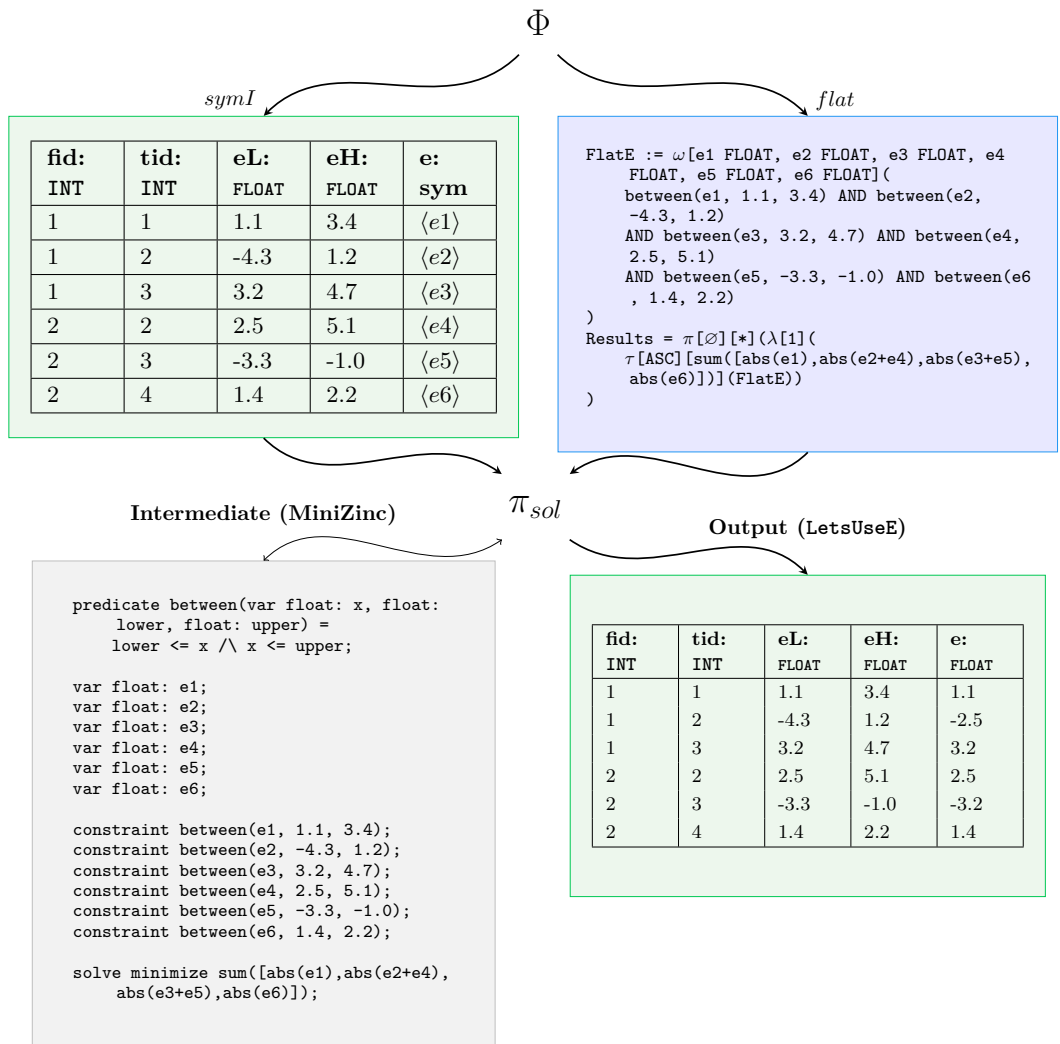
\begin{figure}[!htbp]
    \centering
    \resizebox{\textwidth}{!}{
    \begin{tikzpicture}[
        node distance=5cm,
        every node/.style={minimum height=1.5cm, font=\normalsize},
        transformbox/.style={ellipse, draw=black!0, fill=gray!0, inner sep=3pt, font=\LARGE\bfseries},
        flatbox/.style={rectangle, draw={rgb,255:red,33;green,150;blue,243}, fill={rgb,255:red,232;green,232;blue,255}, inner sep=10pt, minimum width=7.5cm, minimum height=5.25cm},
        symIbox/.style={rectangle, draw={rgb,255:red,0;green,200;blue,83}, fill={rgb,255:red,237;green,247;blue,237}, inner sep=10pt, minimum width=7.5cm, minimum height=5.25cm},
        intermediatebox/.style={rectangle, draw=gray!50, fill=gray!10, inner sep=10pt, minimum width=7.5cm, minimum height=8cm},
        outputbox/.style={rectangle, draw={rgb,255:red,0;green,200;blue,83}, fill={rgb,255:red,237;green,247;blue,237}, inner sep=10pt, minimum width=7.5cm, minimum height=5.25cm},
        labelstyleleft/.style={font=\normalsize\bfseries, anchor=south west},
        labelstyleright/.style={font=\normalsize\bfseries, anchor=south east},
        arrow/.style={->, thick, >=stealth}
    ]

    \node[transformbox] (phi) at (0,0) {$\Phi$};

    \node[above=0.5cm of phi, font=\normalsize\itshape] {(Continued from Figure \ref{fig:energy-balance-specification})};

    \node[symIbox, below left=1cm and 0cm of phi] (symi) {
        \scalebox{1.2}{\usebox{\exEBSymIboxSavebox}}
    };
    \node[labelstyleleft, above left=-0.5cm and 0cm of symi.north] {$symI$};

    \node[flatbox, below right=1cm and 0cm of phi] (flat) {
        \scalebox{1.0}{\usebox{\exEBFlatboxSavebox}}
    };
    \node[labelstyleright, above right=-0.5cm and 0cm of flat.north] {$flat$};

    \node[transformbox, below=.25cm of $(symi.south)!0.5!(flat.south)$] (pisol) {$\pi_{sol}$};

    \node[intermediatebox, below=2cm of symi] (minizinc) {
        \begin{minipage}[c][4.8cm][c]{6.8cm}
        \centering
        \scalebox{1}{\usebox{\exEBIntermediateboxSavebox}}
        \end{minipage}
    };
    \node[labelstyleleft, above =0cm and 0cm of minizinc.north] {Intermediate (MiniZinc)};

    \node[outputbox, below=2cm of flat] (output) {
        \scalebox{1}{\usebox{\exEBOutputboxSavebox}}
    };
    \node[labelstyleright, above =0cm and 0cm of output.north] {Output (\codett{LetsUseE})};

    \draw[arrow] (phi.south west) to[out=-135, in=45] (symi.north);
    \draw[arrow] (phi.south east) to[out=-45, in=135] (flat.north);
    \draw[arrow] (symi.south) to[out=-45, in=145] (pisol.north west);
    \draw[arrow] (flat.south) to[out=-135, in=35] (pisol.north east);
    \draw[<->] (pisol.south west) to[out=-145, in=45] (minizinc.north);
    \draw[arrow] (pisol.south east) to[out=-35, in=135] (output.north);

    \end{tikzpicture}
    }
    \caption{Energy Balance optimisation (Part B): Translation through $\Phi$ to relational algebra, and evaluation via MiniZinc to final solution.}
    \label{fig:energy-balance-translation}
\end{figure}

\FloatBarrier
\subsection{Pareto-optimal Subset Selection} \label{subapp:kursawe}
\begin{figure}[!htbp]
    \centering
    \includegraphics[width=0.8\linewidth]{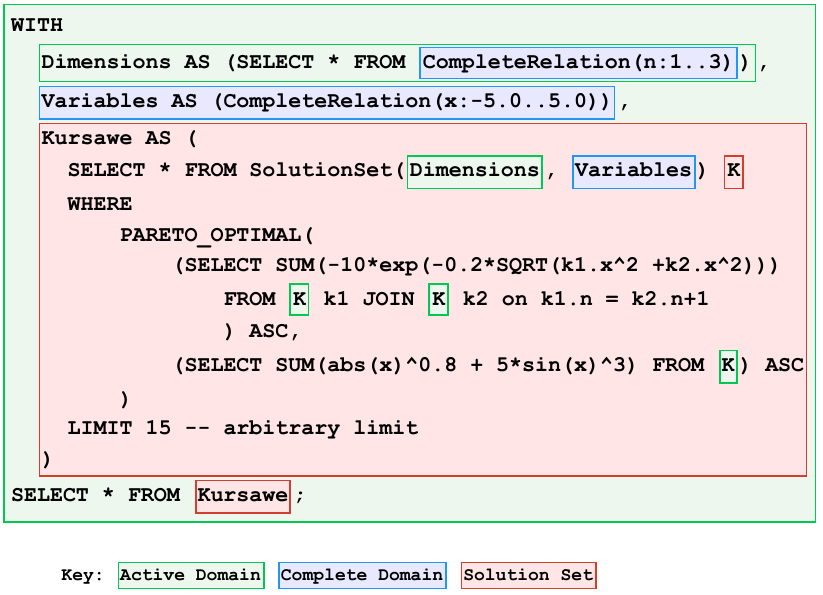}
    \caption{The Kursawe Test Problem Illustration SQL}
    \label{fig:kursawe-sql}
    \end{figure}

Our framework naturally extends to multi-objective problems, like Pareto optimality—demonstrating the separation between declarative specification and evaluation strategy. The Kursawe function~\cite{kursawe_1990_variant.evolution.strategies.vector.optimization} is a classic test problem for Pareto optimal subset selection.  The function may be defined as in \autoref{fig:kursawe}. \autoref{fig:kursawe-sql} is the specification of this problem in polymorphic SQL with $n=3$, ready for an evolutionary or other algorithm to evaluate the \codett{PARETO\_OPTIMAL()} restriction.

\begin{figure}[!htbp]
\centering
\[
\begin{aligned}
&\text{Minimise} \quad f_1(\mathbf{x}) = \sum_{i=1}^{n-1} \left(-10 \exp\left(-0.2\sqrt{x_i^2 + x_{i+1}^2}\right)\right) \\
&\text{Minimise} \quad f_2(\mathbf{x}) = \sum_{i=1}^{n} \left(|x_i|^{0.8} + 5\sin^3(x_i)\right) \\
&\text{subject to} \quad \mathbf{x} \in [-5, 5]^n
\end{aligned}
\]
\caption{The Kursawe test problem for multi-objective optimisation}
\label{fig:kursawe}
\end{figure}

With the Kursawe problem demonstrated, along with the earlier problems, we have shown how the three algebras (active domain relations, complete domain relations, and solution sets) work together to enable the expression of a wide variety of problems in a natural manner, ready for evaluation.

\end{appendix}
\end{document}